\documentclass[journal,twocolumn,10pt,twoside]{IEEEtran}
\normalsize

\ifCLASSINFOpdf
   \usepackage[pdftex]{graphicx}
   \usepackage{epstopdf}
\else
\fi

\usepackage{amsmath}
\usepackage{amssymb}
\usepackage{cite}
\usepackage{graphicx}
\usepackage{algorithm}
\usepackage{algpseudocode}
\usepackage{subfigure}
\usepackage{multirow}
\usepackage{longtable}
\usepackage{threeparttable}
\usepackage{caption3}
\usepackage{array}
\usepackage{cases}
\usepackage{color}
\usepackage[T1]{fontenc}
\usepackage[utf8]{inputenc}
\usepackage{authblk}
\usepackage{subfigmat}
\usepackage{fixltx2e}

\newtheorem{observation}{Observation}

\newcommand{\out}{\mathrm{out}}

\newcolumntype{C}[1]{>{\centering\arraybackslash}m{#1}}
\newcommand{\R}{\textnormal{R}}
\newcommand{\E}{\textnormal{E}}
\newcommand{\dc}{\mathrm{dc}}
\newcommand{\rf}{\mathrm{rf}}
\newcommand{\ant}{\mathrm{ant}}
\usepackage{tabularx}

\begin{document}
\title{Foundations of Wireless Information and Power Transfer: Theory, Prototypes, and Experiments}
\author{
\IEEEauthorblockN{Bruno Clerckx, \textit{Fellow, IEEE}, Junghoon Kim, \textit{Member, IEEE}, Kae Won Choi, \textit{Senior Member, IEEE}, Dong In Kim, \textit{Fellow, IEEE}} 

\thanks{Bruno Clerckx and Junghoon Kim are with the Department of Electrical and Electronic Engineering, Imperial College London, London SW7 2AZ, UK
(email: \{b.clerckx,junghoon.kim15\}@imperial.ac.uk). 

Kae Won Choi and Dong In Kim are with the Department of Electrical and Computer Engineering, Sungkyunkwan University, Suwon 16419, South Korea (e-mail:
\{kaewonchoi,dongin\}@skku.edu).

\par This work has been partially supported by the EPSRC of UK under grant EP/P003885/1 and EP/R511547/1, and partially supported by the National Research Foundation of Korea (NRF) Grant funded by the Korean Government (MSIT) under Grants 2020R1A2C1014693 and 2021R1A2C2007638.}
}

\maketitle

\begin{abstract} As wireless has disrupted communications, wireless will also disrupt the delivery of energy. Future wireless networks will be equipped with (radiative) wireless power transfer (WPT) capability and exploit radio waves to carry both energy and information through a unified wireless information and power transfer (WIPT). Such networks will make the best use of the RF spectrum and radiation as well as the network infrastructure for the dual purpose of communicating and energizing. Consequently those networks will enable trillions of future low-power devices to sense, compute, connect, and energize anywhere, anytime, and on the move. In this paper, we review the foundations of such future system. We first give an overview of the fundamental theoretical building blocks of WPT and WIPT. Then we discuss some state-of-the-art experimental setups and prototypes of both WPT and WIPT and contrast theoretical and experimental results. We draw a special attention to how the integration of RF, signal and system designs in WPT and WIPT leads to new theoretical and experimental design challenges for both microwave and communication engineers and highlight some promising solutions. Topics and experimental testbeds discussed include closed-loop WPT and WIPT architectures with beamforming, waveform, channel acquisition, and single/multi-antenna energy harvester, centralized and distributed WPT, reconfigurable metasurfaces and intelligent surfaces for WPT, transmitter and receiver architecture for WIPT, modulation, rate-energy trade-off. Moreover, we highlight important theoretical and experimental research directions to be addressed for WPT and WIPT to become a foundational technology of future wireless networks.
\end{abstract}
\begin{IEEEkeywords}Wireless power transfer,
wireless information and power transfer, rectenna, harvester, beamforming, reconfigurable intelligent surface, intelligent reflecting surface, waveform, multi-antenna, prototypes, experiments
\end{IEEEkeywords}

\IEEEpeerreviewmaketitle

\section{Introduction}
\IEEEPARstart{W}{ireless} information and power transfer (WIPT) is an emerging paradigm in future Internet of Things (IoT) and wireless sensor networks (WSN). Radiofrequency (RF) waves have
traditionally been used for wireless information transfer (WIT)
or wireless communications. However, radio waves carry
both energy and information. Thanks to the reduction
in the power consumption of electronics and the increasing need to energize a massive number of low-power autonomous devices, far-field/radiative RF wireless power transfer (WPT) has attracted attention as a feasible and promising power supply technology for remotely energizing low-power devices, such as sensors, radio-frequency identification (RFID) tags, and consumer electronic \cite{Popovic:2013,Visser:2013,Clerckx:2018}. WPT has numerous advantages for low-power wireless applications since it eliminates the hassle of connecting cables, is controllable with on-demand energy delivery, preserves the environment by avoiding massive battery disposal, can operate in dangerous (e.g., in hazardous environment) or even impossible (e.g., for biomedical implants) environments, delivers power over long distances, and can be implemented in a small form factor at the receiver compared to other technologies such as inductive coupling or magnetic resonance \cite{Zeng:2017}. In addition, WPT can be integrated with WIT so as to exploit radio waves for the dual purpose of communicating and energizing. 
\par For over 100 years, energy and information have been studied separately and have evolved as two separate fields \cite{Varshney:2008}. WIT was pioneered by Marconi and is currently in its fifth generation (5G), while WPT was pioneered by Tesla but is not even in its first generation (1G) \cite{Clerckx:2018}. This separation has two main consequences. First of all, today's wireless networks blast RF energy in the air for communication but do not make use of it to charge devices. Second, delivering ubiquitous wireless power would require the deployment of a separate network of dedicated power transmitters \cite{Huang_Lau:2014}. 
\par
\begin{figure}
	\centering
	\includegraphics[width=\columnwidth]{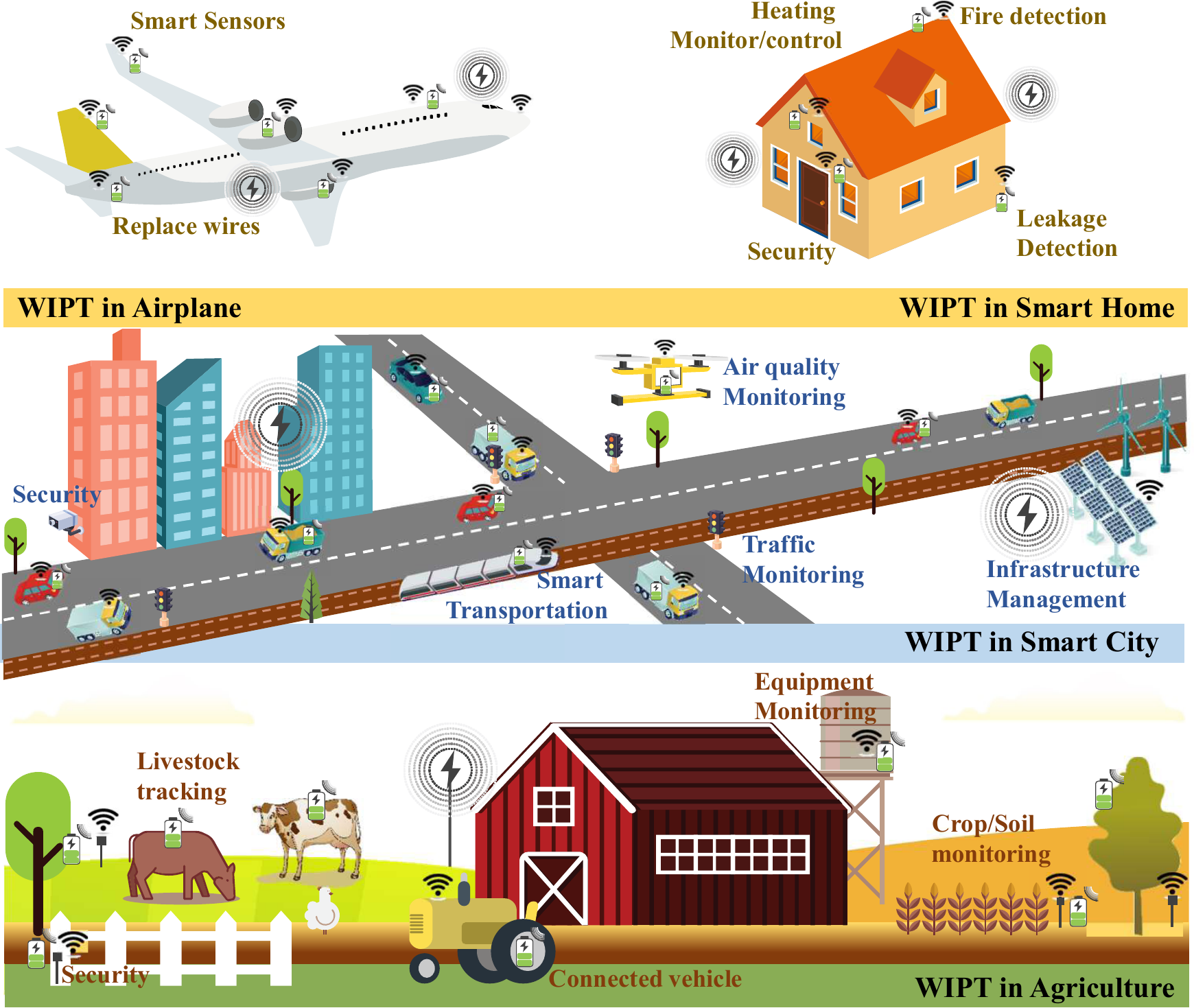}
	\caption{Applications of WPT and WIPT.}
	\label{fig_applications}
\end{figure}

 Imagine instead future networks that reconcile Tesla and Marconi visions into a new and unified network paradigm where radio waves are exploited to the full potential to carry both energy and information. Information and energy would flow hand in hand via the wireless medium. WIT and WPT would refer to two extreme strategies targeting communication and power supply, respectively. A unified WIPT design could smoothly evolve between the two extremes to make the most of the RF spectrum/radiation and the network infrastructure for communications and power delivery, thereby surpassing traditional systems that separate communications and power. Such a capability would be a game changer for future networks such as sixth generation (6G) and beyond. Future network operators will be telecommunication and energy providers and offer a true ubiquitous wireless connectivity such that trillions of future low-power devices will sense, compute, connect, and energize anywhere, anytime, and on the move. Such a capability will also address the growing need for high energy efficiency, green connectivity, and energy-neutral/zero-power IoT. As illustrated in Fig. \ref{fig_applications}, numerous applications will emerge in many sectors such as smart homes (monitoring and detection of fire, carbon monoxide, leak, moisture, temperature, motion), smart city (management of infrastructure, air quality, traffic, parking, waste), healthcare (in-body, around the body), transportation (connected car, vehicle-to-vehicle), automotive and airline (replacing cables), agriculture (forest/soil/crops monitoring), logistics, emergency, security, prevention, defence, etc.

\par Such vision comes with numerous engineering requirements and design challenges \cite{Zeng:2017,Clerckx:2019,Clerckx:2021}: 1) Charging range of 5-100 meters (m) in indoor and outdoor deployments; 2) End-to-end power transfer efficiency of up to a fraction of a percent to a few percents; 3) Operate in line of sight (LoS) and non-line of sight (NLoS) to broaden the applications of WPT-aided networks; 4) Support mobility at pedestrian speed; 5) Deliver power ubiquitously within the network coverage area; 6) Ensure WPT is safe; 7) Reduce the energy consumption of wireless powered devices; 8) Unify communication and power into WIPT; 9) Integrate WPT with sensing, computing and communication in 6G and beyond networks. 

\par To address those challenges, integrated efforts across societies must be made, including circuit, sensors, systems, antenna, propagation, communication, microwave design, signal processing, information theory, machine learning, sensing, and computing.

\par Challenges (1)--(7) are studied in various review papers emphasizing separately RF, circuit and antenna solutions \cite{Hemour:2014,OptBehaviour,Valenta:2014,Costanzo:2016}, communications, signal and system design solutions \cite{Zeng:2017,Clerckx:2021}, as well as integrated solutions bridging RF, signal and system designs to get a better grasp of the basic building blocks of an efficient WPT architecture \cite{Clerckx:2018}. This integration has led to the noticeable and welcomed trend to depart from naive and oversimplified models and has triggered the need to design efficient WPT signal strategies accounting for the various sources of nonlinearity (transmitter and energy harvester) in the system \cite{Clerckx:2021}.

\par Challenge (8) is reviewed in \cite{Clerckx:2021,Krikidis:2014,Ding:2015,Lu:2015,Clerckx:2019} with the objective to lay the fundamentals of WIPT from energy harvester models to signal and system designs. This challenge requires the characterization of the trade-off between the amount of information and the amount of energy that can be delivered in a wireless network and the design of (single-user and multi-user) signal strategies to achieve this trade-off. Importantly, \cite{Clerckx:2019} demonstrated how crucially the trade-off, the signaling and resource allocation strategy depend on the energy harvester model and related nonlinearity. 

\par Challenge (9) is reviewed in \cite{Clerckx:2021} where it is shown that WPT, sensing, computing, edge learning, and communication need to be jointly controlled so as to optimize the efficiency of a system supporting specific applications. In particular, there exist trade-offs between transferred energy and energy consumption of sensing/computing and communication. Quantifying and exploiting such trade-offs can substantially improve system performance.

\par Various methodologies have also been used to tackle the above challenges and optimize the system, including the model-based optimization approach and the data-driven machine learning approach. The pros and cons of each approach, especially when accounting for various nonlinearities in wireless-powered networks, and interesting emerging opportunities for the approaches to complement each other have been discussed and identified in \cite{Clerckx:2021}.

\par This paper complements the existing literature and contributes to the literature by discussing one key missing and timely piece of the puzzle, namely real-world prototype and experimentation to validate theory. This paper gives an overview\footnote{This is by no means a comprehensive review of the literature.} on the fundamental theoretical building blocks of WPT and WIPT, discusses some state-of-the-art experimental setups and prototypes of both WPT and WIPT, and contrasts theoretical and experimental results. We draw a special attention to the integration of RF, signal and system designs in modern WPT and WIPT. We stress how such integration leads to new theoretical and experimental design challenges for both microwave and communication engineers and highlight some promising solutions. Topics and experimental testbeds discussed include closed-loop WPT and WIPT architectures with beamforming, waveform, channel acquisition, and single/multi-antenna energy harvester, centralized and distributed WPT, reconfigurable metasurfaces and intelligent surfaces for WPT, transmitter and receiver architecture for WIPT, rate-energy trade-off. Moreover, we highlight important theoretical and experimental research directions to be addressed for WPT and WIPT to become a foundational technology of future wireless networks.

\emph{Organization:} Section \ref{WPT_section} discusses WPT architecture, prototypes and experiments. Section \ref{Section_WIPT} builds upon the WPT section to design, prototype and experiment WIPT. Section \ref{conclusions} draws some conclusions and discusses promising future research directions.

\emph{Notation:} In this paper, $j$ denotes the imaginary unit, i.e., $j^2=-1$. Scalars are denoted by italic letters. Vectors and matrices are denoted by boldface lower- and upper-case letters, respectively. $\mathbb{E}[\cdot]$ denotes statistical expectation or direct current (DC) averaging and $\Re\{\cdot\}$ represents the real part of a complex number. $|.|$ and $\left\|.\right\|$ refer to the absolute value of a scalar and the 2-norm of a vector. The complex conjugate, transpose, Hermitian transpose, and Frobenius norm of an arbitrary-size matrix $\mathbf{A}$ are denoted as $\mathbf A^*$, $\mathbf{A}^{T}$, $\mathbf{A}^{H}$, and $\|\mathbf{A}\|_F$, respectively.

\section{Wireless Power Transfer}\label{WPT_section}
\subsection{Architecture} \label{modern_WPT_subsection}

\par
\begin{figure}
	\centering
	\includegraphics[width=0.9\columnwidth]{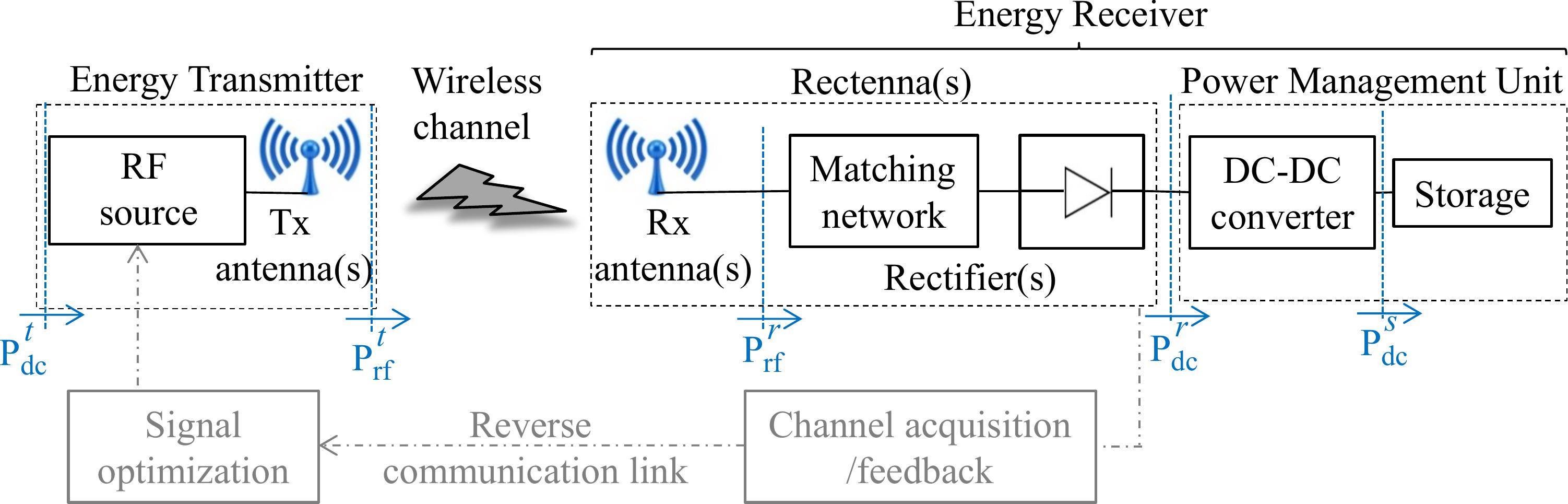}
	\caption{A block diagram of a modern closed-loop and adaptive WPT architecture.}
	\label{fig_sys}
\end{figure}

 Fig. \ref{fig_sys} shows a modern WPT system architecture where an RF energy transmitter (ET) is separated from an energy receiver (ER) by the wireless channel. A signal generated by a DC power source is upconverted to the RF domain at the ET, is then broadcasted over the air by one or multiple transmit antennas, and is finally collected at an ER in the RF domain before being converted to DC. The ER consists of one or multiple receive antennas combined with one or multiple rectifiers (the combination of antenna and rectifier being denoted as rectenna) and a power management unit (PMU). Since the electronics conventionally require a DC power source, a rectifier is needed to convert RF to DC. 

A low power device is then directly supplied from the recovered DC power. The recovered DC power can also be stored in a battery or in a super capacitor for high power low duty-cycle operations or be fed to a DC-to-DC converter before being stored. Importantly, the ET and ER in the WPT system of Fig. \ref{fig_sys} can be fully optimized/adjusted in a closed-loop and adaptive manner as a function of the wireless channel. Such adaptation requires a channel acquisition/feedback module and a signal optimization module. Therefore, WPT offers full control of the design to maximize as much as possible the end-to-end power transfer efficiency defined as
\begin{align}\label{e_equation}
e=\frac{P_{\dc}^s}{P_{\dc}^t}=\underbrace{\frac{P_{\rf}^t}{P_{\dc}^t}}_{e_1}\underbrace{\frac{P_{\rf}^r}{P_{\rf}^t}}_{e_2}\underbrace{\frac{P_{\dc}^r}{P_{\rf}^r}}_{e_3}\underbrace{\frac{P_{\dc}^s}{P_{\dc}^r}}_{e_4},
\end{align}
where $e_1$, $e_2$, $e_3$, and $e_4$ denote the DC-to-RF, RF-to-RF, RF-to-DC, and DC-to-DC power conversion efficiency, respectively. 

\subsubsection{Energy Transmitter}
\par We consider a \textit{multi-antenna} ET equipped with $M$ antennas (with a single antenna being a special case) that transmits an unmodulated multisine waveform over multiple $N$ frequencies (with a single frequency being a special case)  $f_n$, $n=0,...,N-1$. The frequencies $f_n=f_0+n \Delta_f$ are evenly spaced by an inter-carrier frequency spacing $\Delta_f$. The transmit signal spanning all $M$ transmit antennas can be written in a vector form as
\begin{equation}
\mathbf{x}_{\rf}(t)=\sqrt{2}\Re\left\{\sum_{n=0}^{N-1} \mathbf{x}_{n} e^{j 2\pi f_n t}\right\}\label{WPT_1}
\end{equation}
where $\mathbf{x}_{n}\triangleq\big[x_{1,n},...,x_{M,n}\big]^T$ denotes the weight vector across the $M$ antennas at frequency $f_n$, with each complex weight scalar $x_{m,n}=s_{m,n}e^{j \phi_{m,n}}$ used to adjust the magnitude $s_{m,n}$ and the phase $\phi_{m,n}$ of the sinewave on antenna $m$ and frequency $f_n$. Such adjustement is controlled by the signal optimization module of Fig. \ref{fig_sys} so as to benefit from a \textit{joint waveform and active beamforming} gain \cite{Clerckx:2016b}. The total average transmit power is expressed as $P_{\rf}^t=\sum_{m=1}^{M}\sum_{n=0}^{N-1}s_{m,n}^2$ and is subject to the constraint $P_{\rf}^t\leq P$. In order to operate at high conversion efficiency $e_1$, efficient high power amplifiers (HPA) have to be used and the transmit signal \eqref{WPT_1} may have to be adjusted by the signal optimization module to satisfy constraints on peak-to-average power ratio (PAPR).

\subsubsection{Wireless Channel}\label{channel}
\par The transmit WPT signal propagates through a multipath channel, characterized by $L$ paths, each path $l$ with a delay $\tau_l$, an amplitude gain $\alpha_l$, and a phase shift $\zeta_{q,m,n,l}$ between transmit antenna $m$ and receive antenna $q$ at frequency $f_n$. The signal at antenna $q$ is then written as 
\begin{equation}
y_{\rf,q}(t) =  \sqrt{2}  \Re \left\{  \sum_{n=0}^{N-1} \mathbf{h}_{q,n} \mathbf x_{n} e^{j 2\pi f_n t}  \right\}+w_{\textnormal{A},q}(t), \label{eq:yt}
\end{equation}
where the antenna noise is denoted as $w_{\textnormal{A},q}(t)$, the channel vector from the $M$ transmit antennas to receive antenna $q$ is denoted as $\mathbf{h}_{q,n}\!\triangleq\!\big[h_{q,1,n},...,h_{q,M,n}\big]$, and the baseband channel frequency response between transmit antenna $m$ and receive antenna $q$ at frequency $f_n$ is written as $h_{q,m,n}=\!\sum_{l=0}^{L-1}\alpha_l e^{j(-2\pi f_n\tau_l+\zeta_{q,m,n,l})}$.

\par Conventionally, we have no control over the wireless channel vector $\mathbf{h}_{q,n}$, i.e., it is dictated by the environment between the transmitter and the receiver. Nevertheless, with the advances in \textit{reconfigurable intelligent surfaces} (RIS) \cite{Huang:2019,Basar:2019} and \textit{intelligent reflecting surfaces} (IRS) \cite{Wu:2019}, one can envision that the wireless channel can be engineered in the future\footnote{RIS and IRS refer to the same thing under different terminologies and can be used interchangeably. For simplicity, we use RIS in the sequel}. RIS is obtained by integrating in the propagation environment a large number of $R$ reconfigurable passive elements that do not require any RF chain. By collaboratively adjusting the impedance of all those passive elements (through the signal optimization module in Fig. \ref{fig_sys}), the transmitted and reflected signals add coherently at the desired receiver, therefore enabling an additional \textit{passive beamforming} gain and increasing the received RF signal power  \cite{Wu:2019,Wu:2019b}. Thanks to the passive structure, RIS is appealing for its relatively low power consumption, light weight, conformal geometry, low profile, low cost, and absence of additive thermal noise during the reflection. 
\par With RIS-aided WPT, the wireless channel vector between the multi-antenna transmitter and receive antenna $q$ on frequency $f_n$ can be written as $\mathbf{h}_{q,n}=\mathbf{g}_{\mathrm{d},q,n}+\mathbf{g}_{\mathrm{r},q,n}\mathbf{\Theta}\mathbf{G}_{\mathrm{i},n}$ where the direct channel between the ET and antenna $q$ of ER is denoted as $\mathbf{g}_{\mathrm{d},q,n}$, the $1\times R$ vector channel between the RIS (equipped with $R$ elements) and antenna $q$ of ER is denoted as $\mathbf{g}_{\mathrm{r},q,n}$, and the $L\times M$ matrix channel between the ET and the RIS is denoted as $\mathbf{G}_{\mathrm{i},n}$. Finally, the scattering matrix of the $R$-port reconfigurable impedance network\footnote{Note that $\mathbf{\Theta}$ is assumed constant across frequencies. More involved designs could enable frequency-dependent scattering matrices $\mathbf{\Theta}_n$.} is denoted as $\mathbf{\Theta}$ and has to satisfy the constraints $\mathbf{\Theta}=\mathbf{\Theta}^T$ and $\mathbf{\Theta}^H\mathbf{\Theta}=\mathbf{I}_L$ \cite{Shen:2020d}. The $R$-port reconfigurable impedance network is constructed with reconfigurable and passive elements that reflect the incident signal in the direction of interest and effectively engineer the wireless channel. 

\subsubsection{Energy Receiver}\label{EH_section}
\par ER can be equipped with a single or multiple antennas, therefore enabling a multiple input-single output (MISO) and multiple input-multiple output (MIMO) WPT, respectively. In both cases, the essential building block of the ER is the \textit{rectenna} \cite{Brown:1984}. 
\par
\begin{figure}
\centering
\subfigure[Single diode rectifier]{\includegraphics[width=0.8\columnwidth]{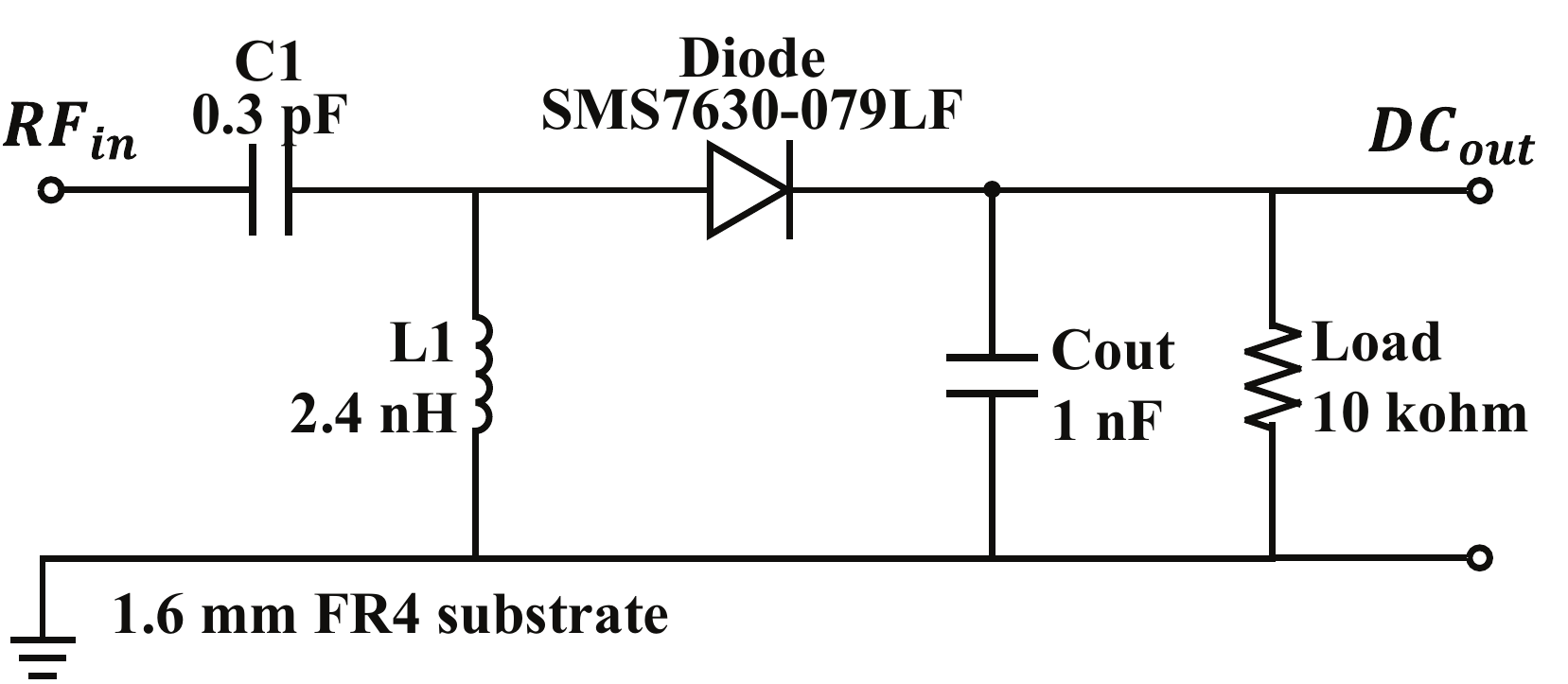}}
\subfigure[Voltage doubler rectifier]{\includegraphics[width=0.8\columnwidth]{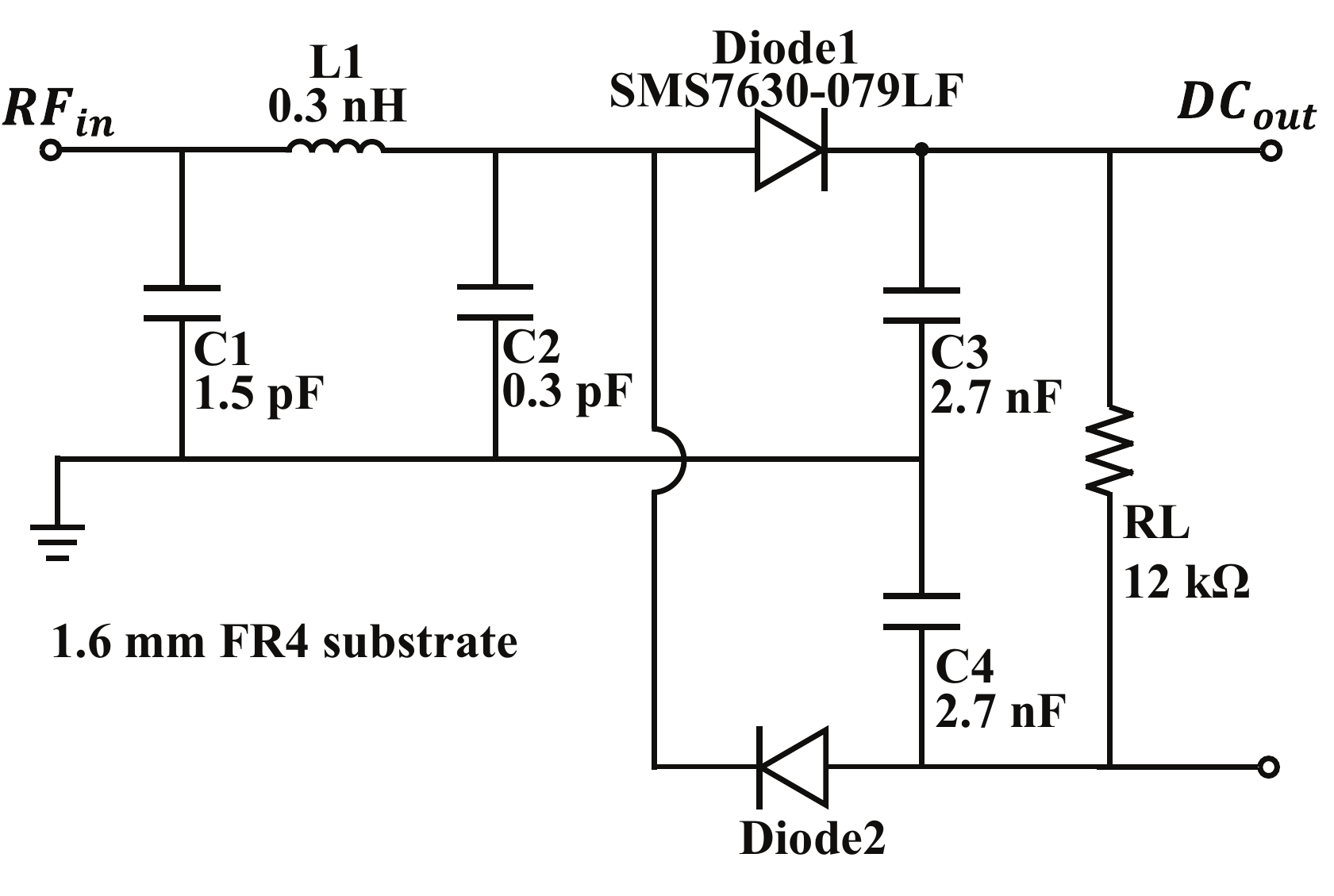}}
\caption{(a) Single series rectifier suited for an average RF input power of -20dBm (10$\mu$W) at 2.45GHz. $v_{\mathrm{s}}$ is the voltage source of the antenna. R1 denotes the antenna impedance. C1 and L1 together form the matching network. SMS-7630 denotes the type of Schottky diode. C\textsubscript{out} and Load together act as the low-pass filter with Load being the output load. (b) Multiple diode rectifier.}
\label{TD_schematic}
\end{figure}

 Let us start with the \textit{single-antenna} case. A rectenna first harvests electromagnetic energy, then rectifies and finally low-pass-filters it. The input power level and input waveforms as well as the operating frequencies influence the design and optimization of the rectenna. Rectifiers are commonly made of a single and multiple (likely Schottky) diodes \cite{Hemour:2014,OptBehaviour,Valenta:2014,Costanzo:2016}. Fig. \ref{TD_schematic} illustrates two different rectifier topologies. The simplest form of rectifier, so-called single diode rectifier, is made of a matching network used to match the antenna impedance to the input impedance of the rectifier, a single diode and a low-pass filter (LPF). It is the general understanding that single-diode rectifiers are more suited for low input power and multiple-diodes rectifiers for higher power.

\begin{figure}
\centerline{\includegraphics[width=0.9\columnwidth]{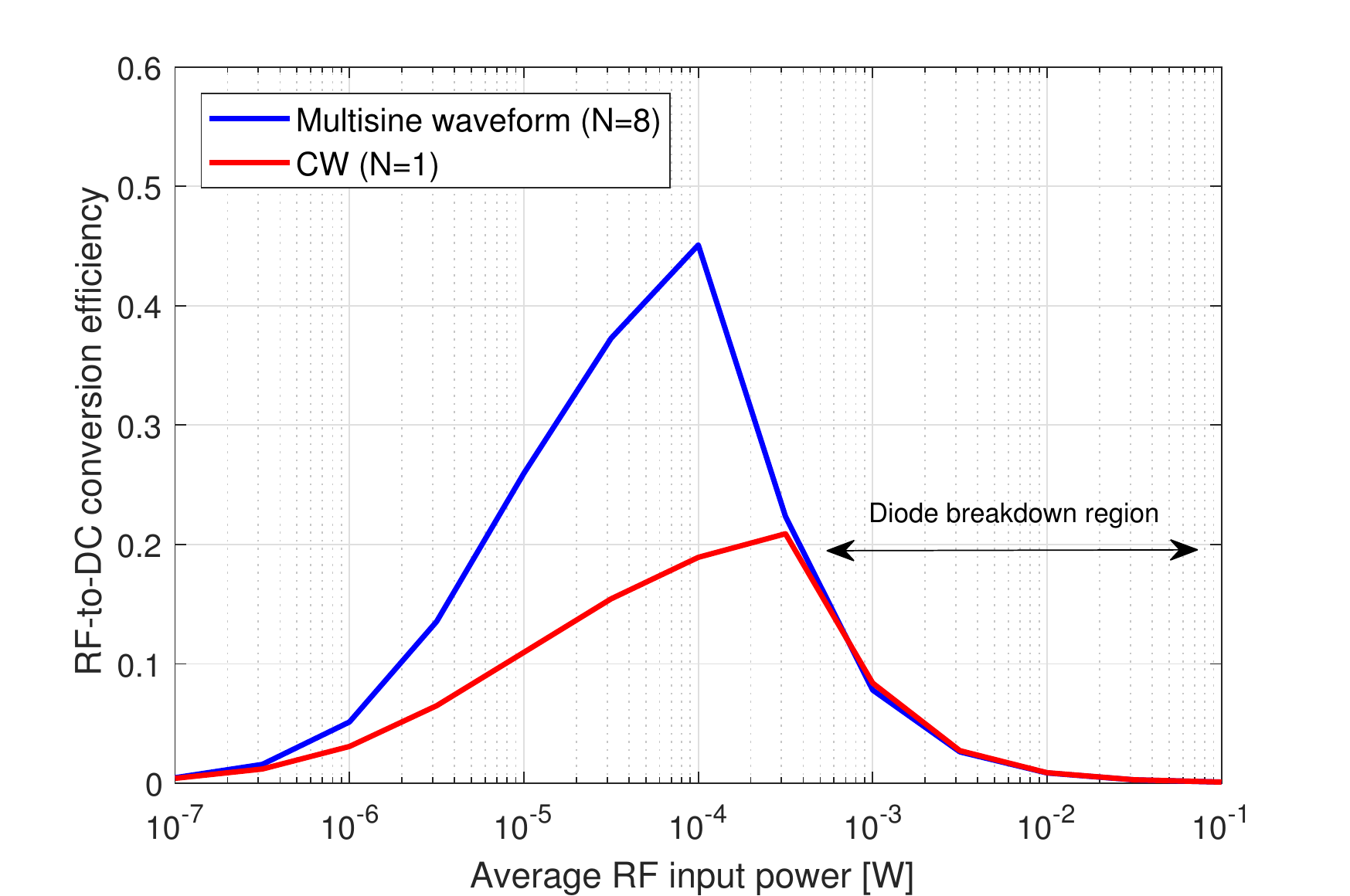}}
  \caption{RF-to-DC conversion efficiency $e_3$ as a function of the average RF input power $P_{\rf}^{r}$ with the single-series rectifier of Fig. \ref{TD_schematic}(a) \cite{Clerckx:2018c}.}
  \label{Pdc_Pin}
\end{figure}
\par Fig. \ref{Pdc_Pin} illustrates the dependency of the RF-to-DC conversion efficiency $e_3$ to the average RF signal power and shape at the rectifier input, when a single diode rectifier (as in Fig. \ref{TD_schematic}(a)) is excited by a continuous wave (CW), i.e., a single sinewave ($N=1$), and a multisine waveform (with uniform power allocation across $N=8$ equispaced carrier frequencies)  \cite{Clerckx:2018c}. The same average RF input power $P_{\rf}^r$ is assumed for those two types of waveforms, but their shape is obviously different. We notice that $e_3$ is low at low input power for both types of excitation due to the rectifier sensitivity to turn on at low input power. Nevertheless, $e_3$ is larger with the multisine waveform than with CW since it conveys bursts of energy that help turning on the diode and increase the rectifier sensitivity in the low power regime. Importantly, for both CW or multisine, $e_3$ increases with $P_{\rf}^r$ whenever the breakdown region is not reached. Beyond a few hundreds of $\mu$W input power, irrespectively of the input waveform, the rectifier enters the diode breakdown region\footnote{The diode SMS-7630 becomes reverse biased at $P_{\rf}^r \approx 500\mu$W to 1mW for CW.}. Consequently the output DC power saturates and $e_3$ significantly drops. To operate beyond such input power and eliminate the saturation problem, multiple-diode rectifier is recommended  \cite{OptBehaviour,Costanzo:2016,Sun:2013}.
\par We observe from Fig. \ref{Pdc_Pin} that $e_3$ is not a constant, but rather depends on 1) the input power level and 2) the shape of the input signal $y_{\rf}$ \cite{OptBehaviour,Trotter:2009,Collado:2014,Valenta:2015,Clerckx:2016b}. This can be reflected mathematically by writing the output DC voltage as $v_{\out}=f_{\mathrm{EH}}\left(y_{\rf}(t)\right)$ with $f_{\mathrm{EH}}\left(y_{\rf}(t)\right)$, a nonlinear function of $y_{\rf}(t)$. Since the harvested DC power $P_{\dc}^r$ is related to $v_{\out}$ through the relationship $P_{\dc}^r=\frac{v_{\out}^2}{R_{\mathrm{L}}}$ (with $R_{\mathrm{L}}$ denoting the output load), $P_{\dc}^r$ is a nonlinear function of $y_{\rf}(t)$. We can therefore write $P_{\dc}^r=e_3\left(y_{\rf}(t)\right)P_{\rf}^r$ to stress that $e_3$ is not a constant but a nonlinear function of $y_{\rf}(t)$. 
The function $f_{\mathrm{EH}}$ can be further investigated by looking at the diode I-V characteristics and taking a polynomial (Taylor) expansion truncated to the fourth order. We then can approximate the output DC voltage of the rectifier $v_{\out}$ across the load (denoted as $R_{\mathrm{L}}$) in Fig. \ref{TD_schematic} as the following nonlinear function of $y_{\rf}(t)$
\begin{equation}\label{vout_def}
v_{\out}=f_{\mathrm{EH}}\left(y_{\rf}(t)\right)= \beta_2 \mathbb{E}\left[y_{\rf}(t)^2\right]+\beta_4 \mathbb{E}\left[y_{\rf}(t)^4\right]
\end{equation}
where $\beta_i=\frac{R_{\ant}^{i/2}}{i!\left(n v_{\mathrm{t}}\right)^{(i-1)}}$ with $R_{\ant}$ the antenna impedance, $v_{\mathrm{t}}$ the thermal voltage and $n$ the ideality factor \cite{Clerckx:2016b}. The operator $\mathbb{E}[\cdot]$ in \eqref{vout_def} takes the DC component of the argument.

\par We note that \eqref{vout_def} is a function of the input signal average power $P_{\rf}^r=\mathbb{E}\left[y_{\rf}(t)^2\right]$ (i.e., the second moment of $y_{\rf}(t)$) but also of its higher moments $\mathbb{E}\left[y_{\rf}(t)^4\right]$. The dependency on the second and higher moments of $y_{\rf}(t)$ explains the results of Fig.\ \ref{Pdc_Pin} for input power levels below the diode breakdown region \cite{Clerckx:2016b}. Indeed, since the I-V characteristics and the polynomial expansion in \eqref{vout_def} are convex, we can apply Jensen's inequality and write 
\begin{equation}\label{Jensen}
\mathbb{E}\left[y_{\rf}(t)^i\right]\geq (\mathbb{E}\left[y_{\rf}(t)^{2}\right])^{\frac{i}{2}} = \left(P_{\rf}^r\right)^{\frac{i}{2}}
\end{equation}
for $i=2,4$, so that
\begin{equation}\label{effect_Pin}
v_{\out}\geq \beta_2 P_{\rf}^r+\beta_4 \left(P_{\rf}^r\right)^{2}.
\end{equation}
From \eqref{effect_Pin}, we can conclude that $e_3$ is an
increasing function of $P_{\rf}^r$. If $y_{\rf}(t)$ is chosen as a multisine waveform with average power $P_{\rf}^r$ uniformly distributed across the $N$ sinewaves, $\mathbb{E}\left[y_{\rf}(t)^4\right]$ can be shown to scale proportionally to $N \left(P_{\rf}^r\right)^{2}$, therefore demonstrating that $\mathbb{E}\left[y_{\rf}(t)^4\right]>\left(P_{\rf}^r\right)^{2}$ for sufficiently large $N$. This explains analytically why multisine (and other types of signals) can outperform CW ($N=1$) \cite{Clerckx:2016b}. Two interesting observations can be drawn from \eqref{Jensen} and \eqref{effect_Pin}, respectively \cite{Clerckx:2021}. 
\begin{observation}\label{higher_order} Two input signals may have the same $\mathbb{E}\left[y_{\rf}(t)^{2}\right]=P_{\rf}^r$ but different $\mathbb{E}\left[y_{\rf}(t)^4\right]$. From \eqref{Jensen}, input signals with large $\mathbb{E}\left[y_{\rf}(t)^i\right]$ are preferred to boost $e_3$ and $P_{\dc}^r$. This explains mathematically the dependence of $e_3$ on the shape of the input signal in Fig.\ \ref{Pdc_Pin}.  
\end{observation}
\begin{observation}\label{higher_input_power} The lower bound \eqref{effect_Pin} highlights that $e_3$ increases with $P_{\rf}^r$. This explains why $e_3$ depends on the input power level in Fig.\ \ref{Pdc_Pin} whenever the rectifier is not in breakdown, and stresses that the strategy that maximizes $P_{\rf}^r$ only maximizes a lower bound on $P_{\dc}^r$ and therefore does not maximize $P_{\dc}^r$ itself.
\end{observation}
\par It is important to note that emphasis is here put on conditions where the rectifier is not in breakdown (i.e., below saturation) since this is the proper operating regime of a rectifier. Experimental results in Section \ref{proto_WPT} also confirm that rectifiers in real wireless deployments are unlikely to enter breakdown whenever designed for the expected range of power levels.
\par
\begin{figure}[t]
	\centering
	\subfigure[]{\includegraphics[width=0.22\textwidth]{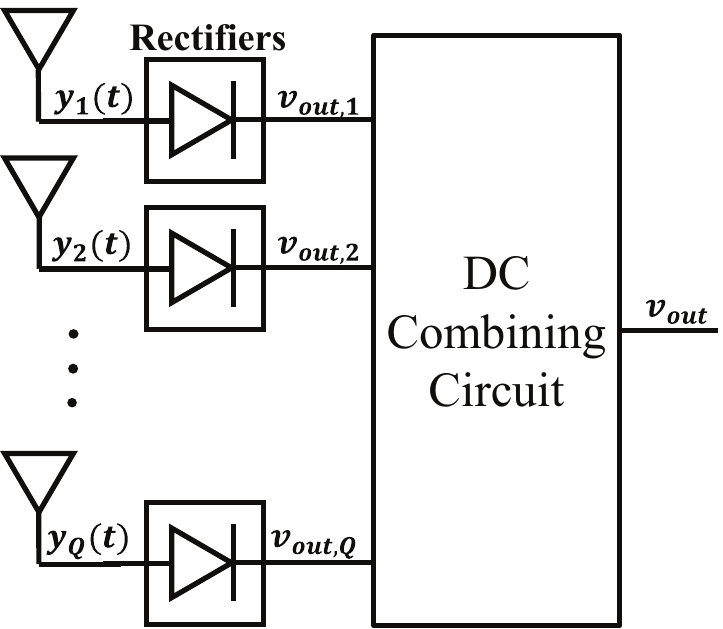}}
	\subfigure[]{\includegraphics[width=0.22\textwidth]{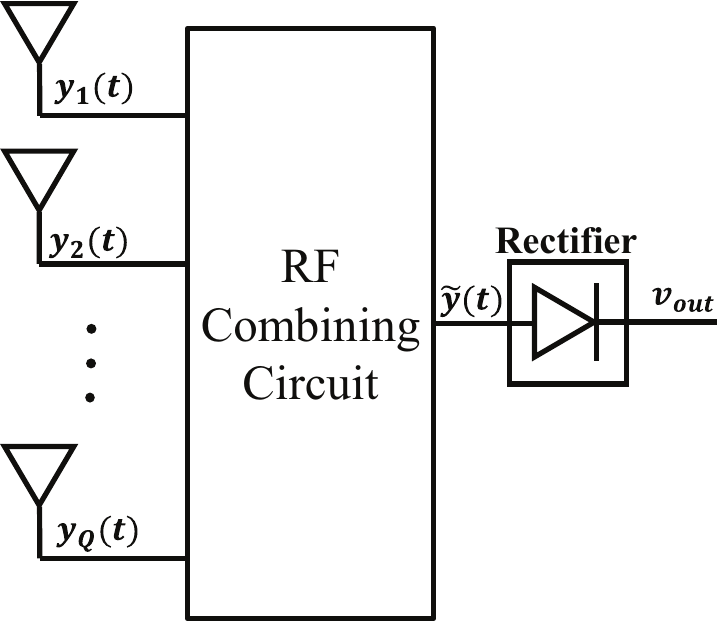}}
	\caption{Two types of multi-antenna energy receivers (a) DC combining receiver (b) RF combining receiver.}
\label{fig_mimo}
\end{figure}

 In \textit{multi-antenna} ER, two main combining strategies (and combination thereof) can be used, namely \textit{DC combining} and \textit{RF combining} as per Fig.\ \ref{fig_mimo} \cite{Shinohara:1998,Hagerty:2004,Olgun:2011,Shen:2020a}. Such combiners open the door to MIMO WPT equipped with multi-antenna arrays at both the ET and ER.

\par In DC combining, there is one rectifier per receive antenna and the number of rectifiers therefore increases with the number of receive antennas $Q$. In contrast, in RF combining, a single rectifier is used to rectify a single combined RF signal obtained by the combination in the RF domain of all the RF signals from all receive antennas. With the DC combiner architecture, $P_{\dc}^r=\sum_{q=1}^Q\frac{v_{\out,q}^2}{R_{\mathrm{L}}}$
with $v_{\out,q}=f_{\mathrm{EH}}\left(y_{\rf,q}(t)\right)$ the output DC voltage of the rectifier connected to receive antenna $q$. With the RF combiner architecture, an  analogue (potentially constant modulus) combiner\footnote{Frequency-dependent analogue combiner $\mathbf{w}_{\mathrm{R},n}$ is also possible but has higher complexity.} $\mathbf{w}_{\mathrm{R}}$ is applied to the received signals \eqref{eq:yt} such that the combined RF signal fed to the single rectifier is given by
\begin{equation}
\tilde{y}(t) =  \sqrt{2}  \Re \left\{  \sum_{n=0}^{N-1} \mathbf{w}_{\mathrm{R}}^H \mathbf H_n \mathbf x_{n}(t) e^{j 2\pi f_n t}  \right\}+\tilde{w}_{\textnormal{A}}(t),
\end{equation}
with the effective combined noise denoted as $\tilde{w}_{\textnormal{A}}$. Due to the passive nature of the RF combining circuit, the RF combining circuit output power cannot be larger than its input power. This is reflected mathematically in the design of the combiner by the constraint $\left\|\mathbf{w}_{\mathrm{R}}\right\|^2\leq 1$.
Finally, the output DC power is given by $P_{\dc}^r=\frac{v_{\out}^2}{R_{\mathrm{L}}}$ where $v_{\out}=f_{\mathrm{EH}}\left(\tilde{y}(t)\right)$. 

\par Notably, since higher RF input power level leads to higher $e_3$ (recall Observation \ref{higher_input_power} and Fig.\ \ref{Pdc_Pin} for power levels below saturation) and the rectifier in RF combining operates on a higher RF power input signal compared to DC combining, RF combining theoretically achieves a higher RF-to-DC conversion efficiency than DC combining. Expressed differently, $P_{\dc}^{r}$ increases with $Q$, but the increase is more significant with RF combining than DC combining. Hence, in MIMO WPT, increasing $M$ and $Q$ helps increasing $P_{\rf}^{r}$ and therefore $e_2$, but a proper choice of the receiver combiner can provide a further boost in $e_3$ \cite{Shen:2020a}. 

\par Following the rectenna, a \textit{DC-to-DC switching converters} is used to dynamically track the rectifier’s optimum load and enhance the DC-to-DC conversion efficiency $e_4$. Indeed, such converter dynamically tracks the maximum power point condition so as to convert the output DC voltage $v_{\out}$ of the rectenna to a higher DC voltage level required by the specific application in use \cite{Dolgov:2010,Costanzo:2012}.
Due to the variable load on the rectenna, changes in
diode impedance with power level, and the rectifier
nonlinearity, the rectifier input impedance becomes
variable, which makes matching (as well as any joint optimization of the matching and
load) difficult \cite{Bolos:2016}. Nevertheless, multisine signals lead to more efficient DC-to-DC voltage boost \cite{Ouda:2018} and benefit from \textit{configurable} DC-to-DC converter to trace the optimal efficiencies as a function of the input waveform shape and power level \cite{Ouda:2019}. 

\subsubsection{Signal Optimization}

\par The above discussion sheds a new technical challenge and opportunity in modern WPT (and consequently in WIPT) system design. WPT system should not be designed by maximizing $e_1$, $e_2$, $e_3$, $e_4$ independently from each other as such approach would not maximize $e$. Indeed, since $e_1$, $e_2$, $e_3$, $e_4$ are coupled due to rectenna and HPA nonlinearities, the transmit signal and the channel state influence the input signal shape and power to the rectifier and therefore $e_3$ but also influence $e_2$ and $e_1$. This motivates the design of new and efficient signal optimization and channel acquisitions modules to adapt as a function of the wireless channel state, the transmit signals, the RIS and the receive combiner. Such entire link optimization can maximize $e$ while accounting for the nonlinearities.  

\par A common formulation of such optimization consists in finding the signaling strategies that maximize $P_{\dc}^{r}$
\begin{align}\label{WPT_opt_problem}
\max_{\mathbf{x}_0,...,\mathbf{x}_{N-1},\mathbf{w}_{\mathrm{R}},\mathbf{\Theta}} \hspace{0.3cm}& P_{\dc}^r(\mathbf{x}_0,...,\mathbf{x}_{N-1},\mathbf{w}_{\mathrm{R}},\mathbf{\Theta}) \\
\mathrm{subject\,\,to} \hspace{0.3cm} &P_{\rf}^t\leq P,
\end{align}
where maximization is here performed over the transmit vector on all $N$ frequencies $\mathbf{x}_0,...,\mathbf{x}_{N-1}$, the receive combiner $\mathbf{w}_{\mathrm{R}}$ (for RF combining in multi-antenna ER), and the scattering matrix $\mathbf{\Theta}$ (if RIS is present as part of the wireless channels) under the average transmit power constraint $P_{\rf}^t\leq P$.  Additional constraints can also be added to problem \eqref{WPT_opt_problem} to account for $e_1$ and HPA nonlinearity - for instance in the form of peak-to-average power (PAPR) constraints \cite{Clerckx:2016b}.

\par
\begin{figure}
	\centerline{\includegraphics[width=0.9\columnwidth]{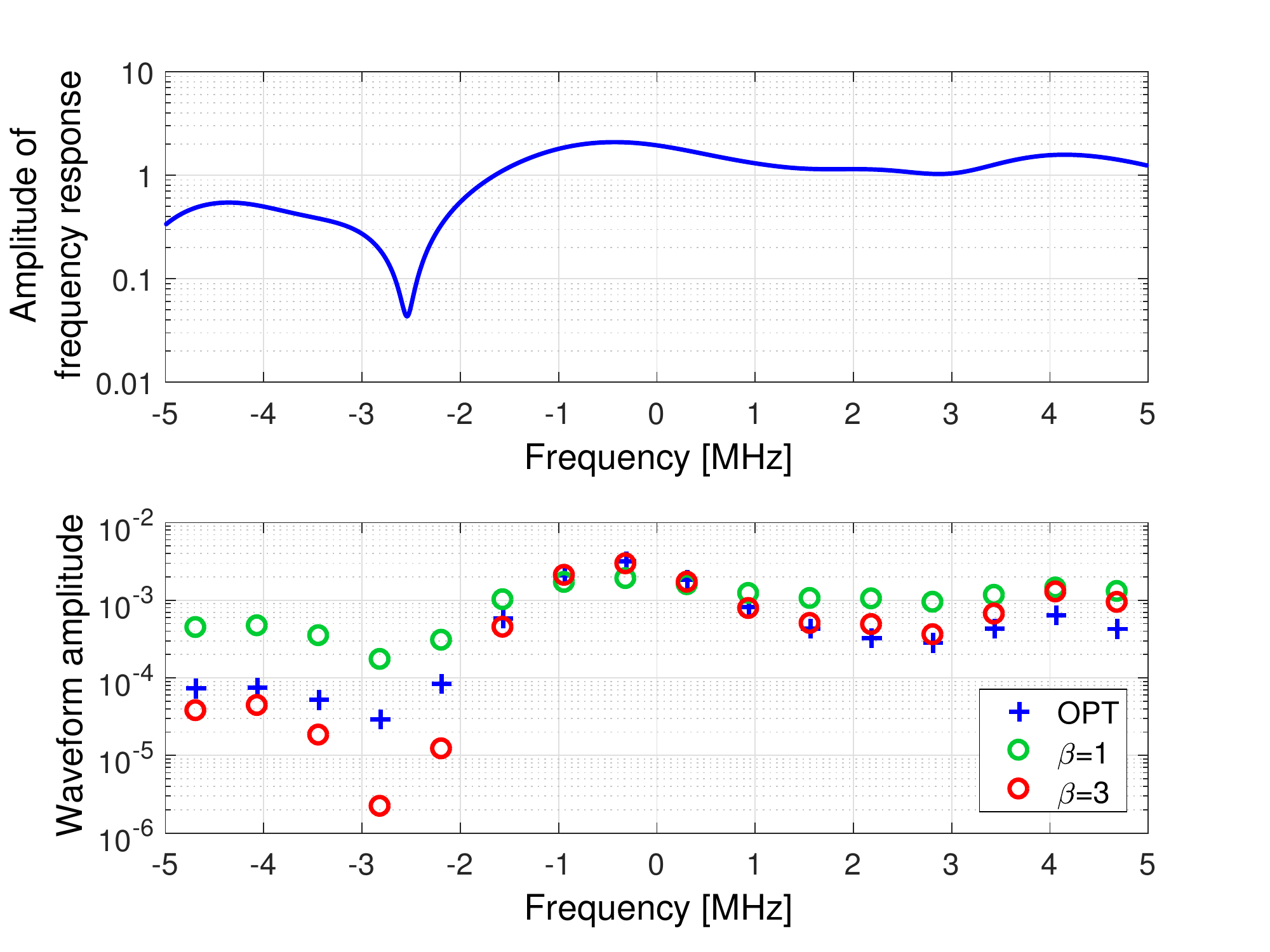}}
	\caption{Frequency response of the wireless channel and WPT waveform
magnitudes (N = 16) for 10 MHz bandwidth \cite{Clerckx:2017}.}
	\label{lowcomplexityWF}
\end{figure}

 The WPT signal optimization of \eqref{WPT_opt_problem} was initiated in \cite{Clerckx:2016b} for joint waveform and active beamforming. Due to the complexity of the optimization, suboptimal low complexity methods, called SMF (scaled matched filter), have also been used in \cite{Clerckx:2017}. SMF designs $\mathbf{x}_n$ by performing maximum ratio transmission (MRT) in the spatial domain and non-uniform power allocation in the frequency domain as
\begin{equation}\label{SMF}
\mathbf{x}_n=\frac{\mathbf{h}_n^H}{\left\|\mathbf{h}_n\right\|}\left\|\mathbf{h}_n\right\|^{\beta}\sqrt{\frac{2P}{\sum_{n=0}^{N-1}\left\|\mathbf{h}_n\right\|^{2\beta}}}.
\end{equation}
We notice the conventional MRT beamforming $\frac{\mathbf{h}_n^H}{\left\|\mathbf{h}_n\right\|}$ on each frequency $n$. Using an exponent $\beta>1$ of the channel norm $\left\|\mathbf{h}_n\right\|$, the power is allocated across all sinewaves/frequencies but more (resp.\ less) power is allocated to the frequency components corresponding to large (resp.\ weak) channel norms, which replicates the behavior of the optimal strategy of \cite{Clerckx:2016b}. This is shown in Fig. \ref{lowcomplexityWF} where the upper graph is the magnitude of the channel frequency response, and the lower figure displays the solution of problem \eqref{WPT_opt_problem} (``opt'') at $N=16$ uniformly spaced frequencies, as well as the SMF solution \eqref{SMF} with $\beta=1,3$. By adjusting $\beta>1$, SMF comes closer to the optimal power allocation. This non-uniform power allocation can be traced back to the nonlinearity in \eqref{vout_def} and originates from a trade-off between maximizing the first term $\beta_2P_{\rf}^r$ (and therefore maximize $P_{\rf}^r$) by allocating power to a single sinewave and leveraging the nonlinearity of the second term $\beta_4 \mathbb{E}\left[y_{\rf}(t)^4\right]$ (and therefore maximize $e_3$) by allocating power across multiple sinewaves. 

\par In the presence of multi-antenna ER with RF combiner, the waveform, transmit beamforming and receive combining need to be jointly optimized as in \cite{Shen:2020a,Shen:2020c}, which offers additional performance enhancements over a transmitter-only optimization. In the presence of RIS, the signal optimization in \cite{Feng:2020,Zhao:2020} leads to an additional passive beamforming on top of the joint waveform and active beamforming gain provided by the transmitter. 

\par The above signal optimization leads to an architecture in which the rectifier is, as much as possible, fixed (for example, with
a fixed load) but the transmit signal (and potentially combiner) is adaptive. Nevertheless, recall the aforementioned discussions on configurable DC-to-DC converter to track the optimum load and trace the optimal efficiency as a function of the input waveform shape and power level. Because the wireless channel changes, it is envisioned that an entire end-to-end optimization of the system should be conducted, likely resulting in an architecture in which the transmit signal, combiner and the rectifier are jointly optimized and adapt themselves dynamically as a function of the channel state \cite{Clerckx:2018}. The challenge is how to operate and design such an adaptive system keeping in mind the energy constraint of the devices. Machine learning techniques may be helpful to that end \cite{Clerckx:2021}, though this research area remains largely underexplored.

\subsubsection{Channel Acquisition}

\par
\begin{figure}
	\centering
	\subfigure[Forward-link training]{\includegraphics*[width=0.9\linewidth]{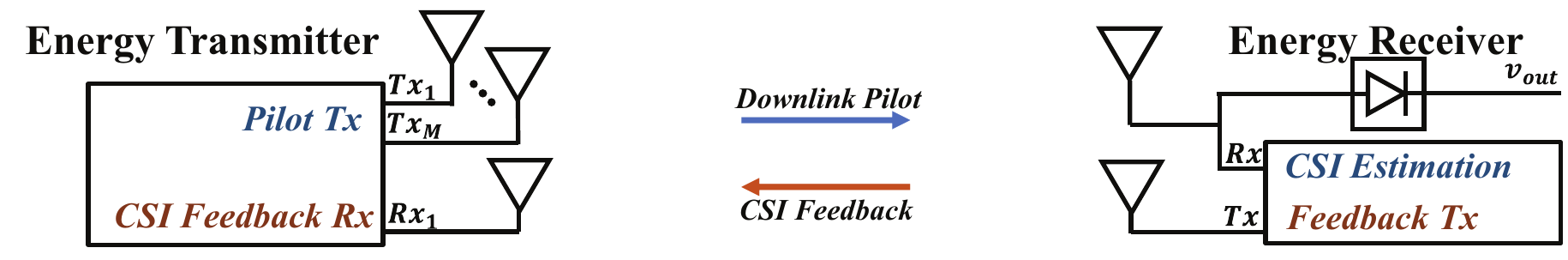}}
	\subfigure[Reverse-link training]{\includegraphics*[width=0.9\linewidth]{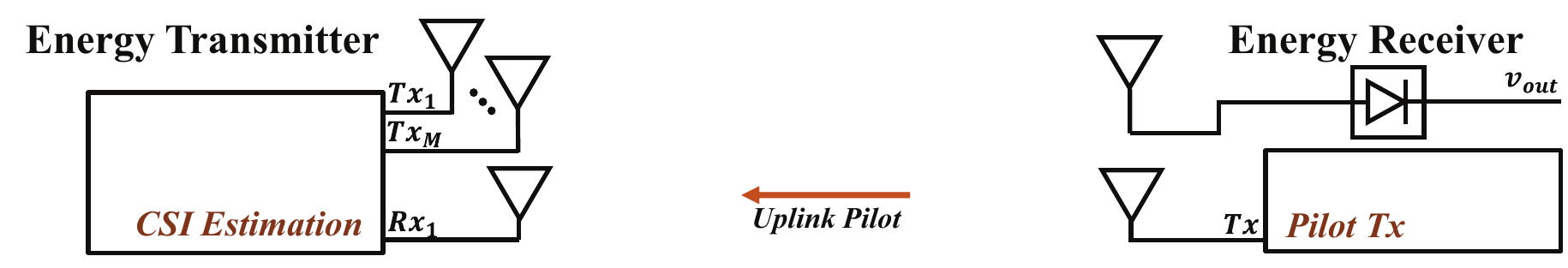}}
	\subfigure[Power probing with limited feedback]{\includegraphics*[width=0.9\linewidth]{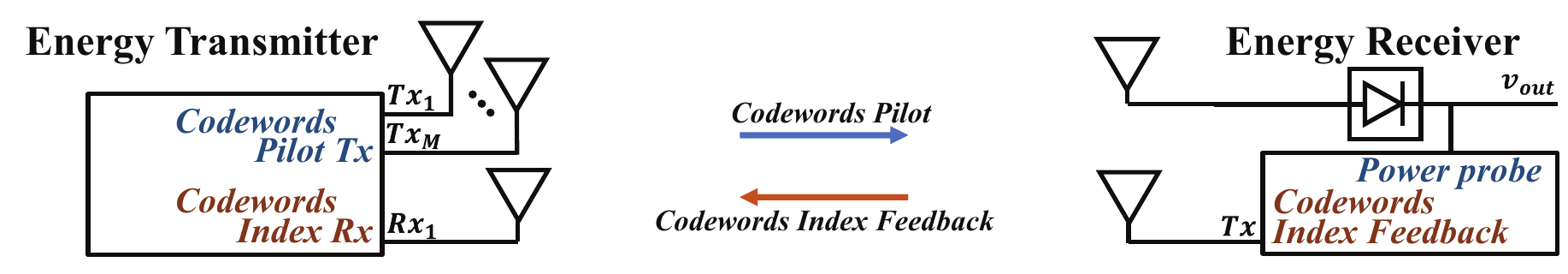}}
	\subfigure[Backscatter-based training]{\includegraphics*[width=0.9\linewidth]{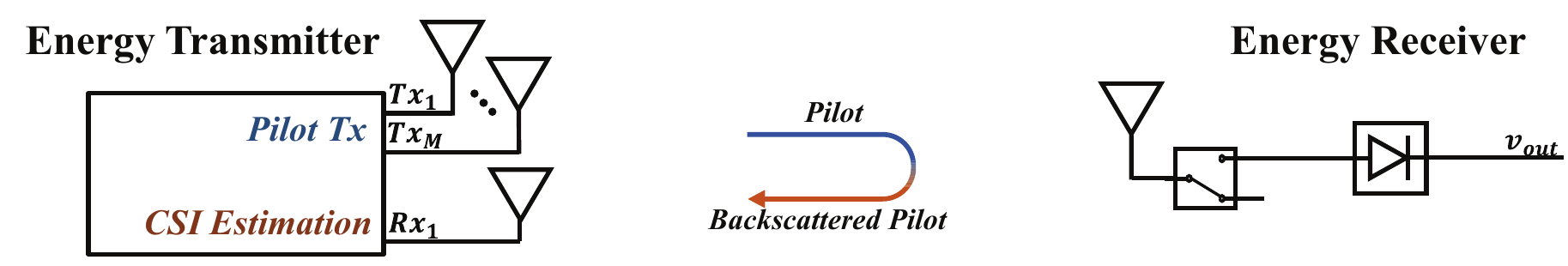}}
	\caption{Four strategies of channel state information acquisition at ET.}\label{channel_acquisition}
\end{figure}

 Optimization of $\mathbf{x}_0,...,\mathbf{x}_{N-1},\mathbf{w}_{\mathrm{R}},\mathbf{\Theta}$ in \eqref{WPT_opt_problem} requires knowledge of the wireless channel state information (CSI) at the transmitter (CSIT), i.e., $\mathbf{h}_{q,n}$ (or $\mathbf{g}_{\mathrm{d},q,n}$, $\mathbf{g}_{\mathrm{r},q,n}$, and $\mathbf{G}_{\mathrm{i},n}$ in RIS-aided WPT). In practice, the CSI should be acquired by the signal optimization module at the ET. Fig. \ref{channel_acquisition} illustrates several CSI acquisition strategies, including forward-link training with CSI feedback, reverse-link training via channel reciprocity, power probing with limited feedback, and channel estimation based on backscatter communications \cite{Zeng:2017,Xu_Zhang:2014,Zeng_Zhang:2015,Abeywickrama:2018,ChoiKim2017b,Huang:2018,Yang:2015}.

 The first two are inspired by strategies used in modern communication systems, but incur complex processing and/or high energy consumption for low power nodes. Indeed the first one requires channel estimation at the low power node and the second one relies on pilot transmission. The third appears more appealing for WPT because it is implementable with low communication and signal processing requirements at the ER and simply relies on the measurement of a DC power level at the output of the harvester and the feedback of some limited information (no channel estimation or pilot transmission is required). The fourth one is also encouraging due to the low energy consumption at the device and relies on the idea that the ET can exploit the backscatter signals to estimate the backscatter channel (i.e., ET-to-ER-to-ET) state information, from which the transmit signal needs to be designed. 

\par The framework of \cite{Huang:2018} revisits optimization problem \eqref{WPT_opt_problem} for the general setup of multi-antenna multisine WPT of \eqref{WPT_1} under power probing with limited feedback. It shows that joint waveform and beamforming benefits can be exploited with just a few bits of CSI feedback. To this end, it relies on measuring the output DC power for a selected set of joint waveform and beamforming weights/precoders and a limited number of feedback bits. Specifically, over a training period made of multiple time slots, the ET sends in each time slot with a different (joint waveform and beamforming) precoding codeword within a codebook, and the ER reports the index of the codeword in the codebook that offers the largest $P_{\dc}^r$.

\par In the presence of multiple receive antennas, RF combining is more complex to implement than DC combining as it requires the knowledge of the Channel State Information at the Receiver (CSIR) for the receive combiner optimization or alternatively needs a communication link between the ET and ER to tell the ER what receive combiner to use.  

\par The smart integration of the ET, wireless channel, ER, channel acquisition and signal optimizations modules ultimately leads to the closed loop, adaptive and optimized WPT architecture of Fig. \ref{fig_sys}.

\subsection{Prototypes and Experiments}\label{proto_WPT}
The previous section highlighted key features such as beamforming, waveform, multi-antenna ER, RIS, channel acquisition and signal optimization. We now detail several state-of-the-art WPT prototypes developed at Imperial College London (ICL) \cite{Kim:2018,Kim:2020,Shen:2021,Shen:2020b} and Sungkyunkwan University (SKKU) \cite{Choi:2019, Aziz:2019,Tran:2021,Choi:2019a,Park:2021} and illustrate the real-world performance benefits of some of those features based on experimental results obtained from over the air measurements in various indoor and outdoor deployments. 
\par Though we here limit to ICL and SKKU prototypes, readers are also referred to other WPT experimental works conducted by other groups around the world, in particular on digital beamforming through baseband precoding using backscattering for feedback \cite{Belo:2019}, analog beamforming through phased array \cite{Yang:2020}, time-modulation array \cite{Masotti:2016}, adaptive beamforming using receive signal strength indicator feedback \cite{Yedavalli:2017, Abeywickrama:2018}, adaptive beamforming using the second and third harmonics for feedback \cite{Zhang:2019,Joseph:2020}.

\subsubsection{920 MHz Long-Range MISO WPT Prototype}\label{SKKU_920}

We have built a long-range WPT prototype testbed for demonstrating the real-time operation of a wireless-powered sensor network (WPSN) \cite{Choi:2019, Aziz:2019}. To enable long-range WPT, a relatively low frequency band (920 MHz) is used, and a large-scale rectangular phased antenna array of 4$\times$16 antenna elements is employed to enable active beamforming. The prototype testbed consists of one ET and one sensor device which relies solely on power supplied by the ET. This WPT testbed is fully autonomous in that the ET is able to adaptively focus its power beam on the sensor device, and the sensor device can dynamically adjust its power consumption according to the stored energy level for long-term survival.

\par
\begin{figure}
    \subfigure[Phased array board]{
        \label{fig:phased_array_board}\includegraphics[width=3.8cm] {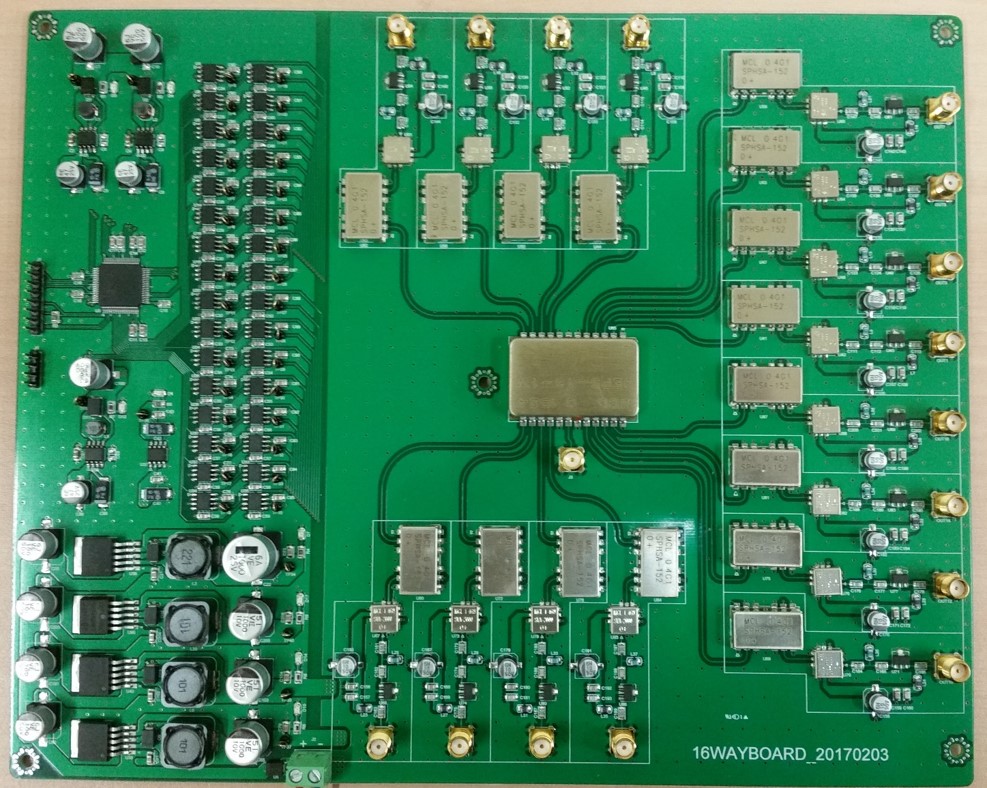}
        }
    \subfigure[Sensor device]{
        \label{fig:sensor_device}\includegraphics[width=4.5cm] {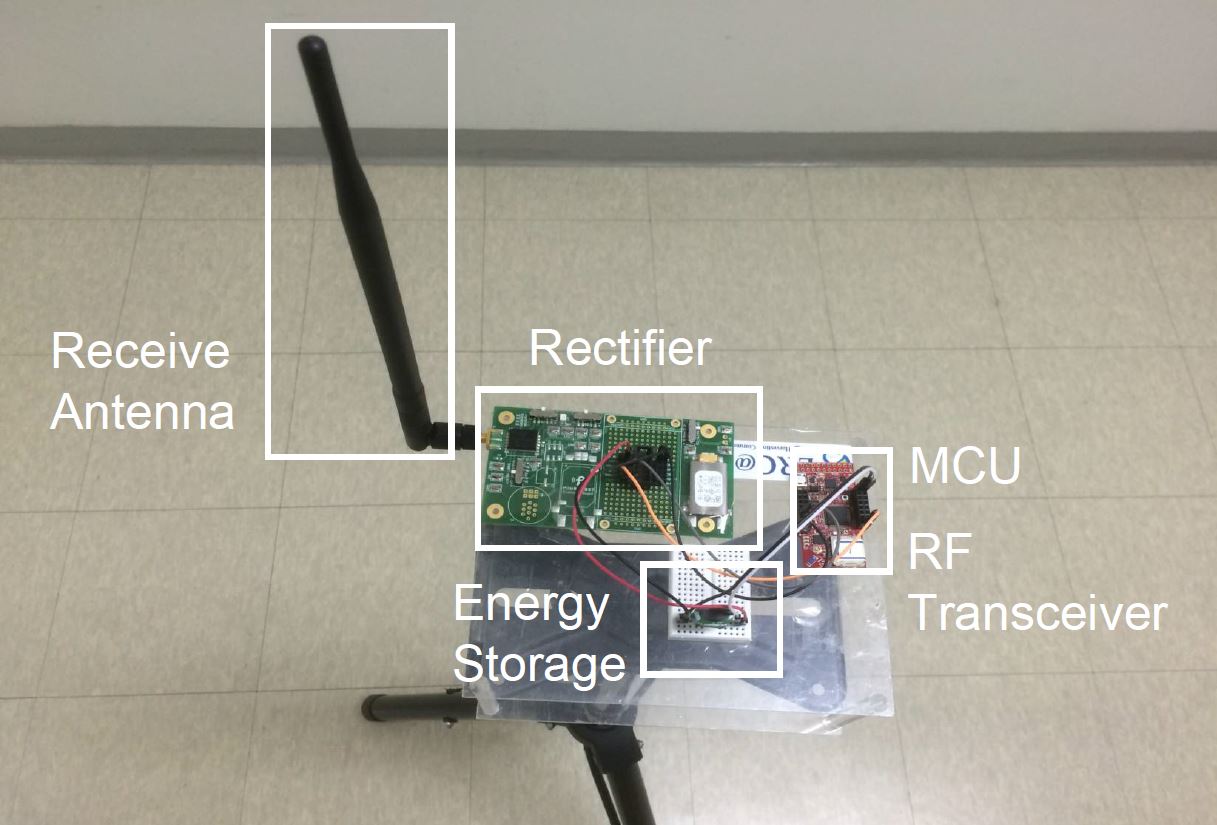}
        }
    \subfigure[Outdoor test environment]{
        \label{fig:outdoor_test_env}\includegraphics[width=5.1cm] {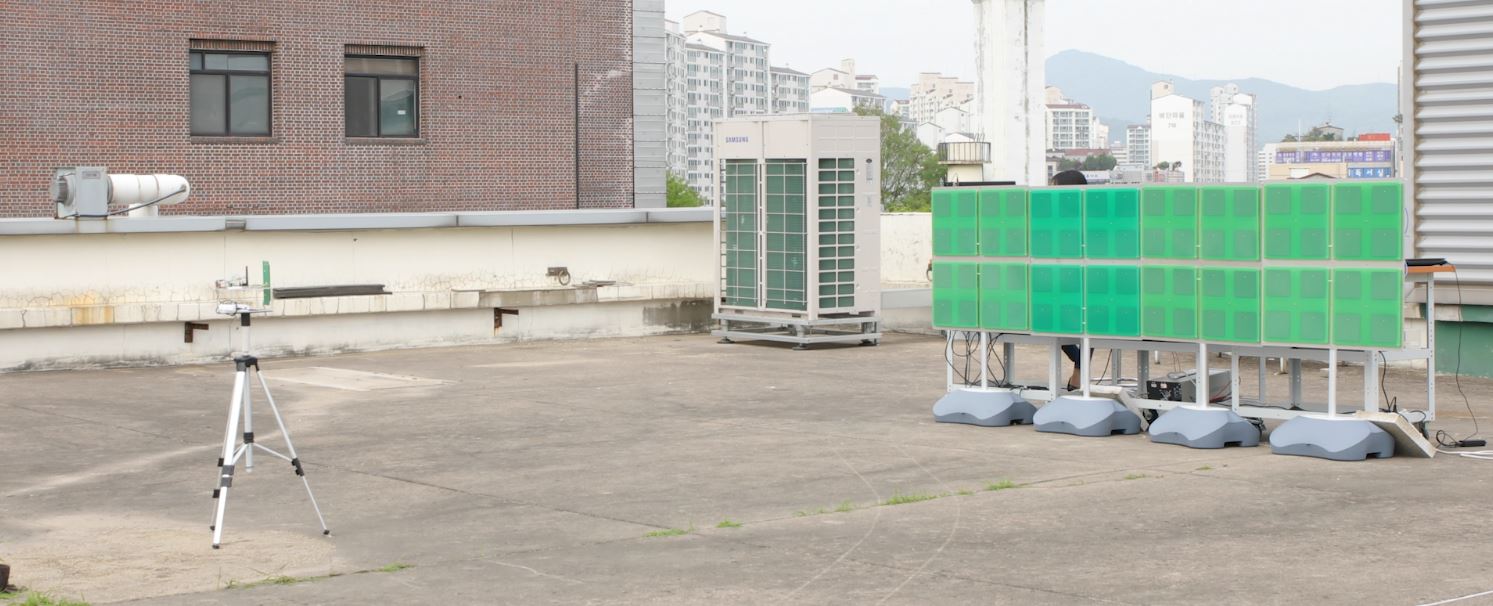}
        }
    \subfigure[50 meter WPT test]{
        \label{fig:50_meter_wpt}\includegraphics[width=2.9cm] {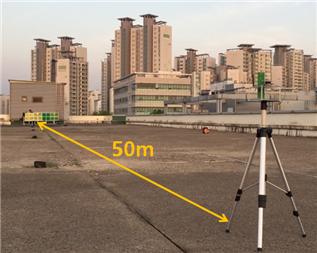}
        }
    \caption{920 MHz WPT prototype testbed.}
\end{figure}

 The ET is equipped with a phased antenna array with 64 antenna elements. We have fabricated a 16-way phased array board in Fig.~\ref{fig:phased_array_board} that outputs 16 RF signals with controllable phase and magnitude.
The phased array board receives a CW source signal and splits it into 16 RF path by using a power splitter.
Each RF path has a variable attenuator, a phase shifter, and an amplifier.
The variable attenuator and phase shifter are all voltage-controlled by an on-board digital-to-analog converter (DAC) chip.
We use a computer and a field programmable gate array (FPGA) device to control the phased array board through the DAC chip.
We have combined four 16-way phased array boards to implement 64 RF paths of the ET.
Four microstrip patch antenna arrays, each of which has 16 antenna elements with 8 dB gain, are fabricated and connected to the phased array boards via coaxial cables.

The sensor device is comprised of a rectifier board, an energy storage, and a sensor board, which are all commercial off-the-shelf (COTS) components (Fig.~\ref{fig:sensor_device}).
The rectifier board is an evaluation board of a Powercast P1110 chip with various functions such as rectification, power management, and receive power measurement.
A supercapacitor with the capacitance of 0.1 F is used as an energy storage, which is connected to the output of the rectifier board.
The sensor board is the Zolertia Z1 mote with TI MSP430 as an MCU and TI CC2420 as an RF transceiver.
The RF transceiver, which implements the 802.15.4 protocol on the 2.4 GHz frequency band, is used for communicating with the ET.
The software platform of the sensor board is the Contiki operating system (OS).

\par
\begin{figure}
    \centering
    \includegraphics[width=6cm, bb=0.8in 0.8in 9.8in 8.1in] {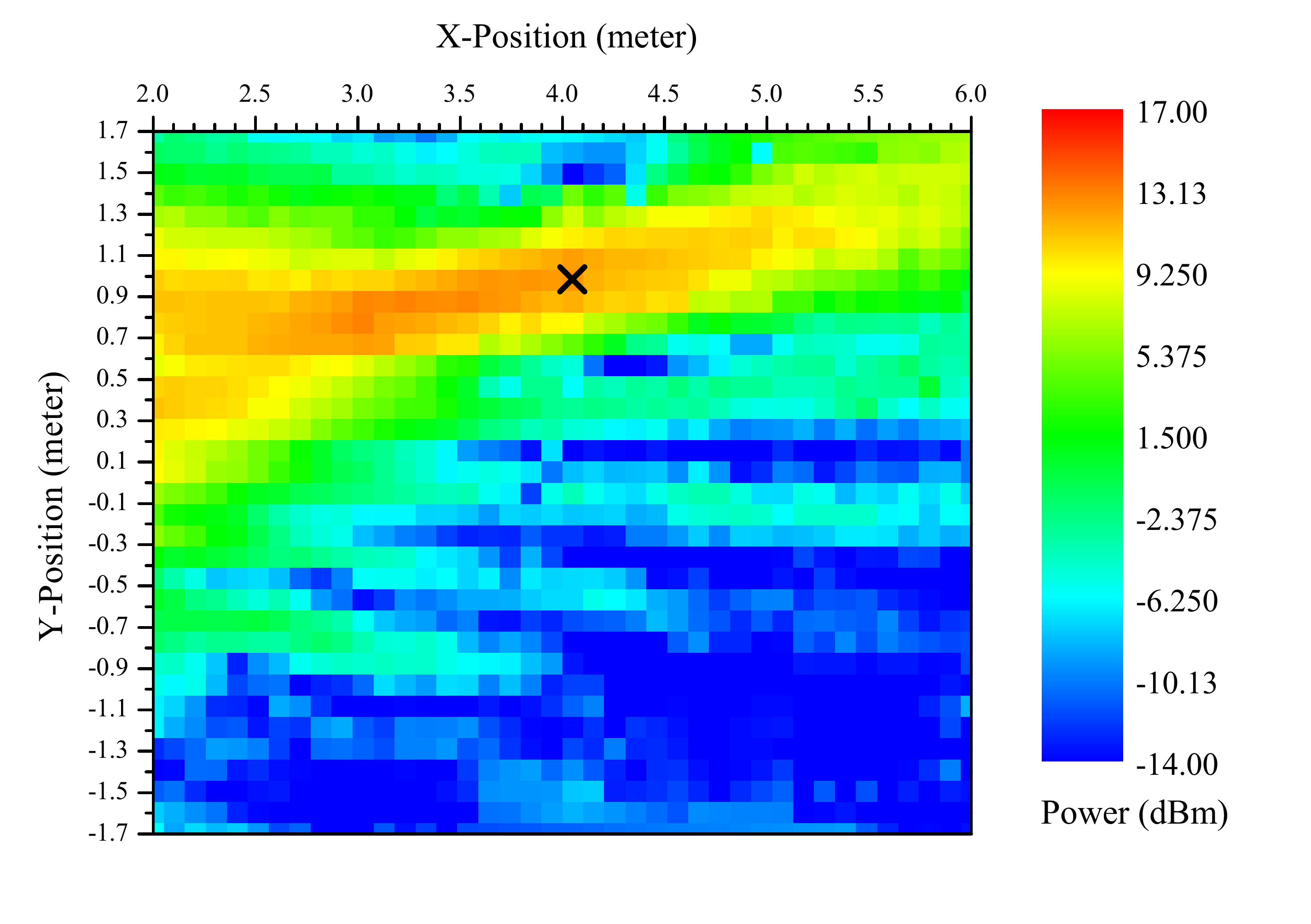}
    \caption{Measured beam shape.}
    \label{fig:beamshape}
\end{figure}

 We have performed beamforming experiments to demonstrate that the ET can focus a beam on the sensor device. The CSI acquisition is based on the power probing strategy where the indices of multiple codewords in the codebook (along with their power levels) are reported and the transmitter adjusts the beamformer based on the collected feedback.
This test is first conducted in an indoor environment where the ET is deployed on one side of a room.
Fig.~\ref{fig:beamshape} shows the heat map of the received RF power inside a room, where the location of the sensor device is indicated by a black X mark. 
We can see that the ET accurately focuses the energy at the sensor device.


\par
\begin{figure}
    \subfigure[2--5 m (indoor)]{
        \label{fig:rxpow_2_5}\includegraphics[width=4.1cm, bb=0.6in 0.3in 9.6in 8.1in] {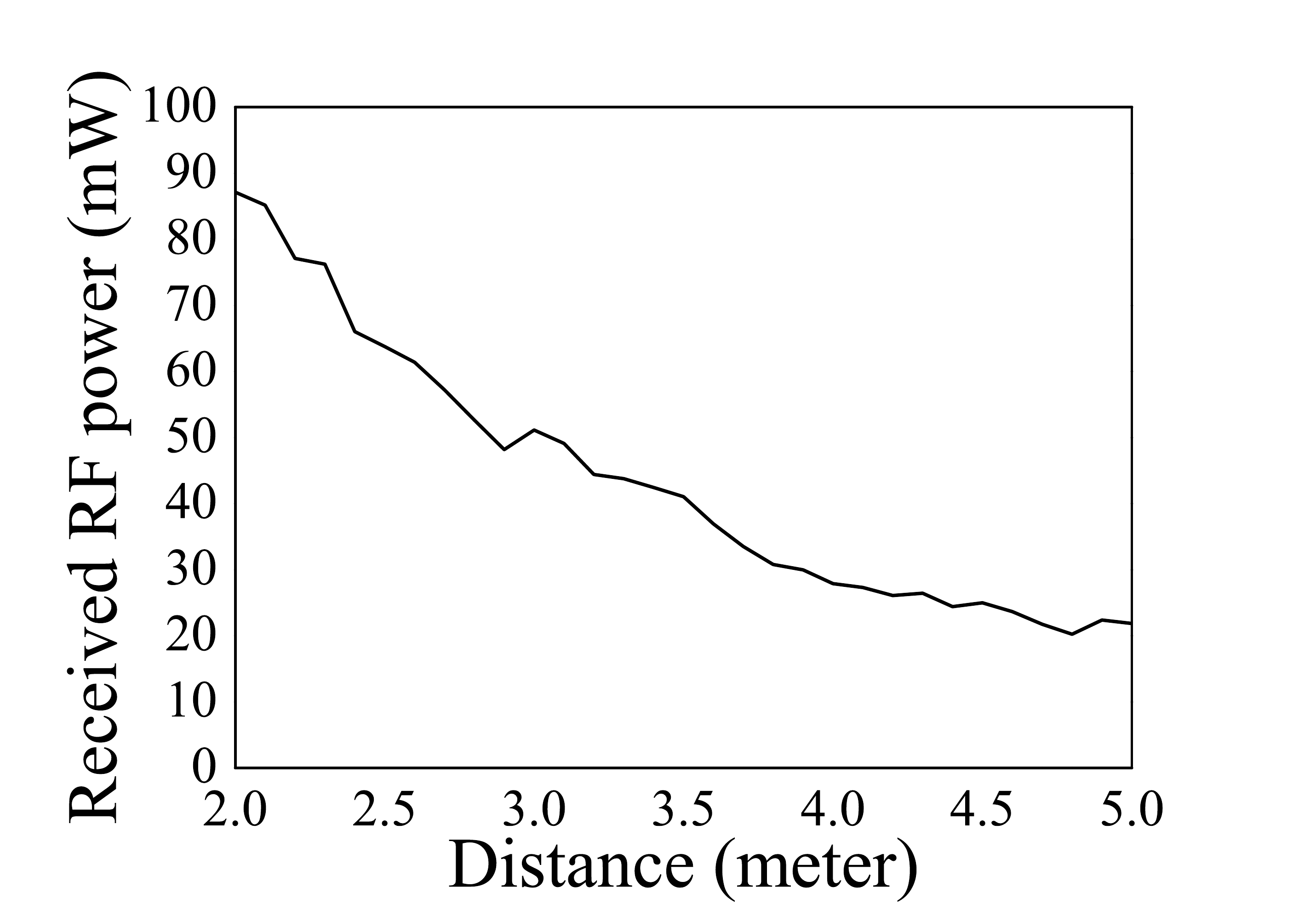}
        }
    \subfigure[15--50 m (outdoor)]{
        \label{fig:rxpow_15_50}\includegraphics[width=4.1cm, bb=0.6in 0.3in 9.6in 8.1in] {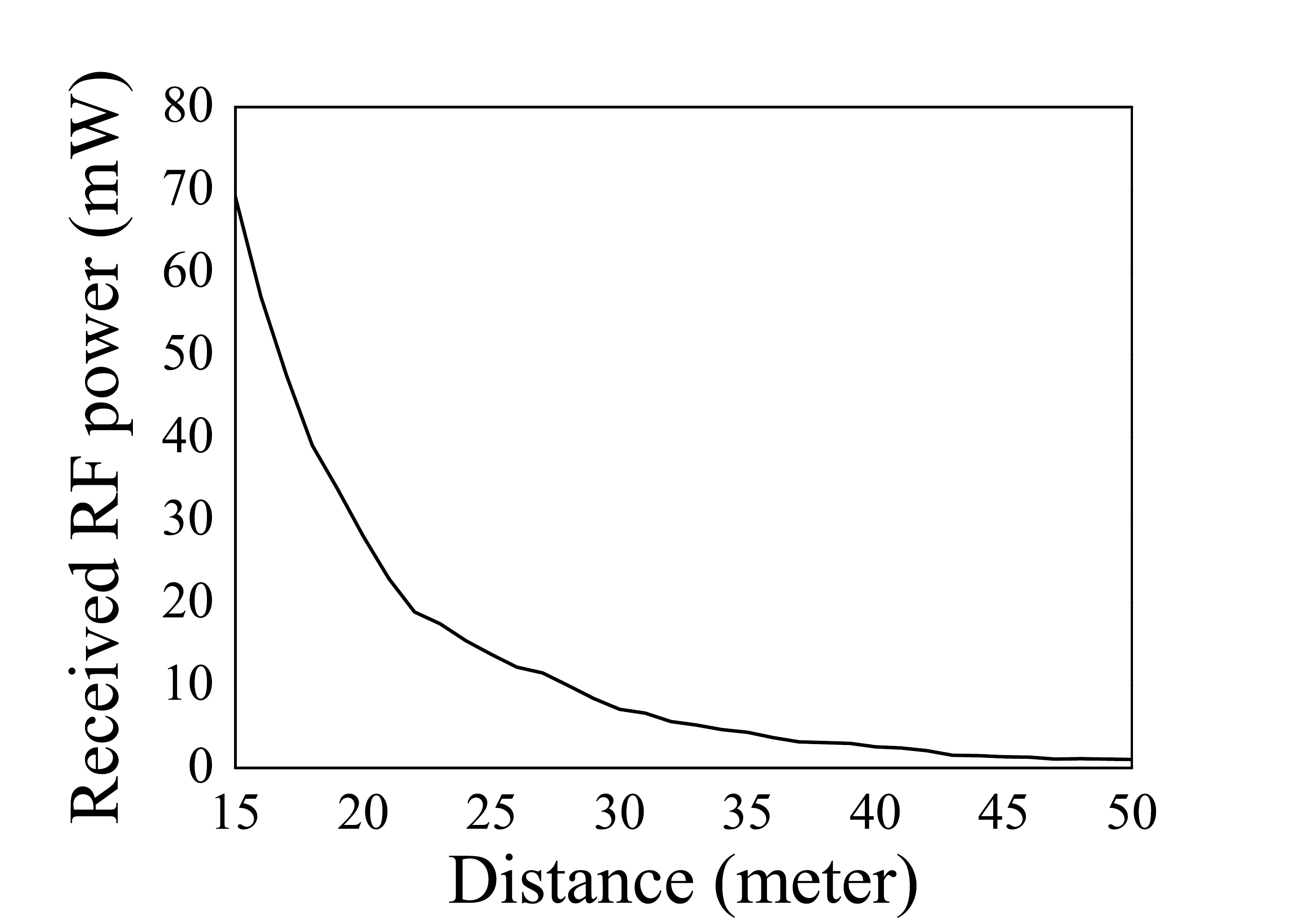}
        }
    \caption{Received RF power over distance.}
\end{figure}

 The received RF power is measured according to the distance in Figs.~\ref{fig:rxpow_2_5} and \ref{fig:rxpow_15_50}.
The indoor test results from 5 to 10 meters are shown in Fig.~\ref{fig:rxpow_2_5}.
For this test, the total transmit power from all antenna elements is limited to 400 mW.
We can see that the received RF power is up to 87 mW at 2 meters distance, at which the power transfer efficiency is 21.8 \%.
The tests from 15 to 50 meters are also conducted in an outdoor environment (Fig.~\ref{fig:outdoor_test_env}), and the received RF power is shown in Fig.~\ref{fig:rxpow_15_50}.
In this test, the total transmit power can go up to 9.5 W.
In this figure, we can see that a few mW can be transferred to tens of meters, which is enough to operate a low-power sensor device.

\par
\begin{figure}
    \subfigure[Received RF power]{
        \label{fig:rxpow_time}\includegraphics[width=4.1cm, bb=0.6in 0.3in 9.6in 8.1in] {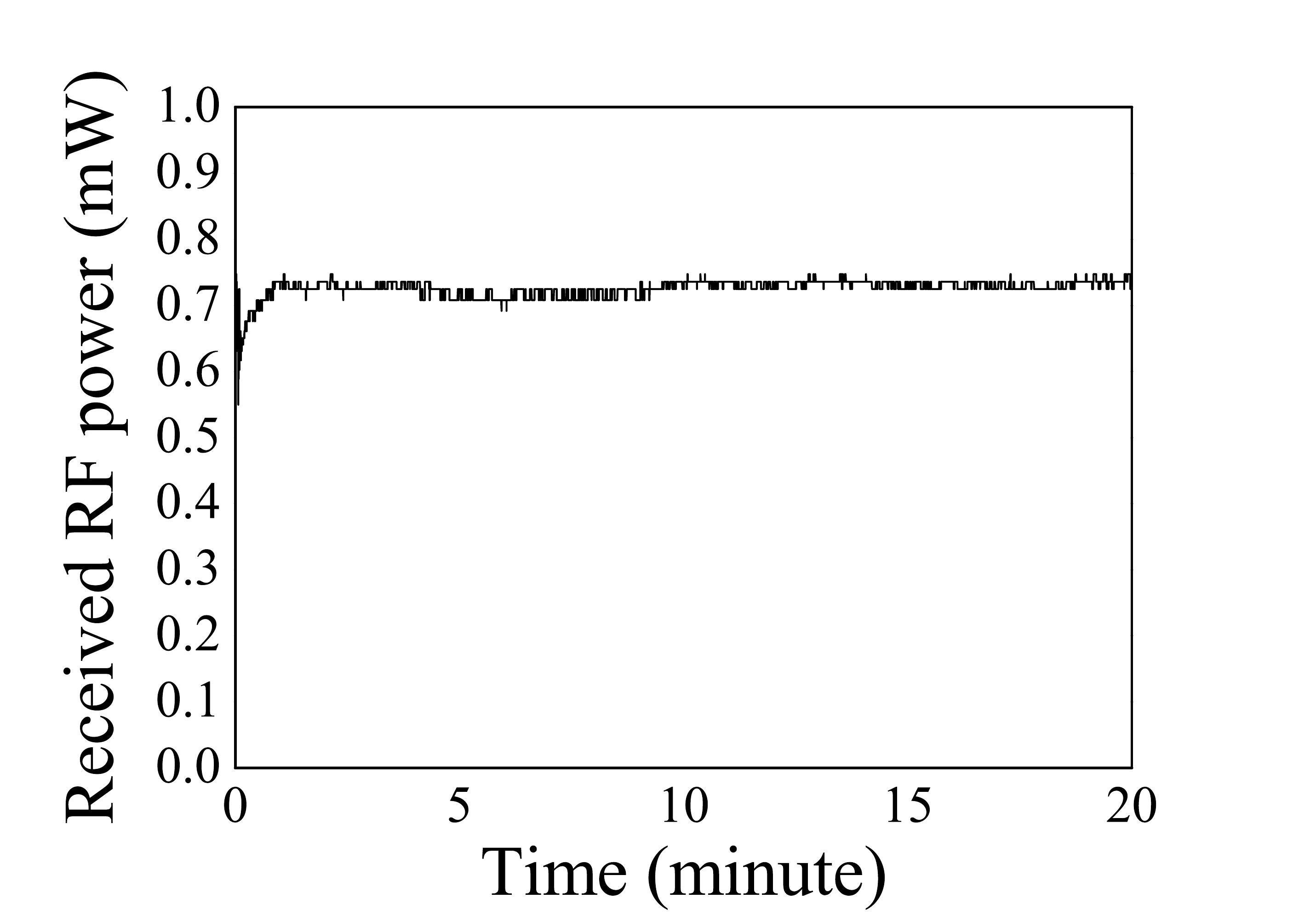}
        }
    \subfigure[Stored energy]{
        \label{fig:stored_energy}\includegraphics[width=4.1cm, bb=0.6in 0.3in 9.6in 8.1in] {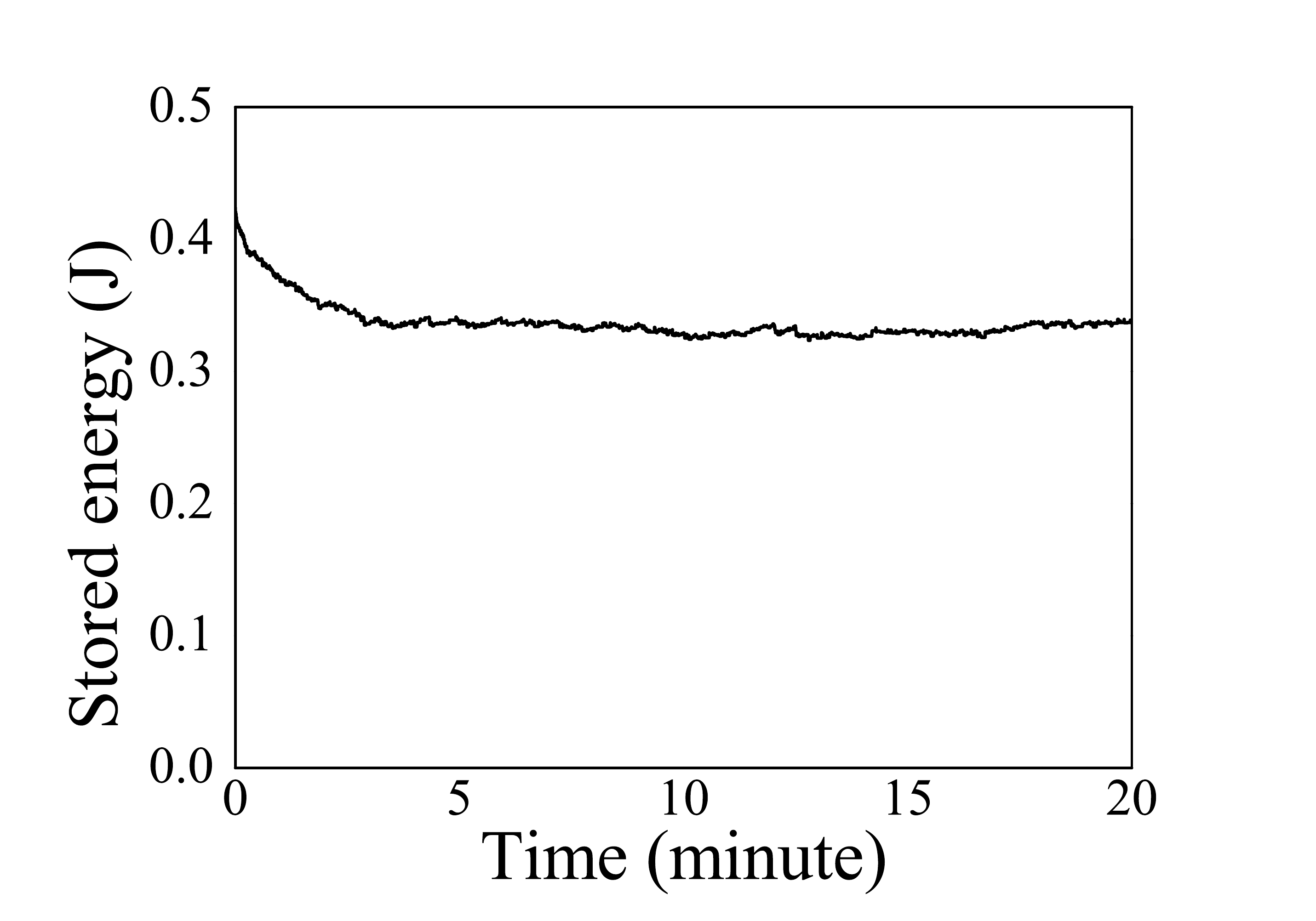}
        }
    \caption{Survivability test results.}
    \label{fig:Survivality}
\end{figure}

 We have placed the sensor device at 50 meters away from the ET (Fig.~\ref{fig:50_meter_wpt}) to test if the sensor device can be kept alive resorting only to the power transferred from the ET.
Fig.~\ref{fig:rxpow_time} shows that the sensor device constantly receives around 0.72 mW from the ET at 50 meters away.
With this power supply, the stored energy in the sensor device is successfully kept to a certain level as long as 20 minutes as shown in Fig.~\ref{fig:stored_energy}.


\subsubsection{2.4 GHz MISO WPT Prototype with Adaptive Waveform}\label{ICL_WPT} 
\begin{figure}[t]
	\centering
	\subfigure[]{\includegraphics[width=0.7\linewidth]{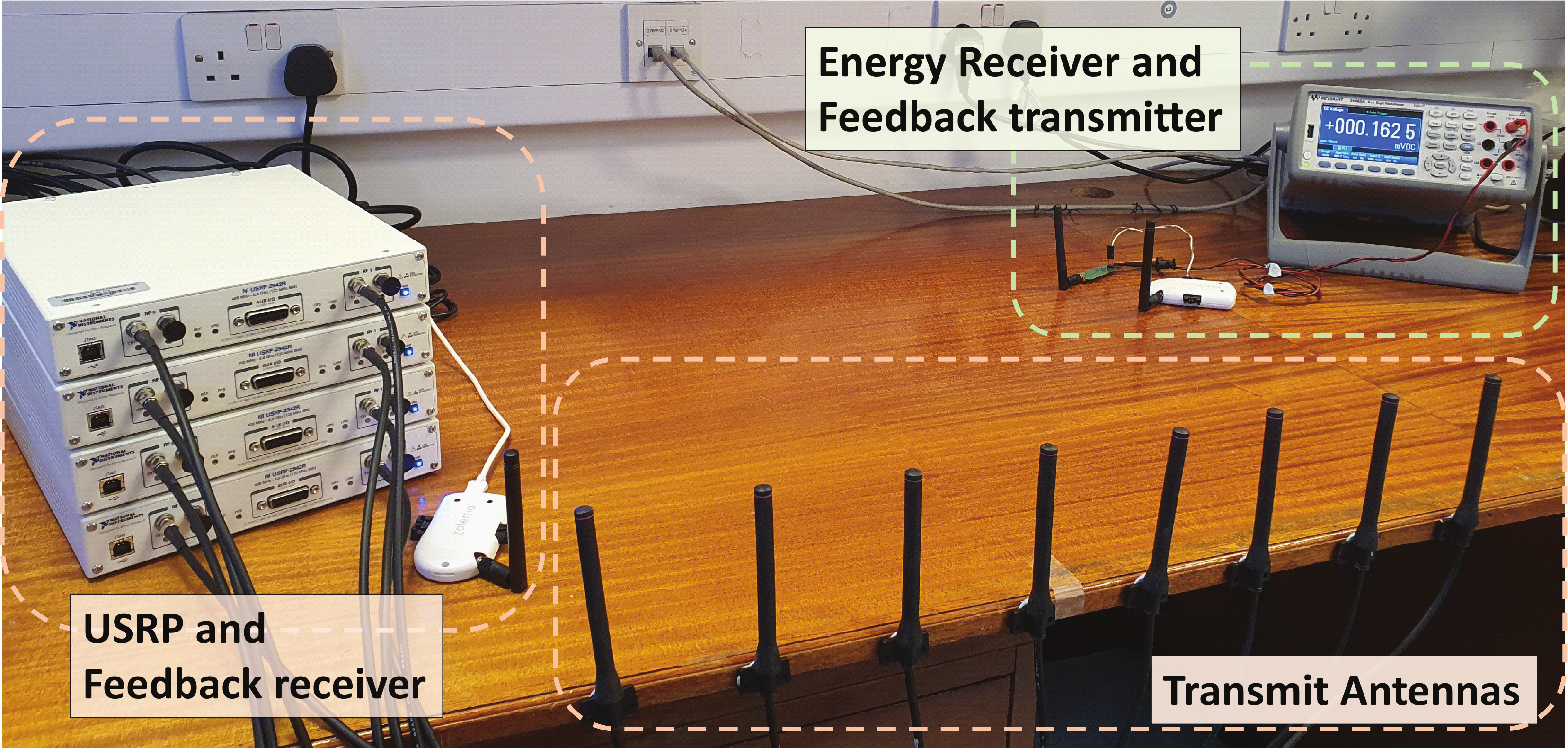}}\\
	\subfigure[]{\includegraphics[width=0.22\linewidth]{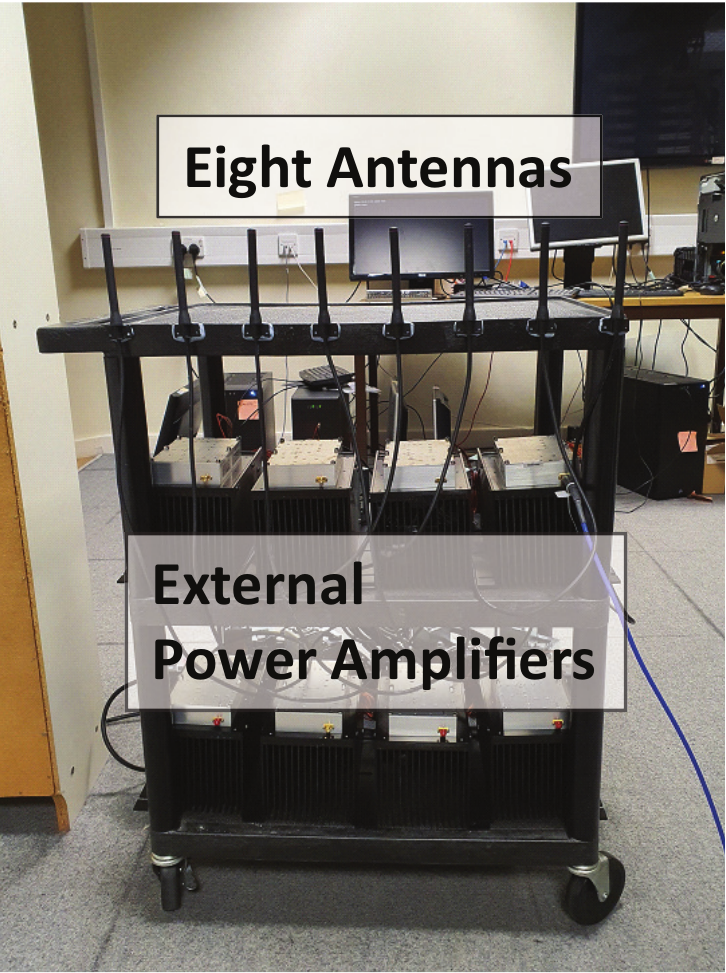}}
	\subfigure[]{\includegraphics[width=0.6\linewidth]{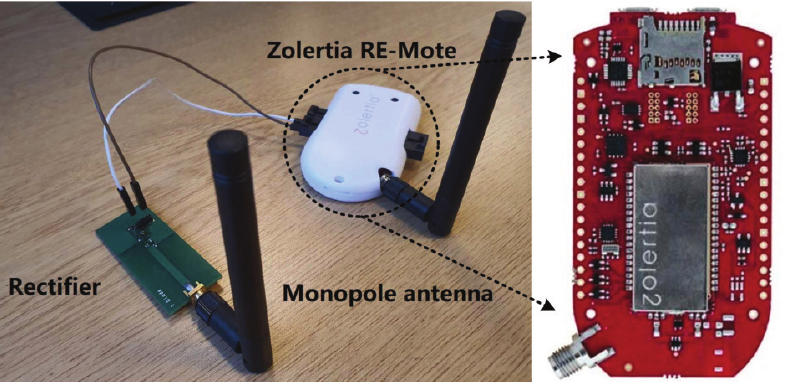}}
	\caption{Closed-loop WPT prototype system with limited feedback: (a) System prototype; (b) Transmit antennas and amplifiers for long-range experiment; (c) ER and feedback system using Zolertia RE-Mote.}
	\label{fig_wpt_prototype}
\end{figure}

\begin{figure}
	\centerline{\includegraphics[width=0.95\columnwidth]{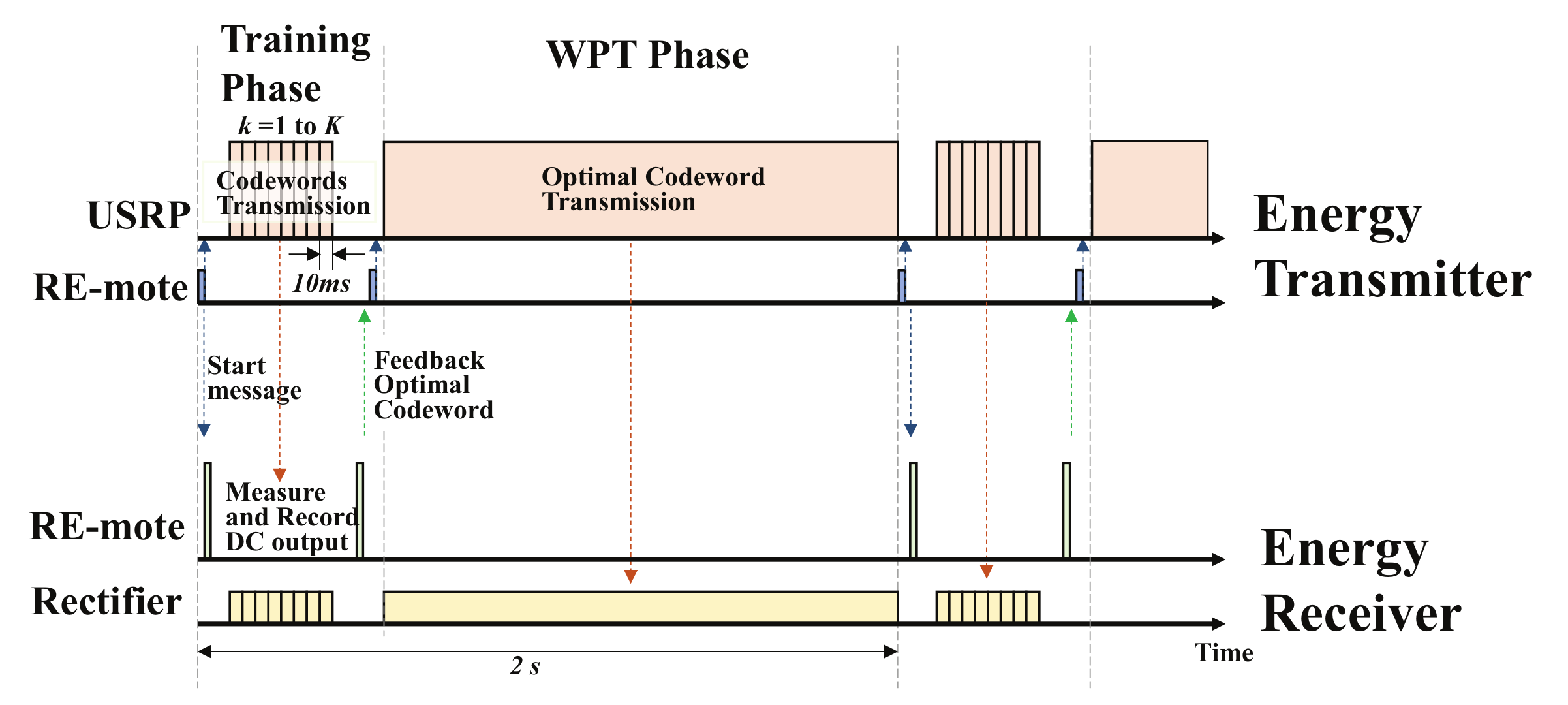}}
	\caption{Frame structure of the WPT prototype system operation.}
	\label{wpt_frame}
\end{figure}

Previous experiment demonstrates the key benefit of multi-antenna ET and active beamforming with CW. We now expand the signal optimization to joint space and frequency and demonstrate the benefits of joint waveform and active beamforming. To that end, we have developed in \cite{Kim:2018,Kim:2020,Shen:2021} a prototype of closed-loop WPT system (Fig. \ref{fig_wpt_prototype}) with adaptive waveform and beamforming that relies on the general modern architecture of subsection \ref{modern_WPT_subsection}. The operation of the prototype follows the frame structure of Fig. \ref{wpt_frame} consisting of a training phase and a WPT phase.


\par The system consists of three important functional blocks: the intelligent ET equipped with an adaptive waveform and beamforming signal generator module, the CSI acquisition and feedback module, and the ER. 
The ET is made up of multiple Software-Defined Radio (SDR) equipments and antennas to generate and radiate adaptive waveform and beamforming RF signals chosen according to CSI. The system operates in the 2.4 GHz band and supports up to eight antennas. All the antennas are identical universal omnidirectional 2.4GHz antennas with 3 dBi antenna gain and 85\% radiation efficiency. The system has external power amplifiers (Mini-Circuits ZHL-16W-43-S+ with 45dB of gain and 41dBm of P1dB) to generate high-power RF signals to experiment WPT performance at longer range of up to 5.5 m. We limited the maximum transmit power of the system below 30 dBm to comply with the maximum Effective Isotropic Radiated Power (EIRP) of 36dBm and conduction power of 30dBm specified in FCC Title 47, Part 15 regulations. The SDR equipment, based on National Instrument (NI) USRP-2942, can generate multi-sine waveforms with different magnitudes and phases at each tone. We consider up to 8-tone signals so that the generated RF signal is in the linear region of the power amplifier. The $N$-tone multisine signal has a uniform frequency gap of $\Delta_f= B/N$ where the bandwidth $B$ = 10 MHz. 

\par The ER is implemented using a single diode rectifier (similar to Fig. \ref{TD_schematic}(a)) which consists of a matching network, rectifying Schottky diode, and low pass filter. The entire circuit is fabricated on a 1.6 mm thick FR-4 substrate and the mounted lumped element values of the matching network were chosen to fit the operating frequency of 2.4GHz. We chose Skyworks SMS7630 Schottky diode as a rectifying diode, which can be operational with low input power because of its low turn-on voltage.

\begin{figure}[t]
	\centerline{\includegraphics[width=0.8\columnwidth]{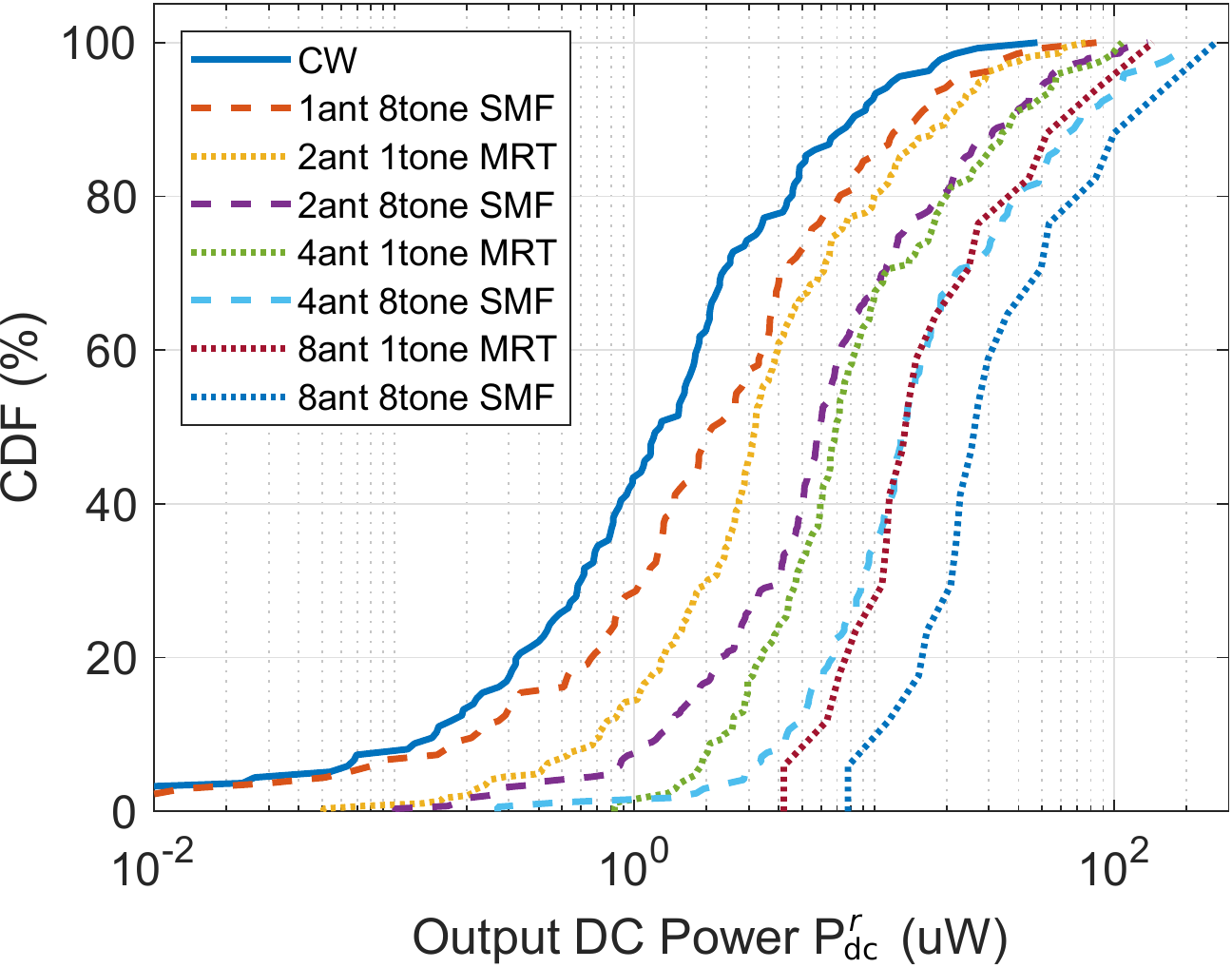}}
	\caption{Cumulative distribution of the output DC power ($P_{\dc}^r$) measurement results at different distances (0.6 m - 5.4 m) with the number of transmit antennas $M$=1, 2, 4, 8, and the number of carriers (tones) $N$=1 and 8 \cite{Kim:2020}.}
	\label{cdf_8ant}
\end{figure}

\begin{figure}[t]
	\centerline{\includegraphics[width=0.8\columnwidth]{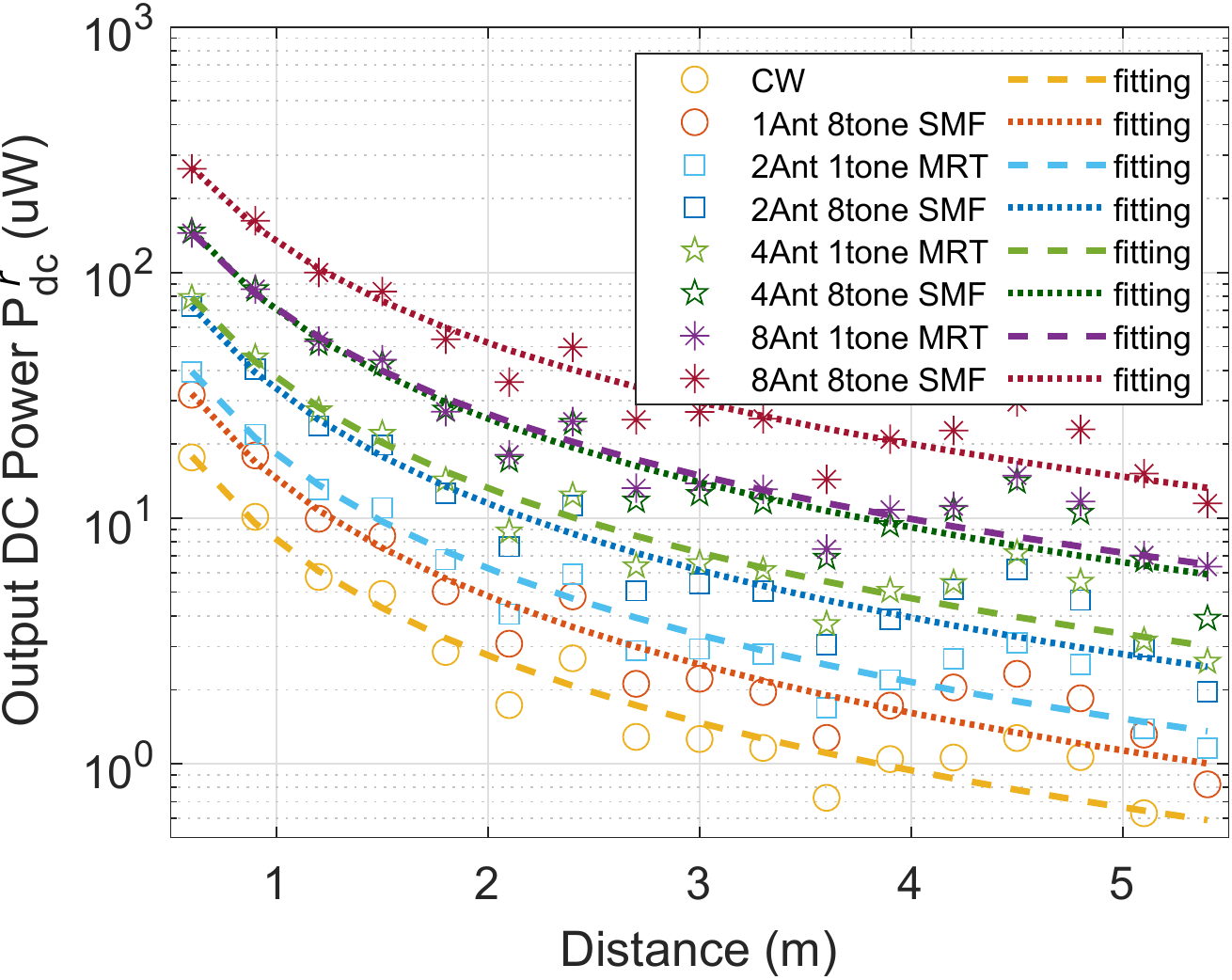}}
	\caption{Measurement results and curve fitting of output DC power ($P_{\dc}^r$) versus distance with $N$=1, 8 and $M$=1,2,4,8 \cite{Kim:2020}.}
	\label{fit_all}
\end{figure}

\begin{figure}[t]
   \centerline{\includegraphics[width=3.3 in]{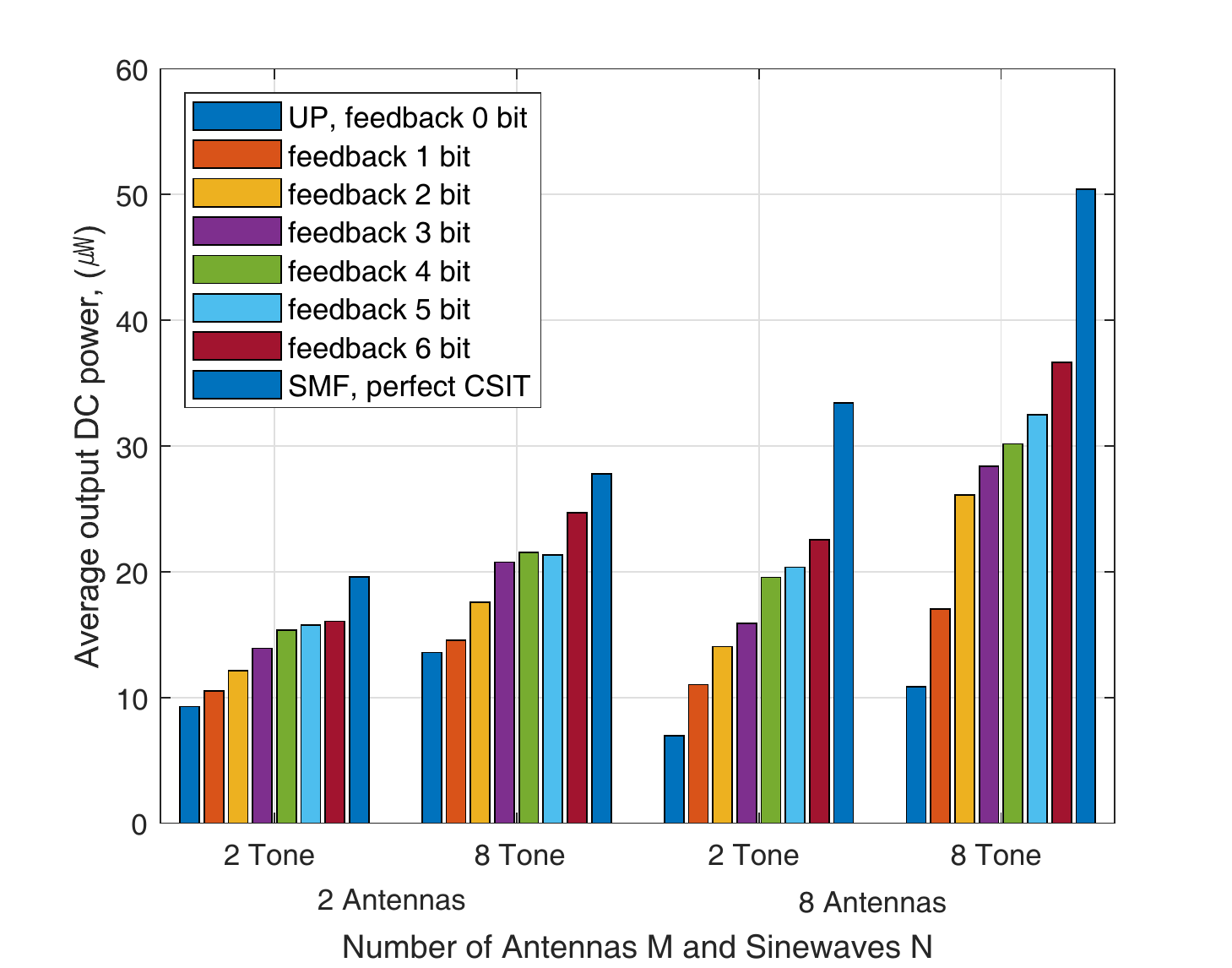}}
  \caption{Measurement results of output DC power $P_{\dc}^r$ of joint waveform and beamforming with limited feedback for $M=2,4$, $N=2,8$ and 1 to 6 bits of feedback \cite{Shen:2021}.}
  \label{LF}
\end{figure}

\par In order to complete the closed-loop configuration, both ET and ER are also equipped with a communication module to enable CSI feedback from the ER to the ET, respectively.
Two CSI acquistion strategies of Fig. \ref{channel_acquisition} are considered, namely forward-link training with very accurate over-the-air feedback and power probing with limited feedback. The former is consuming a lot of energy but is instrumental in providing an upper bound on the performance of this system since the CSIT is very accurate. The latter is very realistic and similarly to the prototype of subsection \ref{SKKU_920}, it is implemented using Zolertia RE-mote (Contiki operating system), which is equipped with ARM-based MCU and IEEE 802.15.4 RF interface. The RE-mote at the ER records the output DC voltage of the rectifier through the built-in ADC and seeks the highest value and corresponding codewords (i.e., finds the codeword in the codebook of joint waveform and beamformers that achieves the highest $P_{\dc}^r$) during the training phase as per Fig. \ref{wpt_frame}. Then, the RE-mote transmits the codeword index to the RE-mote at the ET side using IEEE802.15.4 interface. The ET-side RE-mote receives the codeword index and reports to the ET (USRP) using RS-232C interface. The codeword index fed back from the ER is the basis for signal generation in the consecutive frame. Following the training phase, energy is transmitted in the WPT phase using the optimized signal designed to span space (multi-antenna beamforming) and frequency (waveform) according to the selected codeword. This operation of training phase and WPT phase is repeated periodically with the period depending on the rate of change of the channel (e.g., the velocity of the ER).  

\par Figs.\ \ref{cdf_8ant} and \ref{fit_all} illustrate the output DC power $P_{\dc}^r$ and range of WPT achieved with MRT beamforming with 1, 2, 4, 8 transmit antennas and CW ($N=1$, 1 tone), as well as joint SMF waveform and beamforming ($N=8$, 8 tones) based on experimental data collected in a typical indoor environment at 2.4GHz under an EIRP of 36dBm \cite{Kim:2020}. In those figures, CSI acquisition is based on forward-link training with very accurate over-the-air feedback. Though the optimal design of joint waveform and beamforming results from the solution of an optimization problem \cite{Clerckx:2016b}, the simple and low complexity design of the joint waveform and beamforming based on SMF \eqref{SMF} is shown to significantly boost the output DC power and the range of WPT. We see that WPT efficiency can be enhanced by using either the frequency domain, the spatial domain or both domains together. In addition, a 1-tone (CW) waveform with 8 antennas shows similar performance as an 8-tone waveform with 4 antennas. Similarly, a 1-tone waveform with 4 antennas (1-tone with 2 antennas) offers similar performance as a 8-tone waveform with 2 antennas (8-tone with with 1 antenna). Hence increasing the number of tones from 1 to 8 has a similar effect as doubling the number of antennas. Such behavior shows that it is possible to replace the spatial domain processing (number of antennas) with the frequency domain processing (number of tones) and vice versa. The gains in terms of output DC power and range can also be accumulated using a joint beamforming and waveform strategy.

\par Results in Figs.\ \ref{cdf_8ant} and \ref{fit_all} were based on forward-link training with CSI feedback, which is too high energy  consuming  and/or too  complex for low power nodes. Next, we illustrate the performance of a more practical joint waveform and beamforming architecture based on power probing with limited feedback using off-the-shelf Zolertia nodes. Fig. \ref{LF} illustrates the experimental results obtained with such a strategy for 1 to 6 bits of feedback with $M=2,4$ and $N=2,8$ \cite{Shen:2021}. As $M$ and $N$ increase, we note the significant increase in $P_{\dc}^r$ but also the need for larger codebook sizes (and therefore larger feedback overhead) to come closer to the perfect CSIT performance.

\subsubsection{920 MHz and 2.4 GHz Distributed Antenna WPT Prototypes} The former two prototypes were based on co-located transmit antennas, i.e., antennas are separated by about half a wavelength. Another attractive architecture is to distribute the transmit antennas across the coverage area and enable some form of cooperation or coordination among those antennas \cite{Zeng:2017}. Such distributed antennas system (DAS) distributes energy more evenly in space and makes wireless power accessible ubiquitously compared to a co-located deployment. High energy beams in the direction of users are also avoided, which is desirable from a safety perspective.

\par The prototype of Section \ref{ICL_WPT} has recently been leveraged to demonstrate experimentally the benefits of WPT based on DAS at 2.4 GHz in  \cite{Shen:2020b}. It has been shown that WPT DAS relying on a low cost and low complexity dynamic selection of transmit antenna and frequency can boost $P_{\dc}^r$ by up to 30 dB in a single-user setting and broaden the coverage area. 

\par Another WPT DAS prototype at 920 MHz has been built and tested in \cite{Choi:2019a}. Here all distributed antennas fully cooperate such that the phases of the received signal from all transmit antennas are constructively aligned (based on MRT beamforming) at the receive antenna. Results show that the received RF power distribution with DAS is more evenly distributed than that of the co-located antennas. Consequently, it is shown that the coverage probability (e.g., achieving the received RF power larger than 0 dBm) is higher with DAS than with co-located antennas.

\subsubsection{5.8 GHz RIS-aided MIMO WPT Prototype}

The prototypes so far relied on active antenna arrays (providing an active beamforming gain) and on single-antenna ER. RIS in subsection \ref{channel} was introduced as a promising technology to engineer the channel and provide an additional passive beamforming gain. Moreover, subsection \ref{EH_section} advertised that the use of multi-antenna ER leads to MIMO WPT and can further boost the output DC power. 

\begin{figure}
    \centering
    \includegraphics[trim = {3in 2in 0.5in 2in},clip =true,width = 0.5\textwidth]{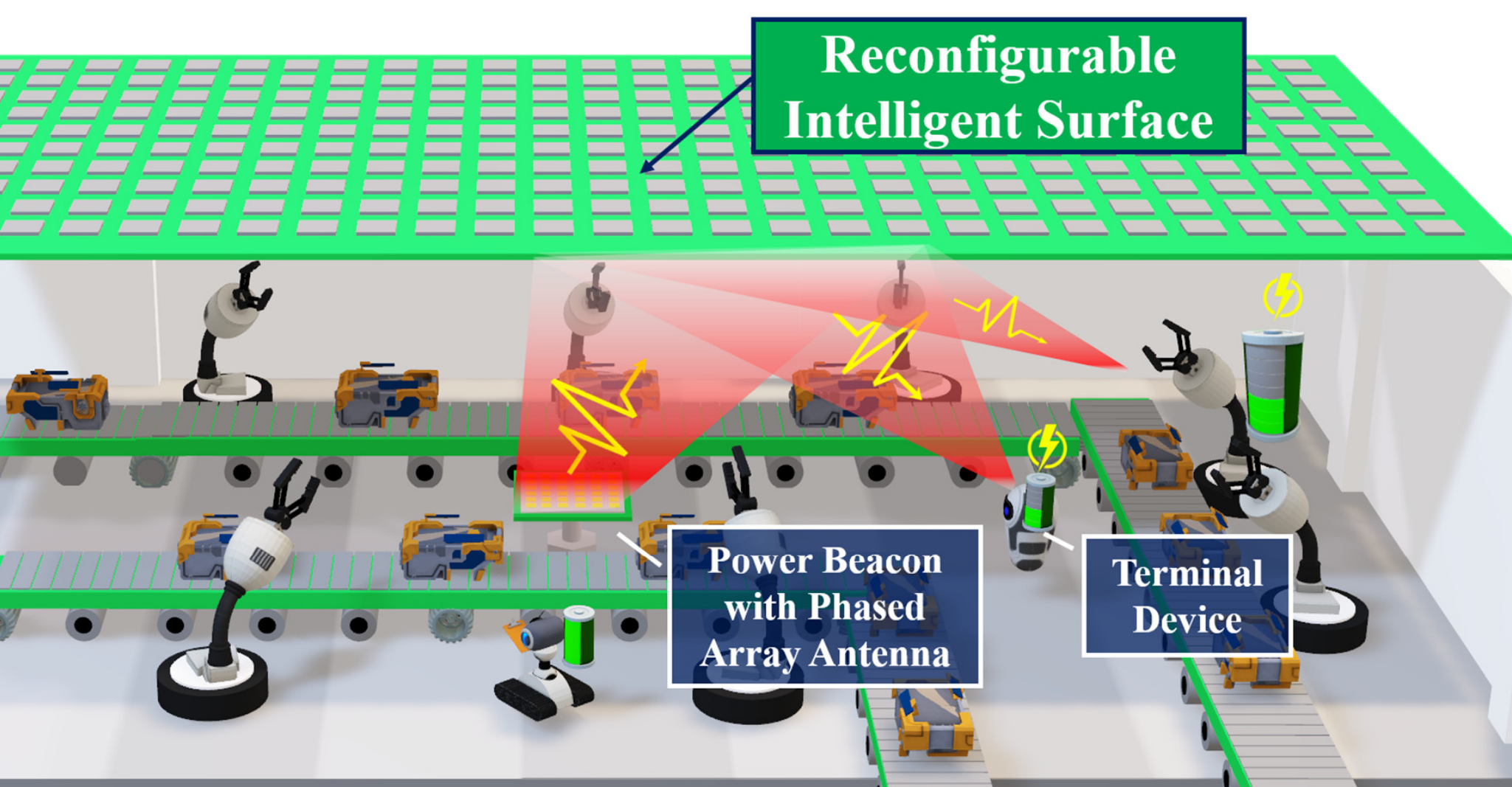}
    \caption{RIS-aided WPT conceptual application scenario.}
    \label{fig:RISconcept_apps}
\end{figure}

We have built a medium-range 5.8 GHz RIS-aided MIMO WPT system prototype using DC combining, jointly providing an active, passive, and combining gains \cite{Tran:2021}. This WPT system is integrated with an RIS with 256 unit cells to enhance the power transfer efficiency and extend the range. A conceptual application scenario is given in Fig.~\ref{fig:RISconcept_apps}, in which an RIS in the ceiling focuses power on the sensor device on the ground by reflecting an EM wave beam emitted by the ET (power beacon) equipped with a phased antenna array.

\par
\begin{figure}
    \centering
    \subfigure[5.8 GHz phased array transmitter]{
        \label{fig:Txphased_array}\includegraphics[trim ={0.1in 1in 0.25in 1in} , clip =true,width=5.5cm]{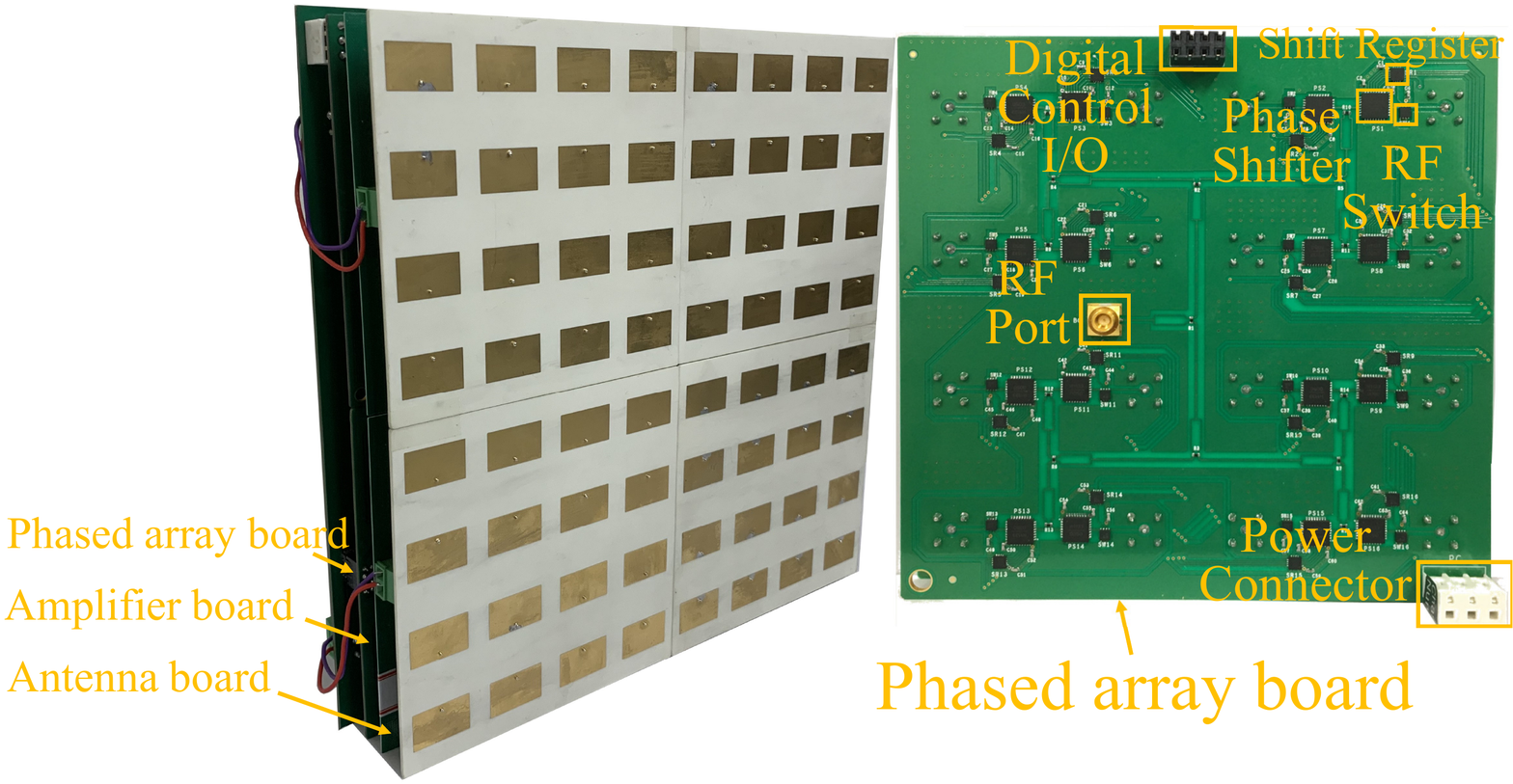}
        }
        \hspace{-0.25cm}
    \subfigure[Sensor device]{
        \label{fig:rx_board}\includegraphics[trim ={2.5in 1in 1.5in 0.8in} , clip =true,width=2.75cm]{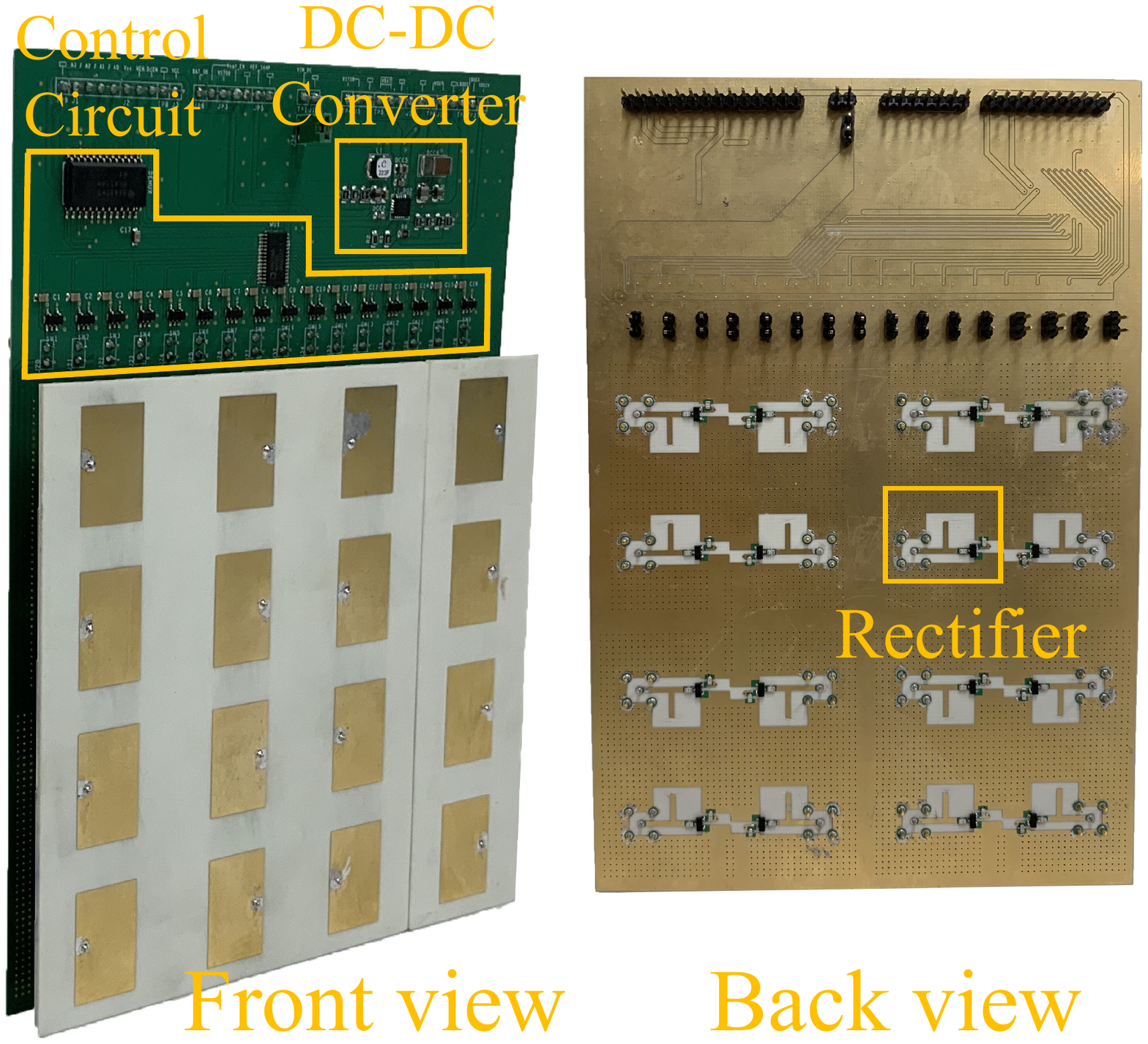}
        }
    \subfigure[RIS board]{
        \label{fig:RIS_board}\includegraphics[trim ={2.4in 0.75in 2in 1in} , clip =true,width=2.9cm]{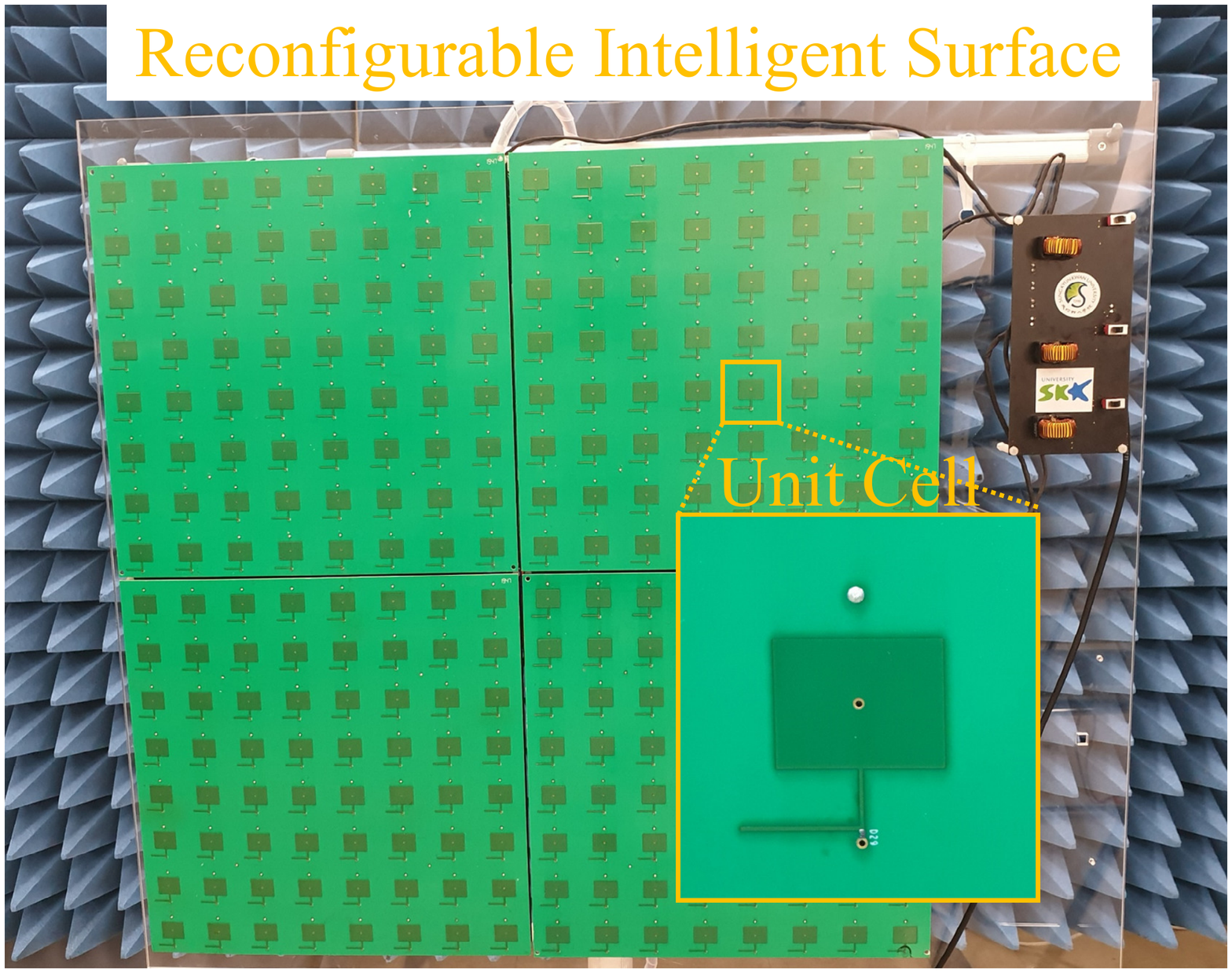}
        }
        \hspace{-0.25cm}
    \subfigure[Experiment set-up]{
        \label{fig:RIS_WPT_testbed}\includegraphics[trim ={0.75in 1in 0.9in 1.5in} , clip =true,width=5.45cm]{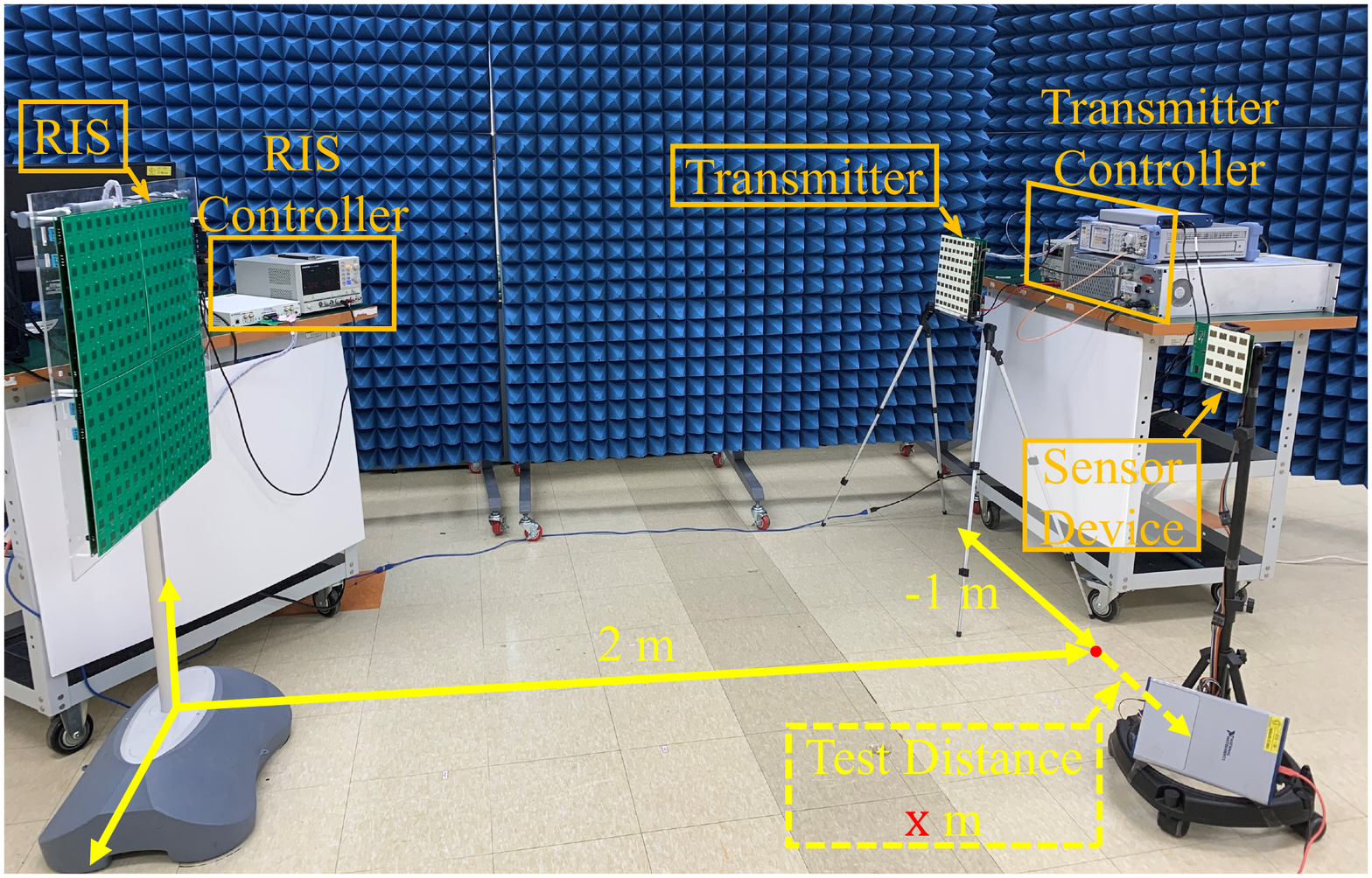}
        }
        
    \caption{5.8 GHz RIS-aided MIMO WPT prototype testbed.}
\end{figure}

 The ET is very similar to that of the 920 MHz WPT system of subsection \ref{SKKU_920}. It is equipped with a rectangular 8$\times$8 phased antenna array.
We have adopted a sandwich structure stacking four boards, which are power divider, phase control, amplifier, and antenna boards, by using board-to-board connectors (Fig.~\ref{fig:Txphased_array}).
The sensor device is equipped with a 4$\times$4 antenna array (Fig.~\ref{fig:rx_board}).
The rectifier attached to each antenna element is a one-stage Dickson charge pump. In this system, we use a DC combining technique which combines the DC outputs of the 16 rectifiers in parallel (see Fig. \ref{fig_mimo} (a)). 
The combined DC power is delivered to a DC-DC converter (TI BQ25570) for the maximum power point tracking (MPPT), voltage regulation, and battery management.

\par We have fabricated a 16$\times$16 RIS with 256 unit cells as shown in Fig.~\ref{fig:RIS_board}. Each unit cell is a rectangular patch resonating at 5.8 GHz, equipped with a single PIN diode for one-bit control. The incident wave on each unit cell travels to the PIN diode, and is reflected back to be re-radiated to the air.
The unit cell is designed in such a way that the phase of the reflected wave has a 180 degree difference according to the ON/OFF states of the PIN diode. The RIS is controlled by an FPGA and a control board for serial-to-parallel conversion, which consists of shift registers, decoders, and D-type flip flops.

\begin{figure}
    \centering
    \subfigure[Phase of the $8\times8$ array transmitter]{
        \label{fig:tx_phase}\includegraphics[trim = {1.5in 3in 1.5in 3in}, clip = true, width=4.1cm]{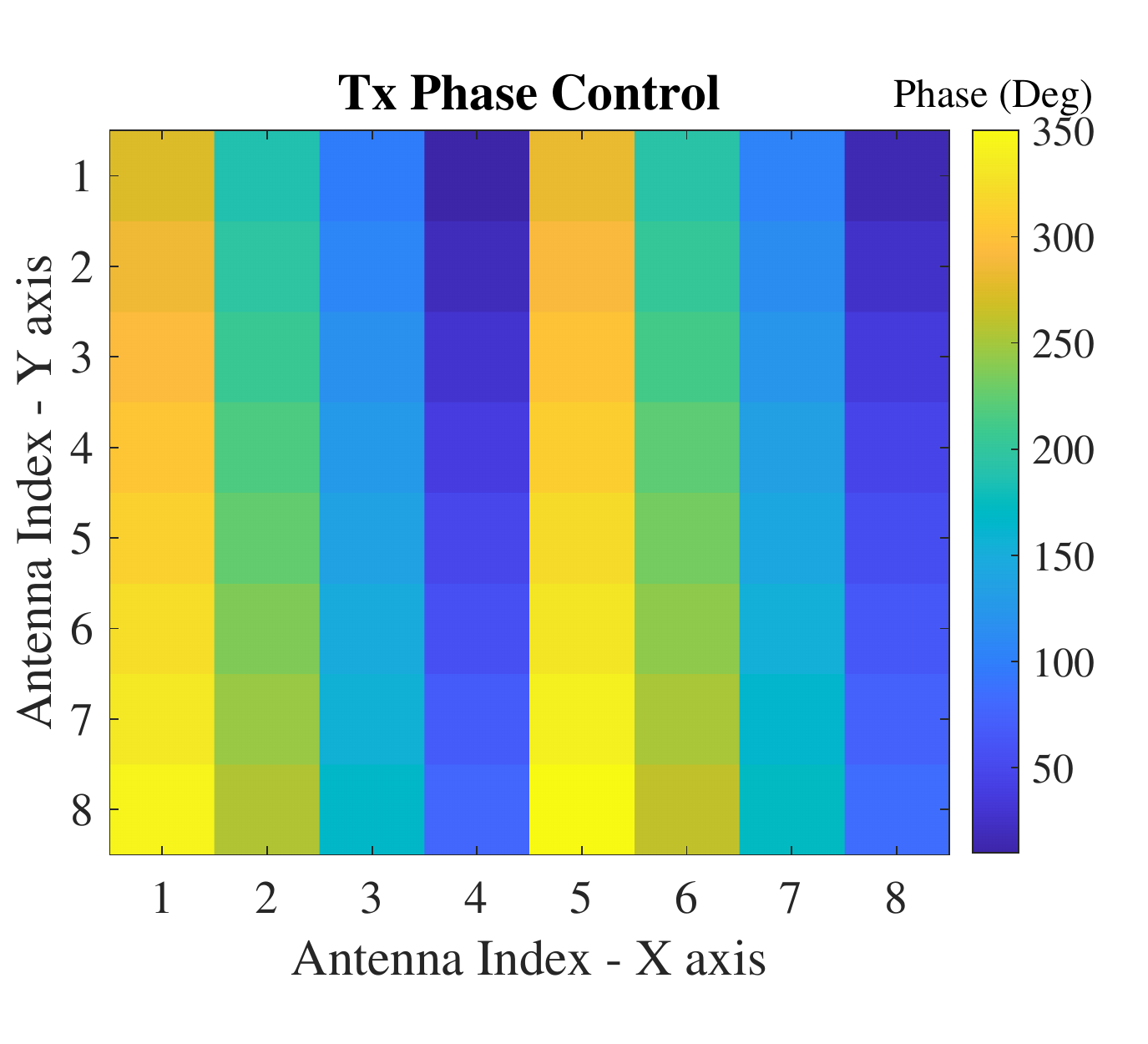}
        }
    \subfigure[Phase control of the $16\times16$ RIS]{
        \label{fig:RIS_pat}\includegraphics[trim = {1.5in 3in 1.5in 3in}, clip = true,width=4.1cm]{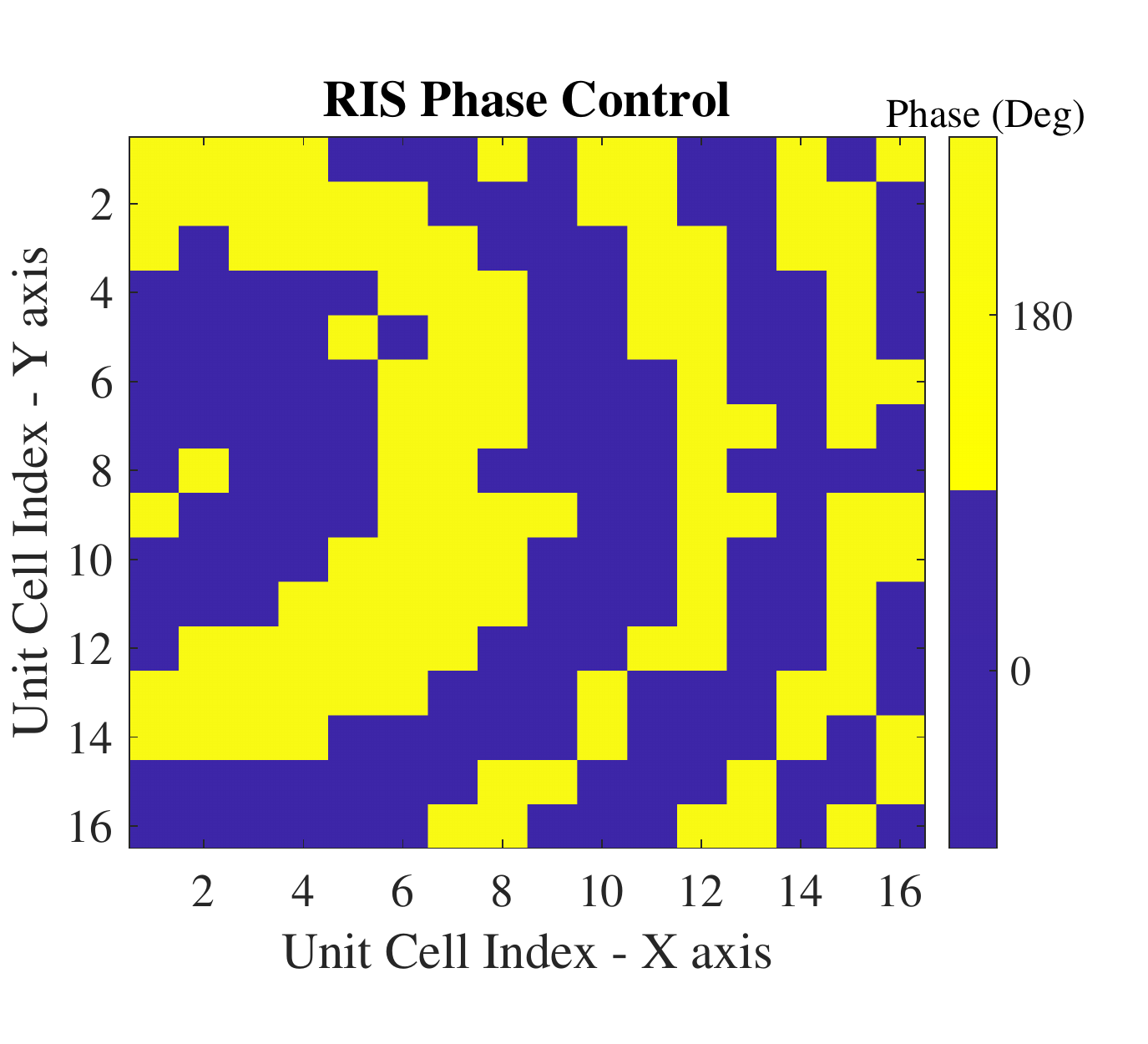}
        }
    \subfigure[Received DC power over distance]{
        \label{fig:rxpower_distance}   \includegraphics[trim = {0.5in 0.25in 0.75in 0.5in},clip =true,width = 0.4\textwidth]{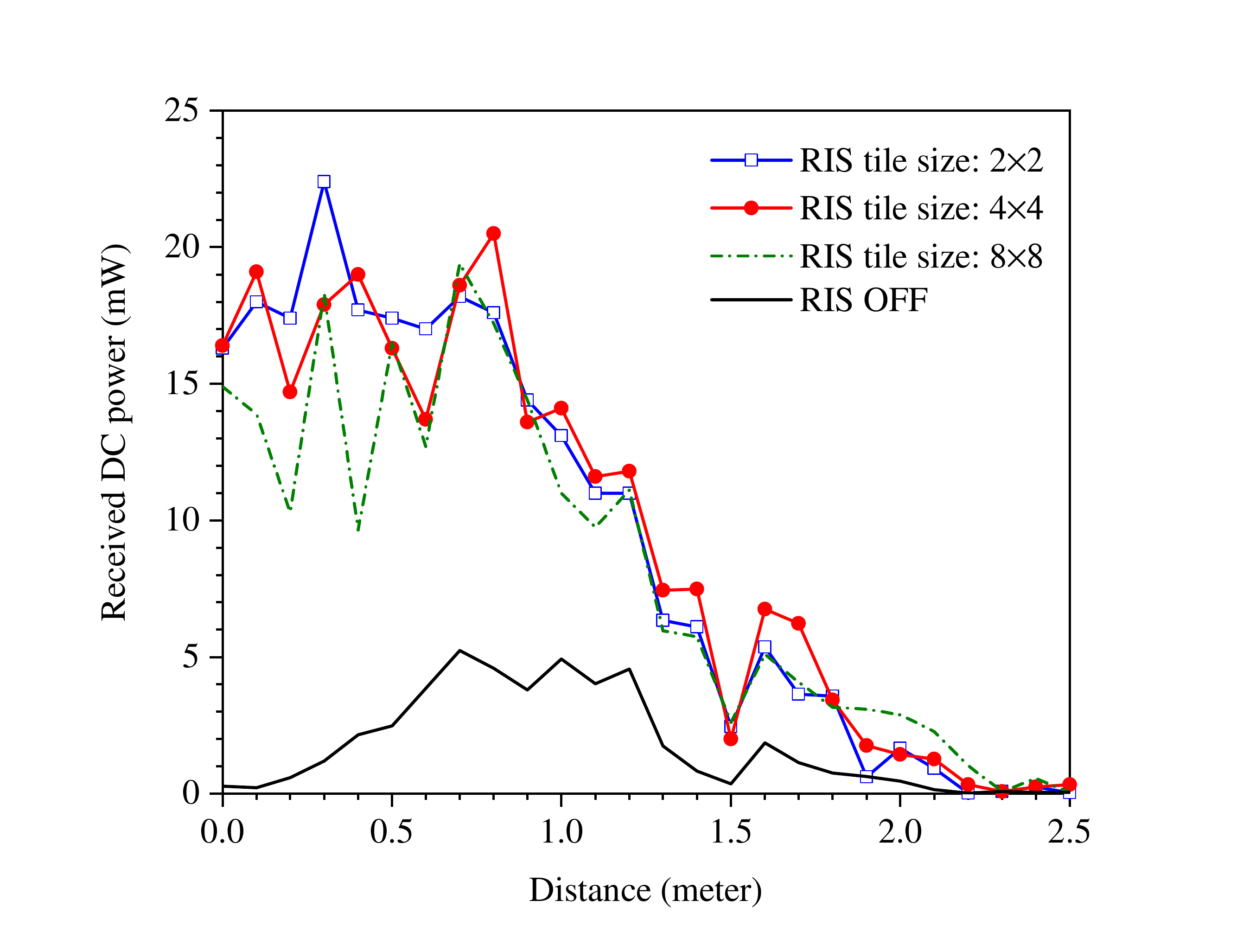}
    }
    \caption{RIS-aided WPT experimental results.}
    \label{fig:exp_results}
\end{figure}

We have proposed a power-based scanning algorithm based on the power probing CSI acquisition strategy specifically designed for RIS, so as to enhance the beam focusing capability and the received power at the sensor device. For the scanning algorithm, the RIS is divided into smaller RIS tiles.
Based on the power level feedback from the device, each RIS tile is iteratively optimized to constructively add up the received power at the sensor device. We have performed experiments to verify the proposed algorithm. The transmitter and sensor device are located at 2 meters away from the RIS as shown in Fig.~\ref{fig:RIS_WPT_testbed}. 
The sensor device is moved to assess the performance at different distances $x$ (see Fig.~\ref{fig:RIS_WPT_testbed}) from 0 meters to 2.5 meters. Fig.~\ref{fig:rxpower_distance} shows the received DC power over the test distance with different RIS tile sizes. 
In this test, the total transmit power from all antenna elements is 2 W. 
It can be observed that around 22 mW is transferred to the sensor when RIS tile size $2\times 2$ (4 unit cells) is used at the test distance of $x = 0.3\text{ m}$, which is significantly higher than the power level collected at the sensor in the absence of RIS.
The corresponding transmitter and RIS phase control states are given in Figs.~\ref{fig:tx_phase} and Fig.~\ref{fig:RIS_pat}, respectively. 
In these figures, we can see that the transmitter directs the beam towards the RIS, and the RIS makes the reflected beam focused onto the receiver by forming a convex lens-like phase pattern.

\section{Wireless Information and Power Transfer}\label{Section_WIPT}
Building upon WPT, we can now investigate WIPT. WIPT can be categorized into three different types.
\par \textit{Simultaneous Wireless Information and Power Transfer (SWIPT)}: The term was first coined in \cite{Zhang:2013} and refers to scenarios where information and power are simultaneously sent from the transmitter(s) to the receiver(s)\cite{Varshney:2008,Grover:2010,Zhang:2013,Zhou:2013,Park:2013}. The information receiver(s) (IR) and ER can be co-located or separated. With co-located receivers, a single (typically low-power) device equipped with an IR and ER is simultaneously wireless powered and receiving data. With separate receivers, the ER and IR refer to different devices, with ER being a low-power device being wireless powered and IR being a device receiving data.
\par \textit{Wirelessly Powered Communication Networks (WPCNs)}: Energy is transmitted in the downlink from a base station to a device and information is transmitted in the uplink \cite{Ju:2014}. The device harvests energy in the downlink and uses the harvested energy to transmit data in the uplink.
\par \textit{Wirelessly Powered Backscatter Communication (WPBC)}: the downlink is used to transmit energy to a device and the uplink is used to transmit information using backscatter modulation from a tag to a reader by reflecting and modulating the incoming RF signal \cite{Han:2017}. Thanks to backscatter communications, tags do not require oscillators to generate carrier signals and power consumption can be decreased by several orders-of-magnitude compared to conventional wireless communications \cite{Boyer:2014}.

\par Moreover, a network could have a mixture of all of these types of transmissions with multiple co-located and/or distributed ETs and information transmitter(s) (IT).
\par In the sequel, we further detail SWIPT architectures, prototytpes and experiments.

\subsection{Architecture} 
We can now build upon Section \ref{WPT_section} to integrate communications and power into a SWIPT architecture. 
\subsubsection{Transmitter} Since both information and power are transmitted simultaneously, the transmit signal \eqref{WPT_1} needs to be upgraded with the weight vector $\mathbf x_{n}$ replaced by a time-varying $\mathbf x_{n}(t)$ (due to modulation)  that contains random complex-valued information and power carrying symbols (drawn from a certain input distribution), instead of  just a deterministic power carrying symbol.
\subsubsection{Receiver}
\par
\begin{figure}
\begin{center}
\subfigure[Ideal Receiver]{\scalebox{0.5}{\includegraphics*{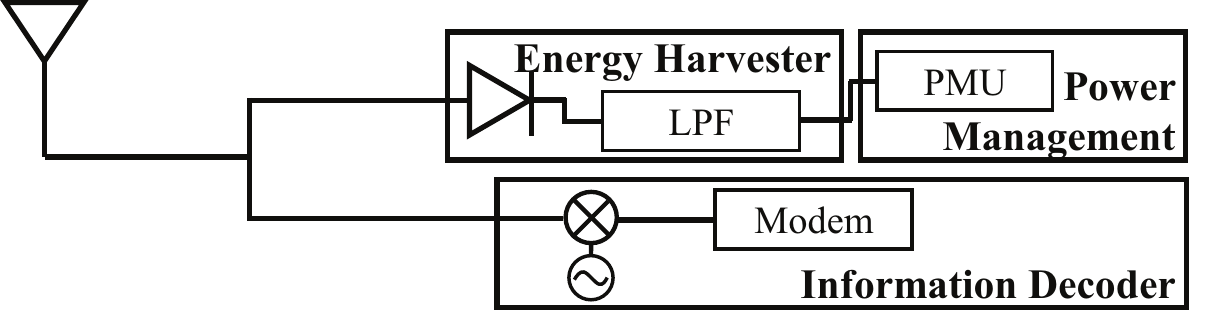}}}
\subfigure[TS Receiver]{\scalebox{0.5}{\includegraphics*{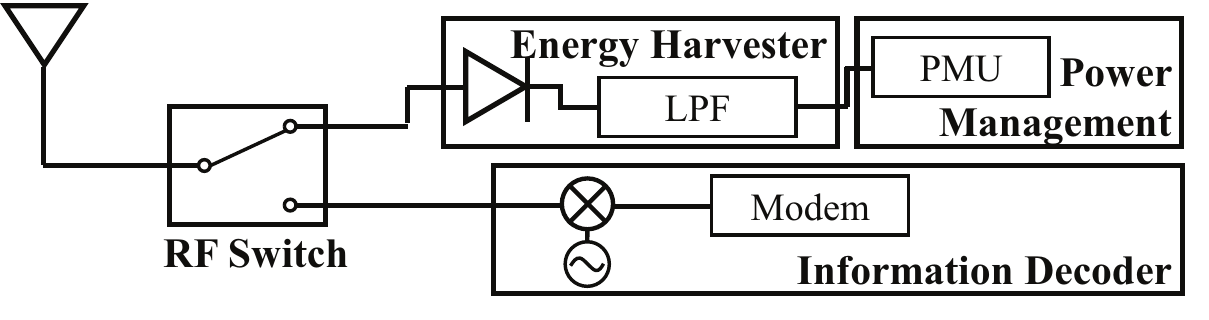}}}
\subfigure[PS Receiver]{\scalebox{0.5}{\includegraphics*{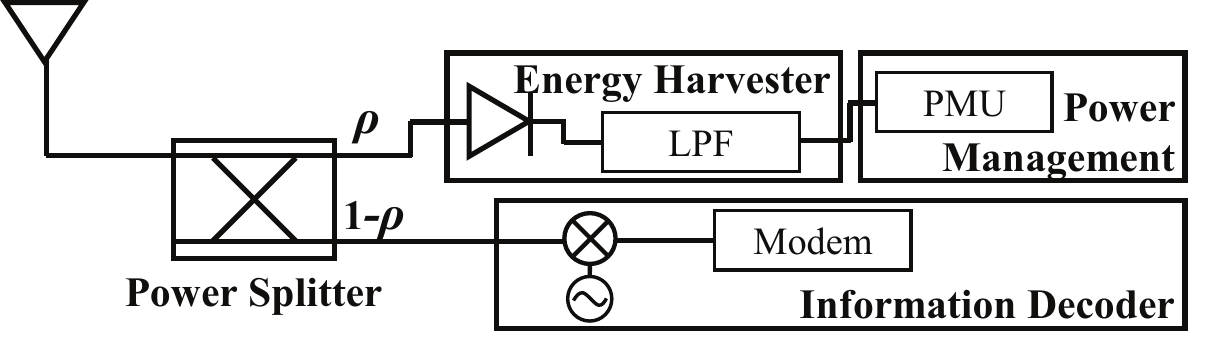}}}
\subfigure[Integrated Receiver]{\scalebox{0.5}{\includegraphics*{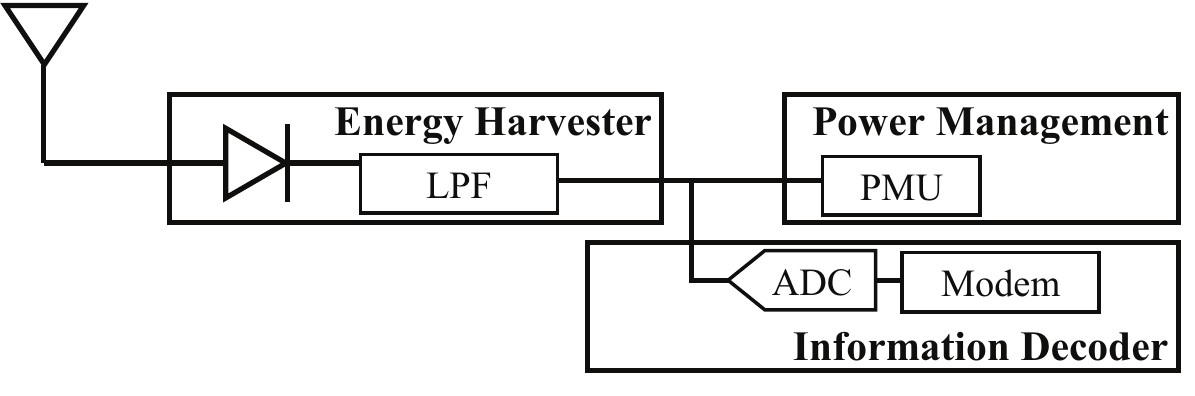}}}
\caption{Four single-antenna receiver architectures for SWIPT: (a) Ideal receiver (using the same signal for both information and energy receivers); (b) TS receiver (switching the signal to either information or energy receiver); (c) PS receiver (splitting a portion of the signal to information receiver and the rest to the energy receiver); and (d) Integrated receiver (demodulation is performed after rectification).}\label{receiver_architectures}
\end{center}
\end{figure}

 Several receiver architectures for SWIPT have been proposed in Fig. \ref{receiver_architectures} that integrate information decoder (ID) and energy harvester (EH). 

\par An \textit{Ideal Receiver} (Fig. \ref{receiver_architectures}(a)) assumes the same signal $y_{\rf}(t)$ is simultaneously conveyed to the harvester and to the decoder \cite{Varshney:2008,Grover:2010}. No practical circuits can currently realize this operation. 

\par A \textit{Time Switching (TS) Receiver} (Fig. \ref{receiver_architectures}(b)) consists of a co-located decoder and harvester (following the structure in subsection \ref{EH_section}) \cite{Zhang:2013,Zhou:2013,Park:2013}. The transmission is divided into two orthogonal time slots, one for WPT and the other for WIT. In each slot, the transmit signal is optimized for either WPT or WIT and the receiver switches accordingly its operation periodically between harvesting energy and decoding information. 

\par In a \textit{Power Splitting (PS) Receiver} (Fig. \ref{receiver_architectures}(c)), the harvester and decoder are the same as those of a TS receiver. The transmit signals are jointly optimized for information and energy transfer and the received signal is split into two parts, where one part with PS ratio $0 \leq \rho \leq 1$ is conveyed to the harvester, and the other with power ratio $1-\rho$ is utilized by the decoder. \cite{Zhang:2013,Zhou:2013,Liu:2013}. Assuming perfect matching, the input voltage signals $\sqrt{\rho R_{\ant}}y_{\rf}(t)$ and $\sqrt{(1-\rho)R_{\ant}}y_{\rf}(t)$ are, respectively, conveyed to the harvester and the decoder.  

\par In the \textit{Integrated Receiver} (Fig. \ref{receiver_architectures}(d)), the received RF signal is first rectified and then the rectified voltage signal is used as an input signal to the ID receiver\cite{Zhou:2013}. Hence, in contrast to prior architectures, the received RF signal is not divided into EH and ID receivers, instead the received RF signal is fully harvested by EH, so it can maximize RF-to-DC conversion efficiency. However, in terms of information decoding, the rectified voltage signal looses some of the characteristics of the original signal during the rectification process so that the information decoding performance (e.g., rate, throughput) are degraded, though a proper signal design can help improving the performance. Such integrated architecture eliminates the need for energy-consuming RF components such as local oscillators and mixers needed for the ideal, TS and PS architectures, therefore making it more suitable for low power applications not demanding high throughput. 

\subsubsection{Signal Optimization}

\par The trade-off between rate and energy in SWIPT is captured by the Rate-Energy (R-E) region $\mathcal{C}_{\R-\E}$ that contains all pairs of rate $R$ and energy $E$ that can simultaneously be achieved/harvested. It is obtained through a collection of input distributions $p(\mathbf{x}_0,...,\mathbf{x}_{N-1})$ that satisfies the average RF transmit power constraint $P_{\rf}^t\leq P$ and can be written mathematically as
\begin{multline}\label{RE_region_def}
\mathcal{C}_{\R-\E}(P)\!\triangleq\!\bigcup_{p}\Bigg\{(R,E):R\leq \sum_{n=0}^{N-1} I_n, E\leq P_{\dc}^r \Bigg\}
\end{multline}
where $I_n$ refers to the rate (measured in terms of mutual information between the channel input and the channel output) on subband $n$. Both $I_n$ and $P_{\dc}^r$ are functions of the input $\mathbf{x}_0(t),...,\mathbf{x}_{N-1}(t)$. 

\par The R-E region can be identified by calculating the capacity (supremization of the mutual information over all possible input distributions $p(\mathbf{x}_0,...,\mathbf{x}_{N-1})$) of a channel subject to an average RF transmit power constraint $P_{\rf}^t\leq P$ and a receiver delivered power constraint $P_{\dc}^r\geq \bar{E}$, for different values of $\bar{E}\geq 0$. Namely,
\begin{align}
\sup_{p(\mathbf{x}_0,...,\mathbf{x}_{N-1})} \hspace{0.3cm}& \sum_{n=0}^{N-1}I_n \label{MI}\\
\mathrm{subject\,\,to} \hspace{0.3cm}  
& P_{\rf}^t\leq P,\\
& P_{\dc}^r\geq \bar{E},\label{MI_2}
\end{align}
where $\bar{E}$ denotes the minimum required or target delivered power. The above problem can be tackled for the receiver architectures of Fig. \ref{receiver_architectures} along with the specific parameters of the receiver \cite{Clerckx:2019}. For instance, with the TS receiver, different R-E trade-offs are achieved by varying the length of the WPT time slot, jointly with the transmit signals; while with the PS receiver, different R-E trade-offs are obtained by adjusting the value of $\rho$ jointly with the transmit signals.
\par There is no communication or information transmission without randomness in the signal. Hence, designing  SWIPT requires  the  transmit  signals to be subject to some randomness, which has an impact on the amount of harvested DC power. This raises interesting  questions  on  how  modulation and waveform need to be re-thought for SWIPT.

\par Starting with the modulation design in single-subband ($N=1$), as a consequence of \eqref{vout_def} and Observation \ref{higher_order}, conventional modulation (such as PSK and QAM) and input distribution (such as circularly symmetric complex Gaussian - CSCG) used in communications are clearly suboptimal for SWIPT. Indeed, \eqref{vout_def} and Observation \ref{higher_order} call for modulation and input distribution with large higher moments so as to boost $\mathbb{E}\left[y_{\rf}(t)^4\right]$. Consequently, to obtain the best trade-off between rate and energy, the modulation should change according to the target delivered power $\bar{E}$. Indeed, it is remarkably shown in \cite{Varasteh:2017b} that the information theoretic optimal solution to Problem \eqref{MI}-\eqref{MI_2} is obtained by adopting a combination of CSCG and on-off-keying inputs with CSCG used for low target delivered power $\bar{E}$ and on-off keying increasingly used as $\bar{E}$ increases. This is due to the fact that on-off keying can be adjusted such that one constellation point is at 0 and is drawn with probability $1-\frac{1}{l^2}$ and the other constellation point has an amplitude $l\sqrt{P}$ and is drawn with probability $\frac{1}{l^2}$, for $l\geq 1$. Such a modulation has the same average power $P$ for all values of $l$ but has a fourth moment that increases with $l$. Consequently, as $l$ increases, such a modulation exhibits a low probability of high amplitude that leads to increasing higher moments $\mathbb{E}\left[y_{\rf}(t)^4\right]$ and therefore $P_{\dc}^{r}$.  

\par
\begin{figure}
	\centerline{\includegraphics[width=0.8\columnwidth]{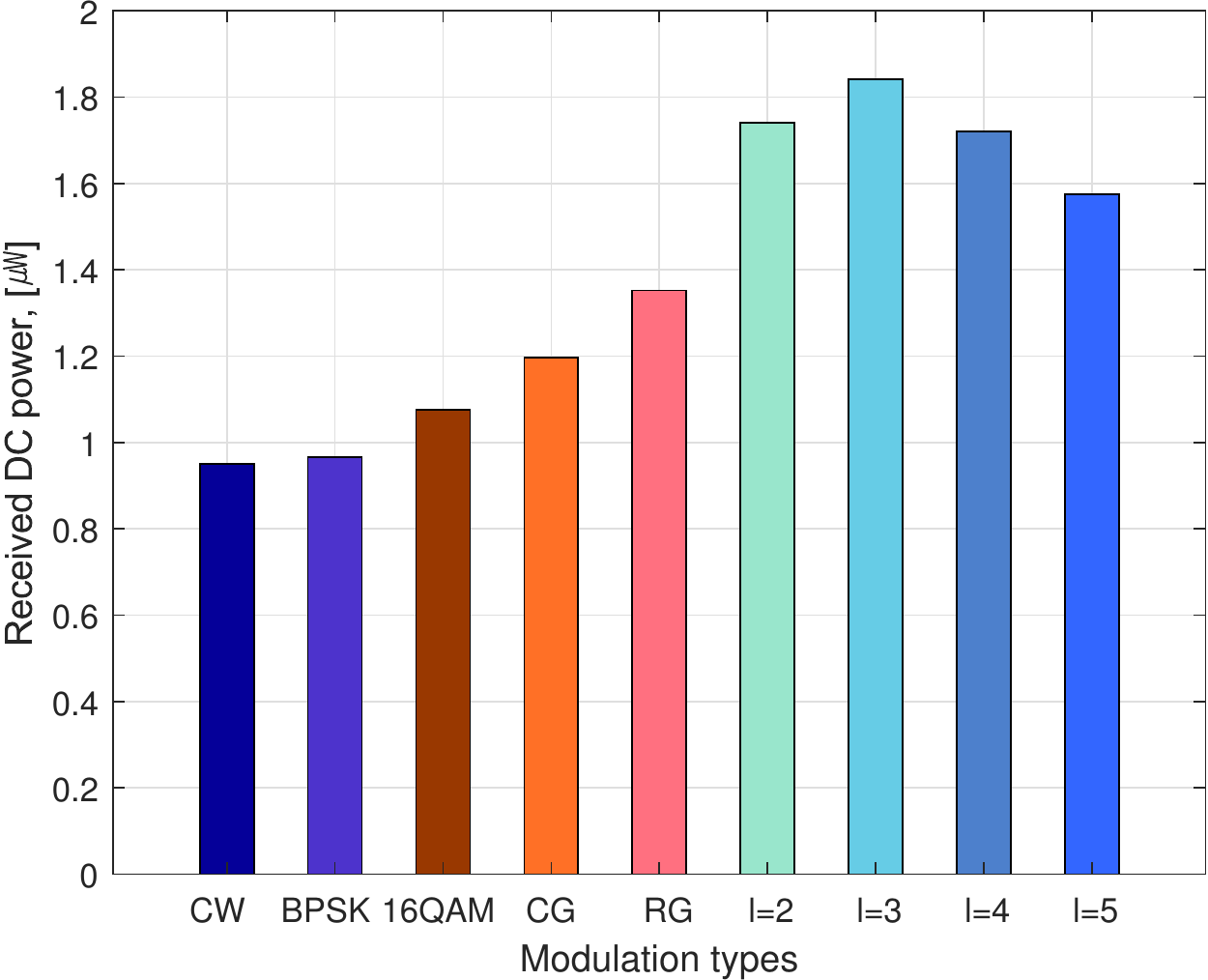}}
	\caption{Measured output DC power ($P_{\dc}^r$) of conventional and WPT-optimized modulations \cite{Kim:2020}. CG refers to CSCG, and RG to real Gaussian. $l$=1,2,3,4,5 refer to on-off keying modulations.}
	\label{modulation_fig}
\end{figure}

 On-off keying has been shown to significantly boost $e_3$ conversion efficiency over various baselines \cite{Kim:2020,Varasteh:2017b}. In Fig.\ \ref{modulation_fig}, we display measurement results of various modulations using the rectifier of Fig.\ \ref{TD_schematic}(a). We note that on-off keying ($l$=1,2,3,4,5) provides almost two times more output DC power ($P_{\dc}^r$) than conventional modulations as BPSK and 16-QAM though all have the same average RF input power ($P_{\rf}^r$). 

\par
\begin{figure}
	\centerline{\includegraphics[width=0.9\columnwidth]{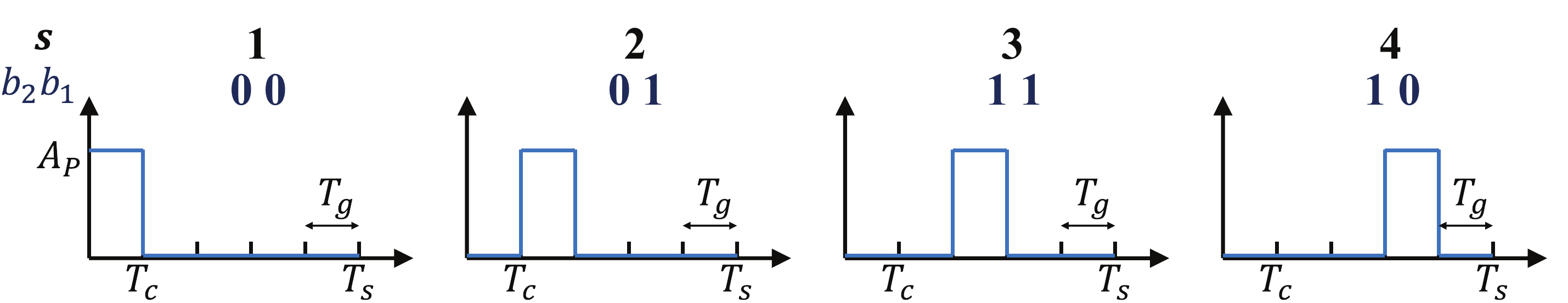}}
	\caption{Example of PPM ($M$=4) symbols and baseband signal \cite{Kim:2021}.}
	\label{ppm_symbols}
\end{figure}

 Such on-off keying can be used in the ideal, TS and PS receiver. Inspired by the benefits of on-off keying, another type of modulation particularly suited for the integrated receiver has been designed in \cite{Kim:2021}. The so-called $M$-PPM is a modified version of conventional pulse-position modulation (PPM) suited for SWIPT integrated receiver architecture ($M$ represents a modulation order) as illustrated in Fig. \ref{ppm_symbols}. The information is encoded on the pulse position during the symbol period of the time domain signal, and all the transmit signal power is concentrated on the corresponding pulse, thus increasing the PAPR of the signal and improving harvested DC power at the receiver (since high PAPR pulse boosts $\mathbb{E}\left[y_{\rf}(t)^4\right]$ in \eqref{vout_def}). In terms of information transfer, the PPM is a non-coherent modulation that does not need a power-consuming local-oscillator at the receiver; thus, it is suitable for low-power integrated SWIPT receiver. 

\par In the multi-subband case ($N>1$), waveform and modulation/input distribution also need to be tailored for SWIPT. Remarkably, as a consequence of Observation \ref{higher_order}, it was shown in \cite{Clerckx:2018b} that the solution to Problem \eqref{MI}-\eqref{MI_2} is such that the input distribution/modulation on each subband, i.e., the entries of $\mathbf x_{n}(t)$, are non-zero mean in general. Specifically, for $\bar{E}=0$, the distribution is zero mean on all subbands and the rate is maximized, and as $\bar{E}$ increases, the waveform becomes more and more deterministic, i.e., the variance of the distribution on each subband decreases, and the rate decreases. Ultimately the waveform reaches the perfectly deterministic multisine waveform of WPT in \eqref{WPT_1} in the limit of large $\bar{E}$ and zero rate. Concretely, what that suggests is that the waveform for SWIPT can be written as a superposition of two waveforms, one deterministic/unmodulated waveform for WPT in the form of a multisine contributing to the non-zero mean (i.e., DC bias), and one modulated waveform for communication in the form of OFDM with zero mean inputs \cite{Clerckx:2018b}.   

\par The benefit of superposed waveform (or equivalently the superiority of non-zero mean inputs over zero mean inputs) can be explained by the fact that a multi-carrier unmodulated waveform, e.g., multisine, is more efficient in boosting $\mathbb{E}\left[y_{\rf}(t)^4\right]$ and therefore $P_{\dc}^r$ compared to a conventional multi-carrier modulated communication waveform \cite{Clerckx:2018b}. 

\subsection{Prototypes and Experiments}
The previous section highlighted key SWIPT features. Though SWIPT prototypes and experiments remain less investigated compared to WPT, we detail and illustrate the real-world performance benefits of several state-of-the-art SWIPT prototypes developed at ICL \cite{Kim:2019, Kim:2021} and SKKU \cite{Choi:2020}. A special emphasis is put on the effect of modulation, superposed waveform and the role of receiver design. A particular attention is drawn to the integrated receiver which is deemed to be more practical for low power applications. Readers are also referred to \cite{Claessens:2018,Rajabi:2018} for other experiments on SWIPT with integrated receivers.

\subsubsection{920 MHz SWIPT Prototype with Integrated Receiver}

We have built and tested a full-fledged SWIPT system prototype with an integrated receiver operating at 920 MHz \cite{Choi:2020}. This prototype is a Single-Input Single-Output (SISO) testbed with single-antenna transmitter and receiver. Uniquely, the transmitter of this SWIPT system sends the power via a high-power CW signal at the carrier frequency while the data is transmitted via a low-power modulated signal around the carrier frequency. The integrated receiver first rectifies the received signal, and then the rectified signal is split into two branches: the power and communication branches. The CW power signal is routed to the power branch for WPT, while the modulated communication signal is fed into the communication branch for decoding. 

The DC-biased OFDM is a common modulation technique for an intensity channel without phase information, for example, visible light communications (VLC). Since we use an integrated receiver, in which the information decoder receives a rectified signal, the channel has only intensity information and the DC bias is provided by the CW power signal. Therefore, the DC-biased OFDM is the most appropriate choice for this experiment. The spectral efficiency is halved since the phase information is lost in the intensity channel. 
The communication in this experiment exploits very wide difference in the power requirements of WPT and communications. Since the communication signal is allocated a much lower power than the CW power signal, the communication signal goes through a linear channel in the rectifier due to the small signal assumption. Thus, we can avoid the intermodulation distortion, and the OFDM modulation is possible. At the rectifier, the received signal amplitude is almost constant since the communication power is very small. Therefore, we do not exploit the non-linearity of the rectifier in this experiment.

\par
\begin{figure}
    \centering
    \subfigure[Transmitter architecture]{
        \label{fig:txarchitecture}\includegraphics[width=6cm] {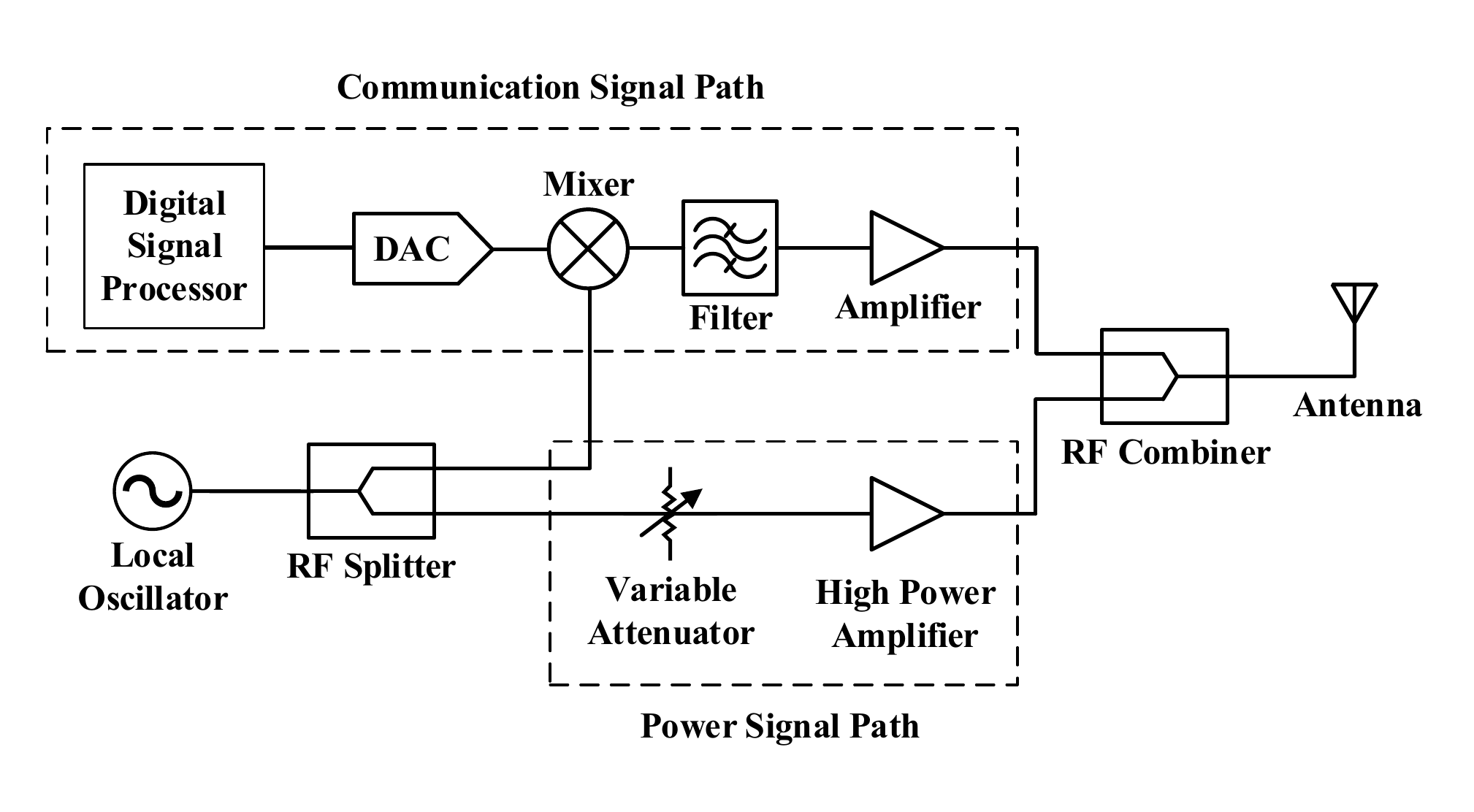}
    }\\
    \subfigure[Receiver architecture]{
        \label{fig:rxarchitecture}\includegraphics[width=6cm] {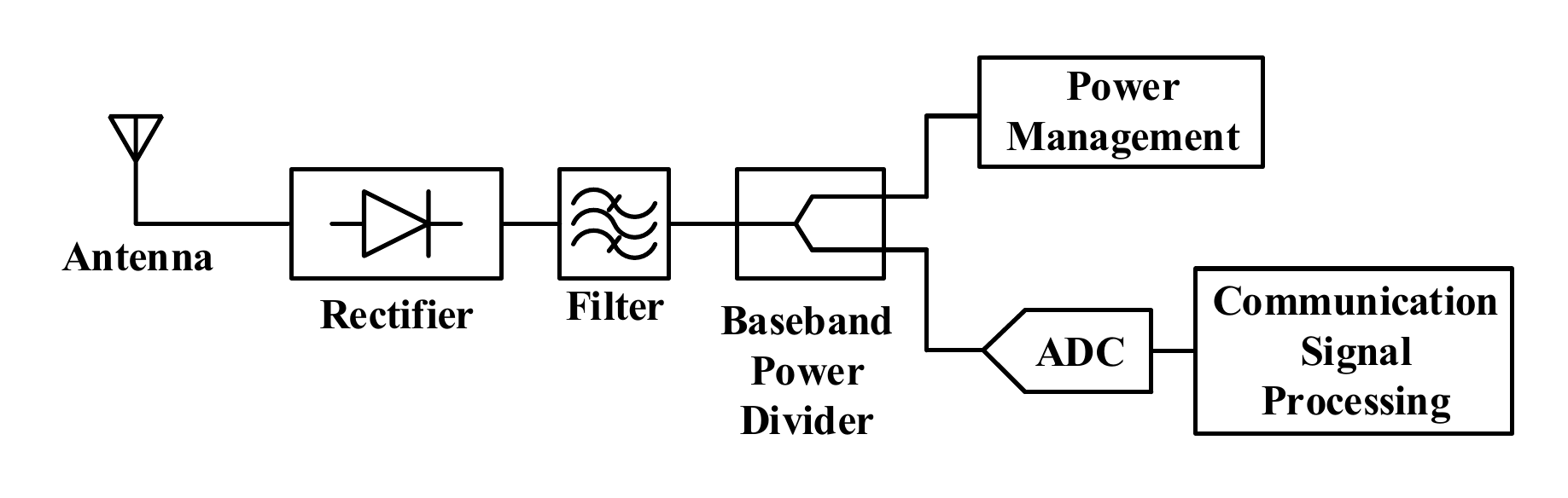}
    }
    \caption{SWIPT architecture.} 
\end{figure}

\begin{figure}
    \centering
    \subfigure[Receiver board photo]{
        \label{fig:receiverphoto}\includegraphics[width=6cm, bb=0.16in 0in 13in 5.5in] {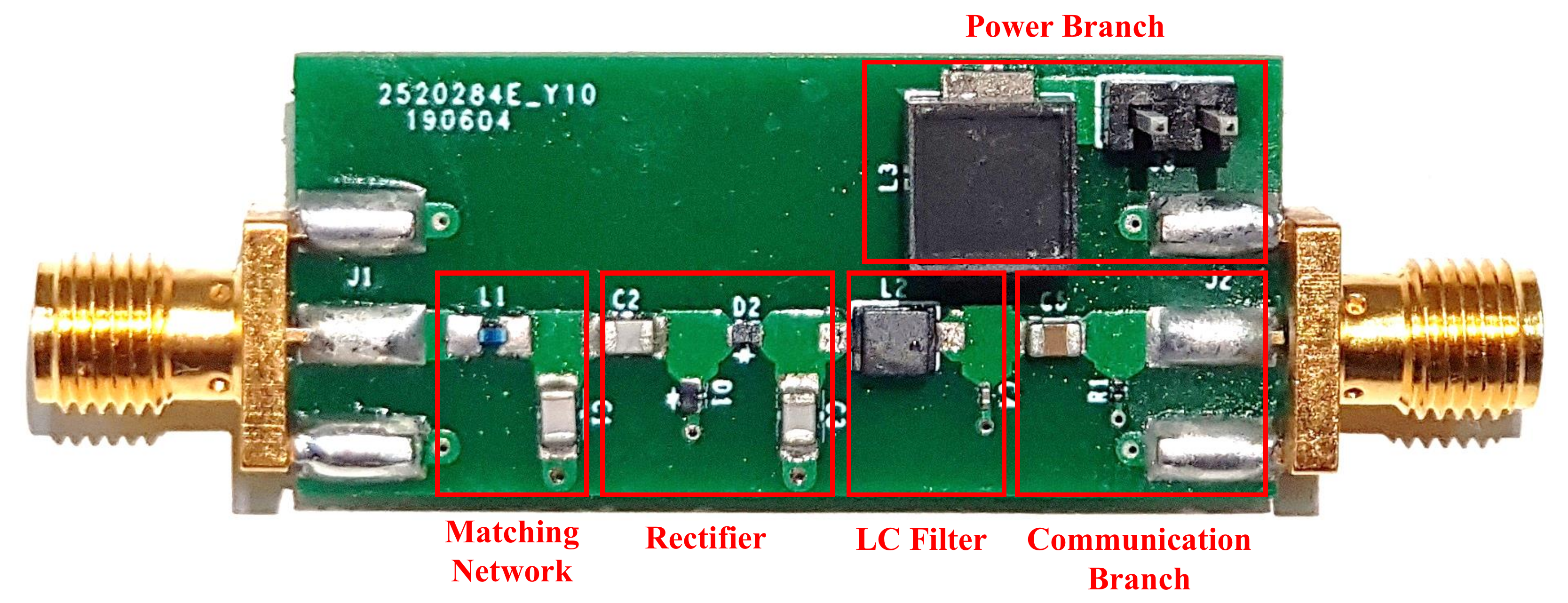}
    }
    \subfigure[SWIPT system testbed photo]{
        \label{fig:swipttestbed1}\includegraphics[width=6cm, bb=0in 0in 11.3in 7.5in] {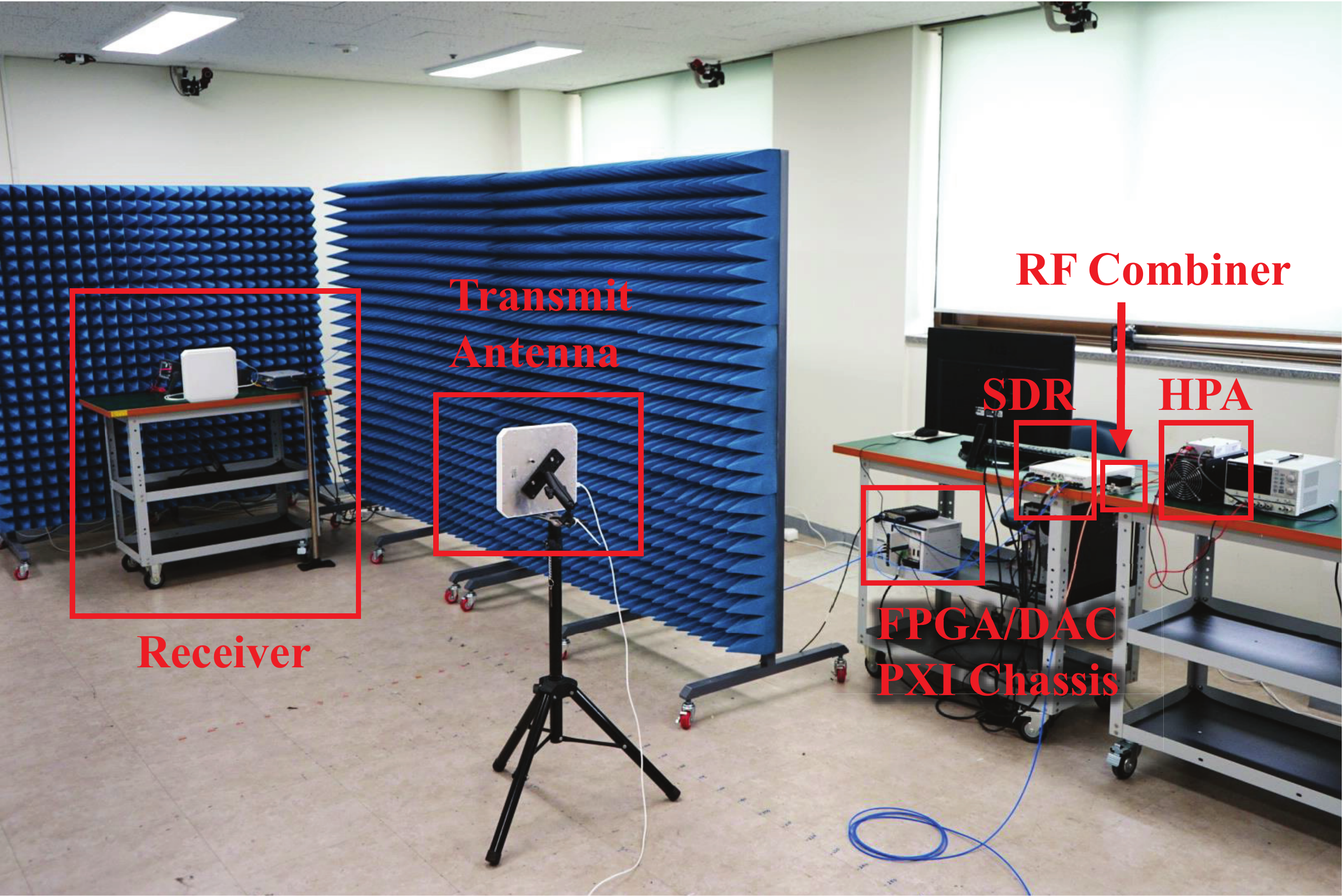}
    }
    \caption{SWIPT prototype testbed.}
\end{figure}

 At the transmitter (Fig.~\ref{fig:txarchitecture}), an SDR platform generates the communication and power signals and transmits them through two different output ports. The power signal is amplified by the high power amplifier (HPA) and combined with the communication signal by using the RF combiner. The combined signal is transmitted and received by circularly polarized panel antennas. At the receiver (Fig.~\ref{fig:rxarchitecture}), the received signal is rectified and decomposed into the power and communication signals by means of the receiver board. We have designed and fabricated the receiver circuit with the one-stage Dickson charge pump as shown in Fig.~\ref{fig:receiverphoto}.
The received power signal is controlled and measured by a source meter.
On the other hand, the received communication signal is amplified by a voltage amplifier, digitized by an ADC, and processed by an FPGA. The transmit and receive codes for the DC-biased OFDM are implemented in the SDR platform and FPGA, respectively, by using the Labview FPGA software for fast signal processing.

\par The sampling rate of the communication signal is 25 MS/s. The FFT size for the DC-biased OFDM is 2048, and the subcarrier spacing is 12.21 kHz. We use 512 subcarriers among the 2048 available subcarriers, and 96 subcarriers around the center and 1440 subcarriers around the edge of the bandwidth are unused. Therefore, the actual bandwidth of the communication signal is about 6.25 MHz.
The transmit power is equally distributed to all active subcarriers. We have implemented the quadrature phase shift keying (QPSK), 16QAM, 64QAM, and 256QAM modulation schemes.

\begin{figure}
    \centering
    \subfigure[QPSK]{
        \label{fig:qpsk4}\includegraphics[width=2.5cm, bb=1in 0.2in 10in 10in] {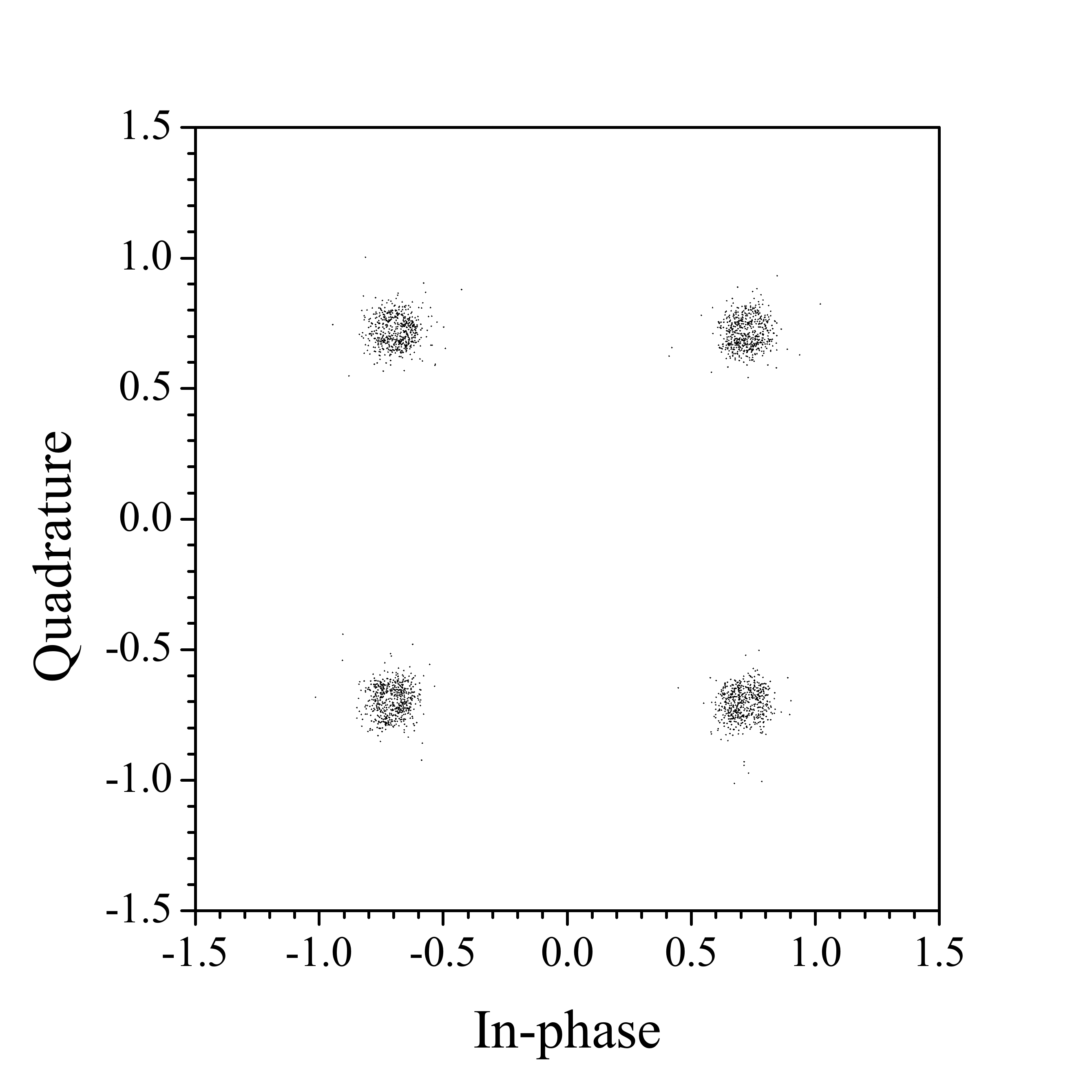}
    }
    \subfigure[16QAM]{
        \label{fig:16qam4}\includegraphics[width=2.5cm, bb=1in 0.2in 10in 10in] {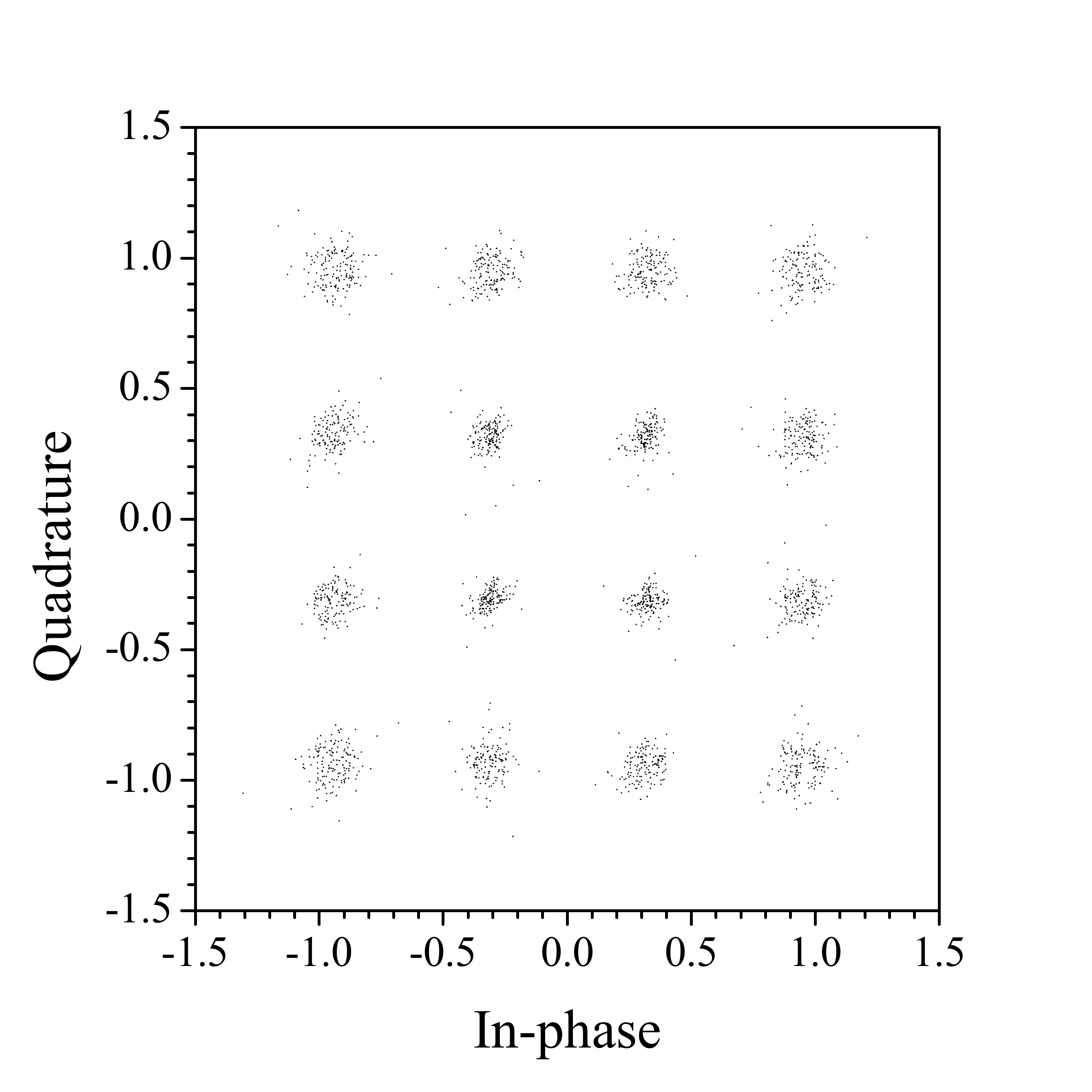}
    }
    \subfigure[64QAM]{
        \label{fig:64qam4}\includegraphics[width=2.5cm, bb=1in 0.2in 10in 10in] {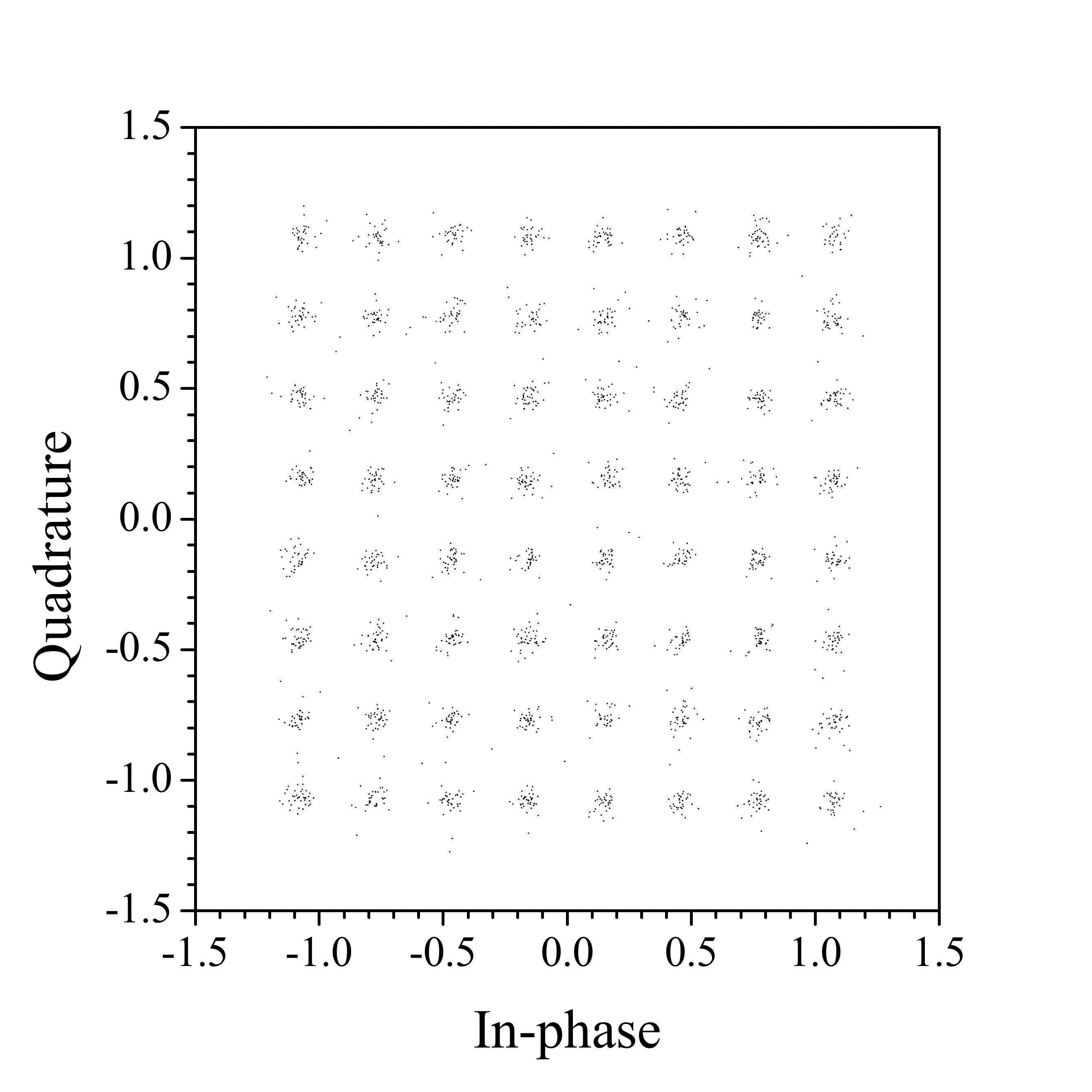}
    }
    \caption{Constellation diagram.}
    \label{fig:constellation}
\end{figure}

We have conducted the experiments of SWIPT with integrated receiver using the testbed shown in Fig.~\ref{fig:swipttestbed1}.
In the experiment, the transmit power of the power signal (i.e., 1 W) is a thousand times higher than that of the communication signal (i.e., 1 mW).
Despite the comparatively small power of the communication signal, the constellation diagrams of the QPSK, 16QAM, and 64QAM modulation schemes at the distance of 4 meters are very clear as shown in Fig.~\ref{fig:constellation}.

\begin{figure}[t]
    \centering
    \subfigure[Bit error rate (BER)]{
        \label{fig:distber}\includegraphics[width=4.1cm, bb=0.6in 0.3in 9.6in 8.1in] {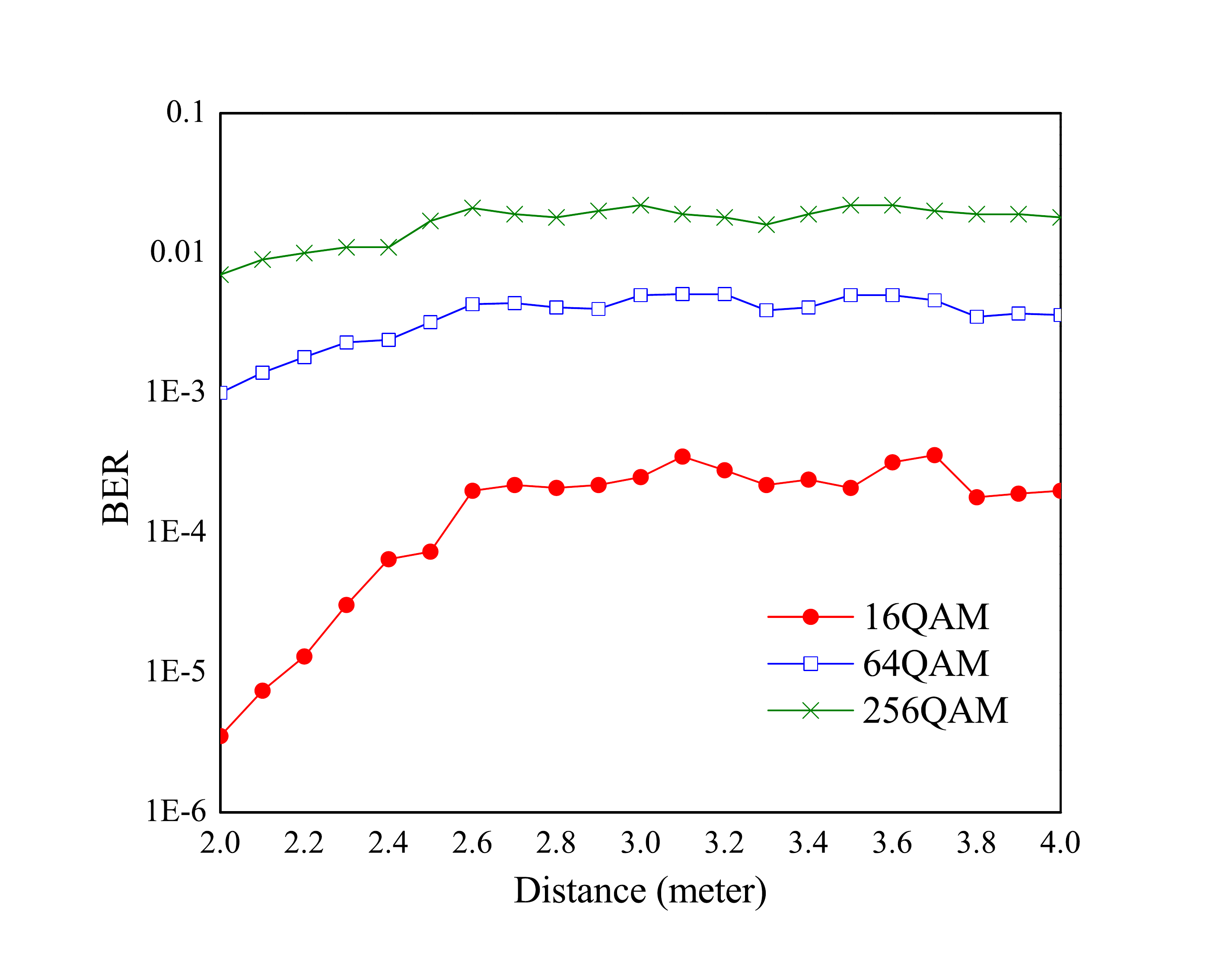}
    }
    \subfigure[Received RF and DC power]{
        \label{fig:distpow}\includegraphics[width=4.1cm, bb=0.6in 0.3in 9.6in 8.1in] {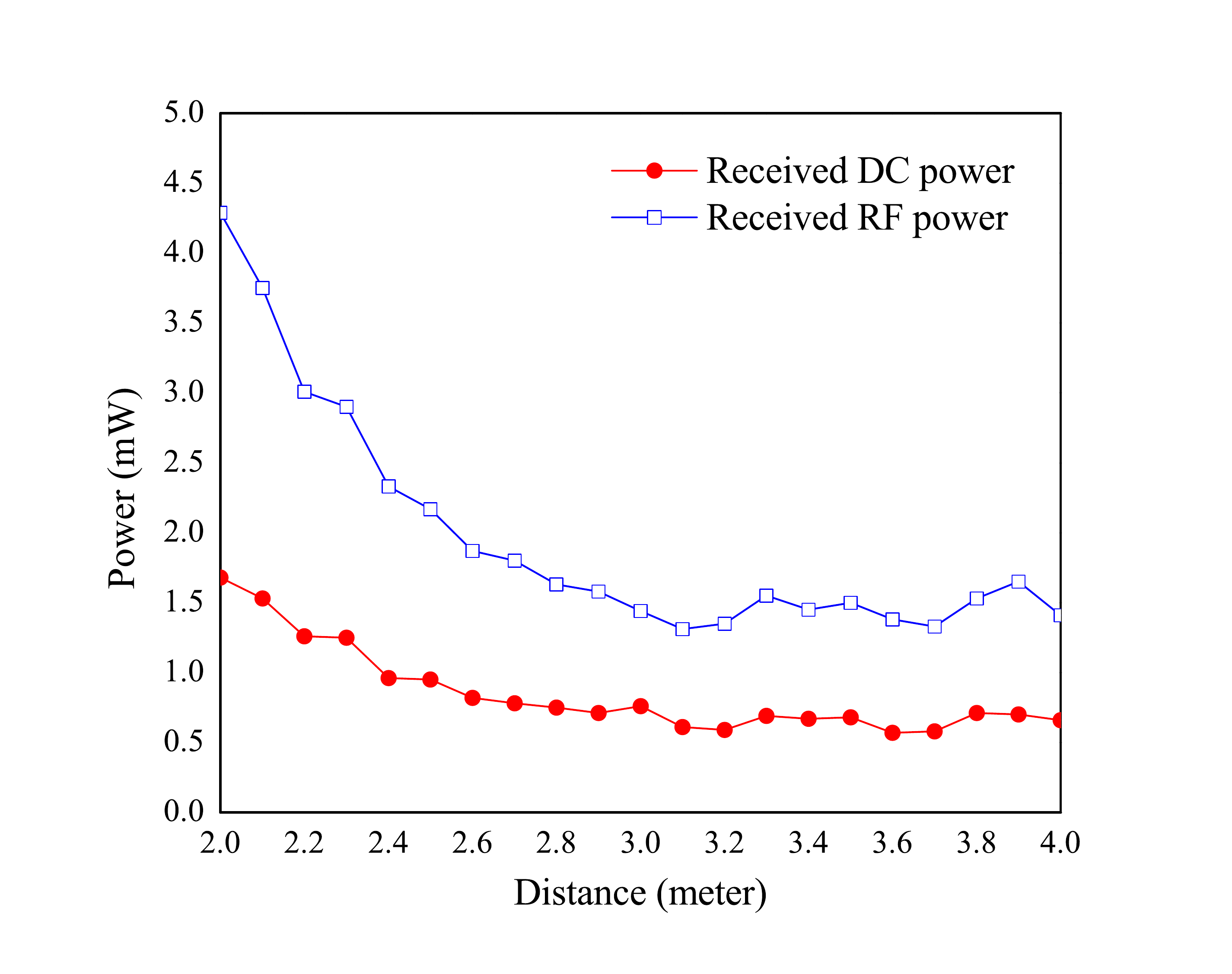}
    }
    \caption{SWIPT performance over distance.}
\end{figure}

The bit error rate (BER) of the modulation schemes over the distance is shown in Fig.~\ref{fig:distber}.
In this figure, we can see that all modulation schemes are usable within the given distance range with just 1 mW transmit power for the communication signal.
In Fig.~\ref{fig:distpow}, we show the received RF and DC power of the power signal over the distance.
The received RF power is measured by the RF power meter directly connected to the receive antenna, whereas the received DC power is the DC power measured by the source meter connected to the power connector of the receiver board.
Fig.~\ref{fig:distpow} shows that the received DC power at 4 m distance is higher than 0.5 mW, which is sufficient power for charging an IoT device.
From Figs.~\ref{fig:distber} and \ref{fig:distpow}, we can see that the SWIPT system testbed is capable of simultaneously transmitting information and power.

\subsubsection{2.4 GHz SWIPT Prototype with TS, PS and Integrated Receivers}

\begin{figure}[t]
	\centering
	\subfigure[]{\includegraphics[width=0.8\linewidth]{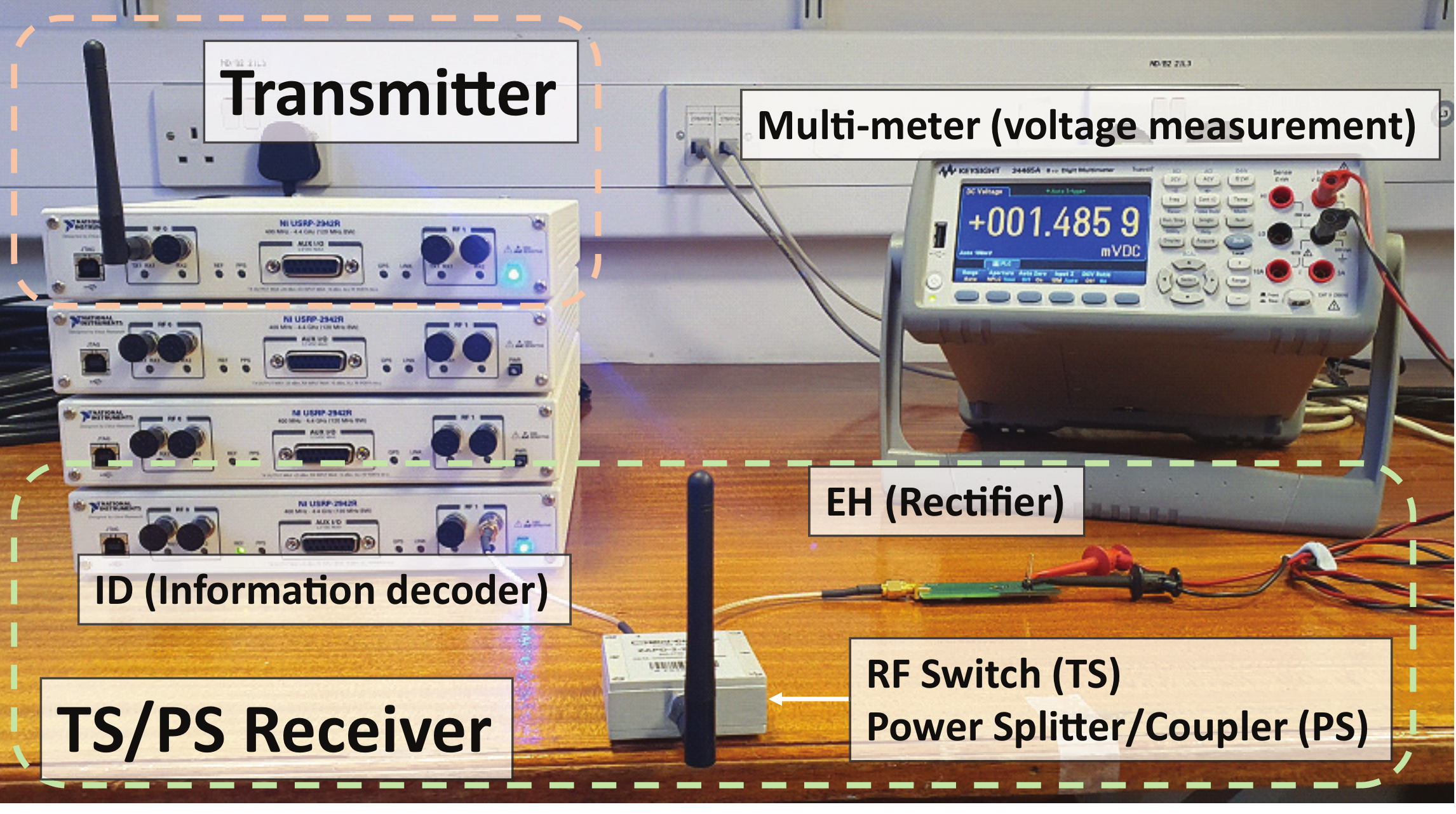}}\\
	\subfigure[]{\includegraphics[width=0.8\linewidth]{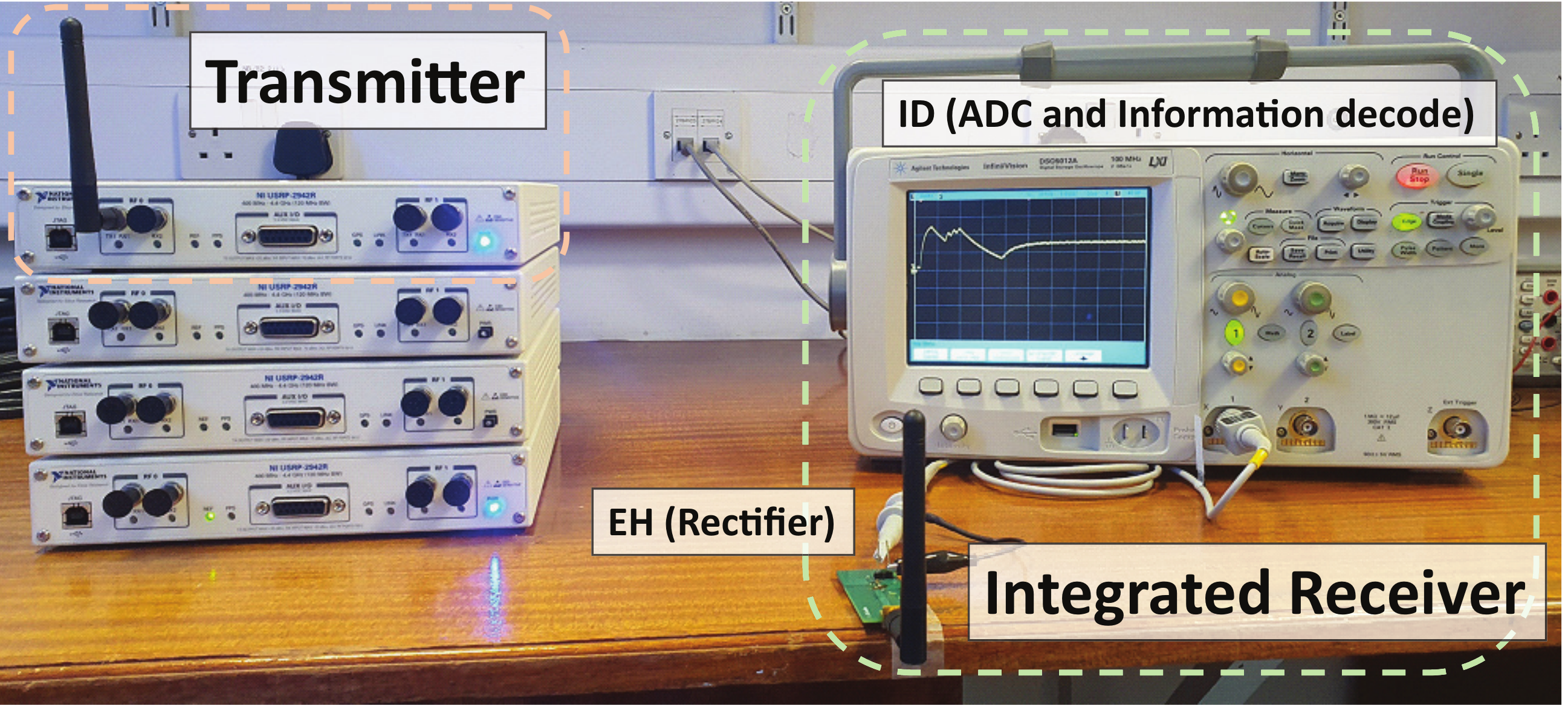}}
	\caption{SWIPT prototype system with (a) PS/TS receiver architecture (b) Integrated receiver architecture.}
	\label{fig_swipt_prototype}
\end{figure}

Previous experiment relied on the integrated receiver architecture. We may wonder how such an architecture compare with the TS and PS receivers. To answer that question, we have built a SISO\footnote{Further enhancements can be obtained by combining with the multi-antenna architectures at both the ET and ER from Section \ref{WPT_section}.} point-to-point SWIPT testbed with three different types of receiver architectures as TS, PS, and integrated receiver. The SWIPT transmitter is implemented using SDR equipment (USRP-2942R), which is able to generate various SWIPT signals from conventional modulations to tailored signal design for SWIPT. The testbed system operates at 2.4GHz carrier frequency, which is the same as WiFi in the ISM band. The transmitter includes features to generate the WPT and communication waveforms and to combine them into SWIPT transmission signals in a time-sharing or superposition manner for the TS and PS receivers and also to generate specially designed SWIPT signal for the integrated receiver. For the TS and PS receivers, a power splitter or an RF switch is located directly behind the receiving antenna to distribute the received signals to both ID and EH blocks in PS and TS manner, respectively. The ID is implemented using additional USRP to demodulate information from the received RF signal and the EH is a simple single diode rectifier to convert RF signals to DC voltage. PS and TS ratios can be adjusted by a control signal for RF switch or by replacing the power splitter/coupler with different coupling ratios. The integrated receiver is implemented using the same EH as for TS/PS receivers and an oscilloscope plays the role of ADC and modem for the ID. We have used Agilent DSO5012A oscilloscope to read DC voltage signal and it supports up to 2GHz sampling rate. Digital input signal to the ID is read from the oscilloscope and processed for demodulation of the information. Meanwhile, the harvested DC power is also measured from the voltage output of the EH. The operation is implemented in LabView, and the entire system is controlled by a host PC. Fig. \ref{fig_swipt_prototype} displays the prototype of SWIPT with different types of receiver architectures.

\par The performance of various signal design methods for each SWIPT receiver architecture was evaluated from both information and power transfer perspectives using the aforementioned prototype and testbed \cite{Kim:2019, Kim:2021}. Here, we present a performance comparison between PS receiver and integrated receiver using various signal designs tailored for each receiver architecture. The PS receiver used in this experiment is equipped with a 10dB power coupler which means 90\% of input RF power is distributed to the EH and the remaining 10\% goes to the ID. In \cite{Kim:2019}, we showed that such a low portion of 10\% of input RF power is enough for information decoding at a coherent information receiver.

\begin{figure}
	\centerline{\includegraphics[width=0.8\columnwidth]{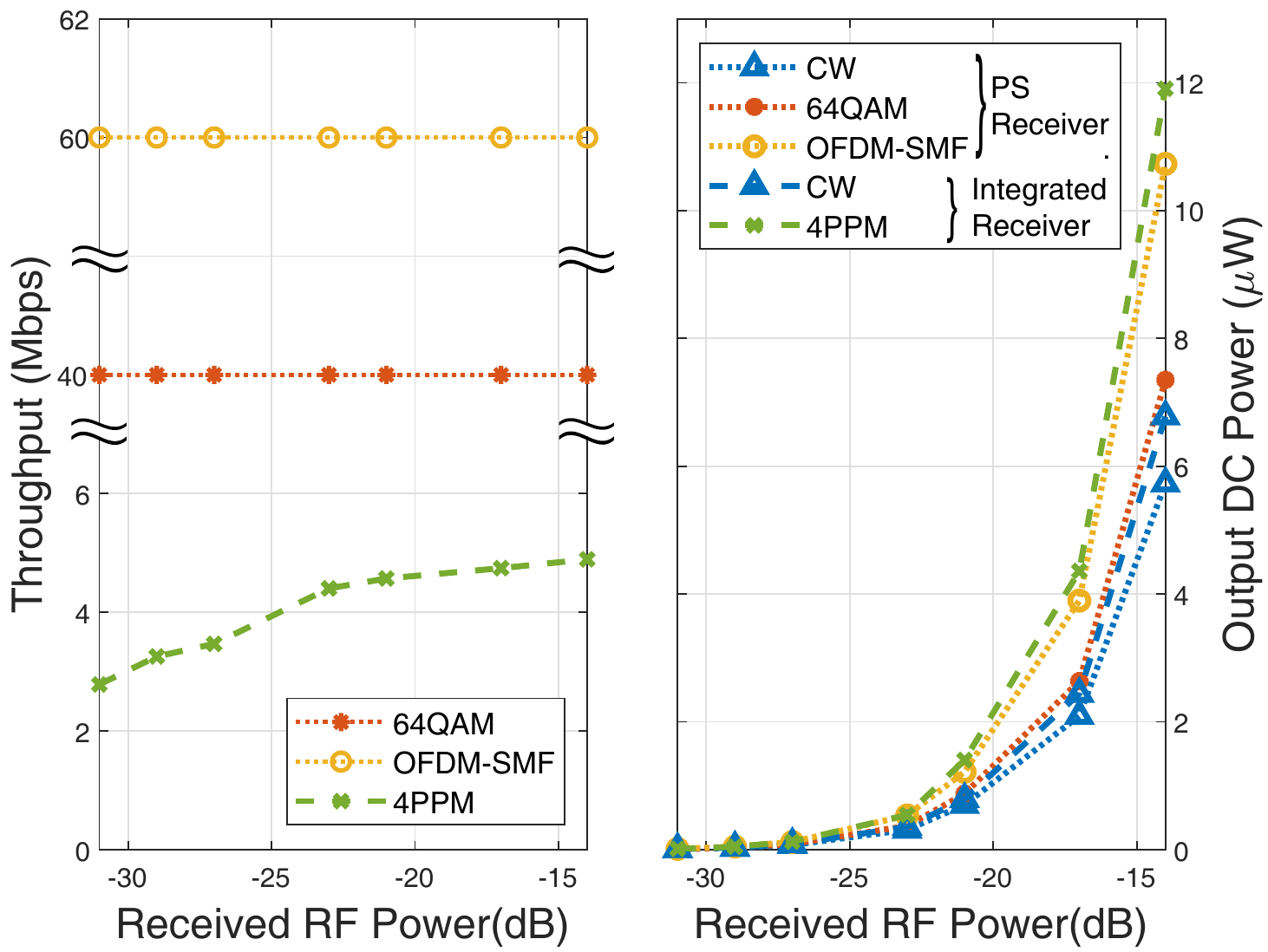}}
	\caption{Throughput and harvested DC power performance of different signals with different receiver architecture in various input RF power levels.}
	\label{perform_swipt}
\end{figure}

Fig. \ref{perform_swipt} displays the performance of throughput and harvested DC power of various signal designs and receiver architectures for several input RF power ranges. We measured the DC output power at the EH and BER at the ID, and the throughput is calculated as $(1-BER) \times maxrate$, where $maxrate$ is the maximum achievable rate of each signal design. The experimented signal's $maxrate$ are 5Mbps for 4-PPM, 40Mbps for 64-QAM, and 60Mbps for superposed OFDM-SMF, with all operating over a 10MHz bandwidth. The experiment ran for one second for each signal and was repeated 100 times, then the measurement results were averaged. Superimposed SWIPT signal (denoted as OFDM-SMF) and conventional 64-QAM modulation signals were used when experimenting the PS receiver architecture. The superimposed SWIPT signal is a combination of WPT and WIT signal with a certain power ratio, and we used OFDM (40subcarriers, 64QAM modulations) as a WIT signal and 8-tone SMF (channel-adaptive low complex multi-tone WPT signal design) as a WPT signal, and the power ratio between WPT and WIT is 9:1. Both signals show no error in the input RF power range of -30 to -15 dBm with the PS receiver, so its throughput is the same as the maximum rate of 40Mbps and 60Mbps, respectively, for both 64-QAM and OFDM-SMF superimposed signals with 10 MHz bandwidth. For the integrated receiver architecture, a 4-PPM signal with 10MHz bandwidth for SWIPT is used. Its maximum rate is 5Mbps and the maximum throughput is achieved at the higher input RF power of -15dBm. The throughput of QAM and OFDM signals is much higher than that of PPM and shows much stable performance in the low-input power region. The ID in the PS receiver (coherent information decoder) for those signals is equipped with LNA, oscillator, and mixers, so it highly outperforms the ID in the integrated receiver, but also consumes significantly more energy. In contrast, the ID in the integrated receiver spends much less energy but its throughput performance is degraded.

\par
To compare more accurately the performance of PS and integrated receivers, CW is added for the power transfer experimental results in Fig. \ref{perform_swipt}. As expected, CW at the integrated receiver shows higher output DC power (with 10 to 20\% gain) over CW at PS receiver. 
The 4-PPM signal at the integrated receiver shows the best power transfer performance gain over CW due to its high PAPR signal design and also does not incur any power loss at ID. The superimposed OFDM-SMF signal at the PS receiver also shows attractive gains but it cannot exceed PPM due to the power loss induced by the PS process. Additionally, CSI is required to generate the OFDM-SMF superimposed signal, so additional power consumption at the ER is required to feedback CSI to the ET.

\begin{figure}
	\centerline{\includegraphics[width=0.8\columnwidth]{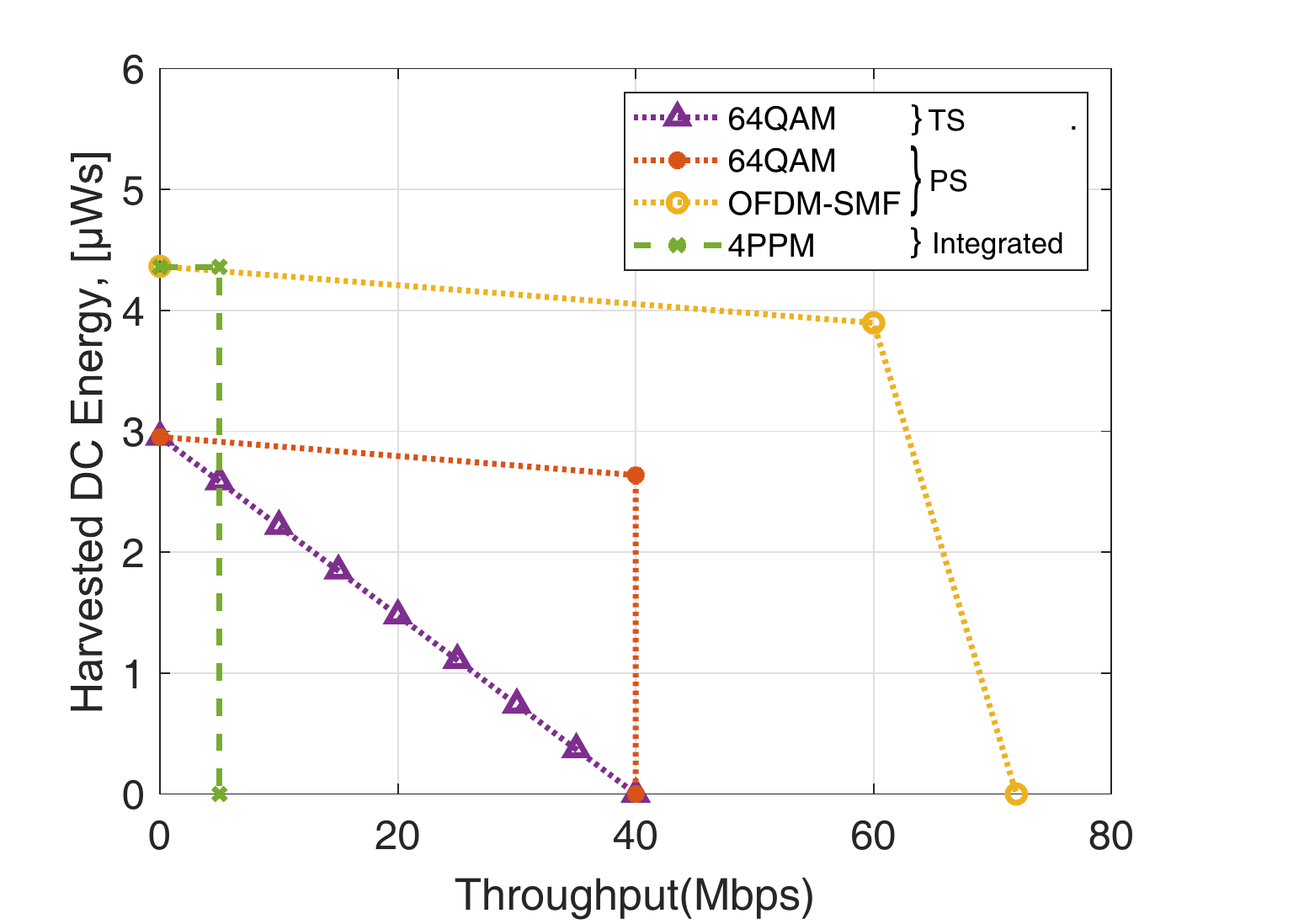}}
	\caption{Throughput-Energy region with various signal designs and TS, PS, and integrated receiver architectures, measured at -17dBm RF input power to the receiving antenna.}
	\label{et_region_swipt}
\end{figure}

\par
Results can also be illustrated using the R-E (or throughput-energy) region of PS, TS, and integrated receivers in Fig. \ref{et_region_swipt}. The throughput-energy region has been measured by sweeping the TS and PS ratios of TS and PS receivers at the fixed RF receiving power level of -17dBm. The same transmit signal of 64QAM modulation has been experimented for both TS and PS receivers. The R-E region of PS is significantly expanded compared to TS due to the PS receiver's capability of information decoding with only a small portion of power to the ID. Additional R-E region expansion is confirmed using superimposed OFDM-SMF relying on the use of multi-carrier and CSIT based channel adaptive design to boost throughput and energy performance, respectively. The PPM signal applied to an integrated receiver exhibits completely different performance characteristics with large power transfer performance but lower throughput. However, it should be noted that the high throughput in TS and PS receivers is made possible by the large power consumption at the ID. Several hundreds of mW of power are consumed for decoding those OFDM WiFi-like information signals at TS and PS receivers \cite{Benali:2016}. Additional power consumption for acquiring CSIT is required for the superimposed OFDM-SMF signal. On the other hand, the non-coherent PPM signal for the integrated receiver demands less than a few mW for information decoding \cite{Pulkkinen:2020}. Overall, the PS receiver can potentially achieve a wider R-E region at the cost of large energy consumption. In contrast, the integrated architecture achieves a smaller region but with less than 1\% of the energy consumption of the PS receiver. 

\begin{figure}
	\centerline{\includegraphics[width=0.8\columnwidth]{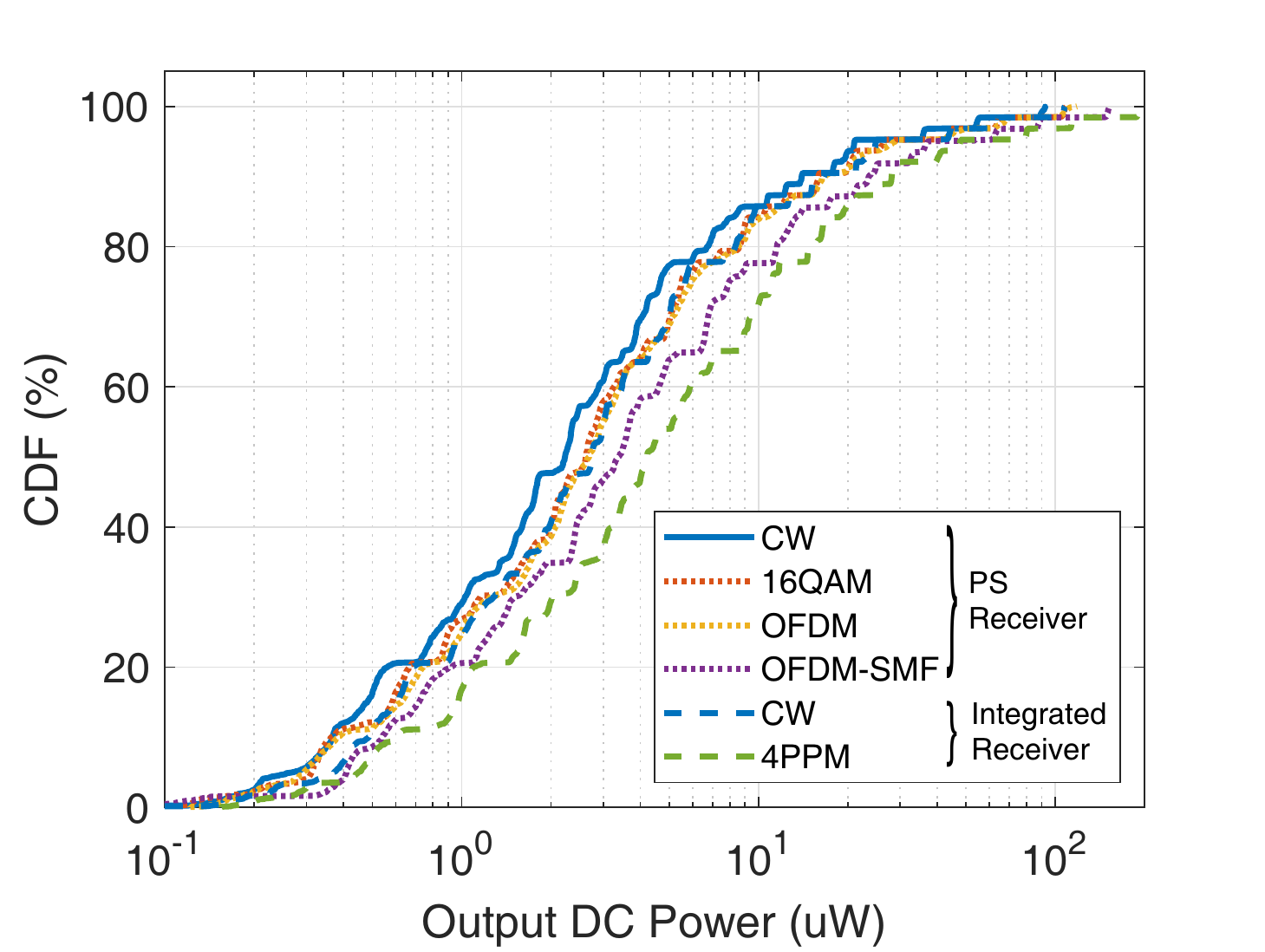}}
	\caption{CDF of output DC power ($P_{\dc}^r$) measurement results at different distances from 0.5 to 5.3 m with two different receiver and various signal designs.}
	\label{cdf_swipt}
\end{figure}

\par
Power transfer performance experiments in many different wireless environments were also carried out using the PS and integrated SWIPT receivers. We have measured the output DC power at 70 different locations in our testbed using various signal design methods such as CW, QAM, OFDM and PPM. The distance between the transmitter and the receiver varies between 0.5 and 5.3 meters, and the transmission power is fixed at 27dBm. Cumulative distribution functions (CDF) of measured DC output power with various modulation signals and receiver architecture are presented in Fig. \ref{cdf_swipt}.
We clearly observe that the CDFs of integrated receiver are shifted to the right compared to the PS receiver, which means that the integrated receiver enables to harvest more energy. These observations are consistent with the results in Fig. \ref {perform_swipt} and \ref{et_region_swipt}. 

\section{Conclusions and Future Research}\label{conclusions}

The paper has provided an overview of fundamental building blocks of modern WPT and WIPT systems and demonstrated their feasibility using prototype and experiments. 

One first conclusion of the paper is that the integration of emerging ideas from RF, communications and signal processing is instrumental in boosting the performance of WPT and in enabling a fully closed-loop and channel-controlled WPT architecture capable to adapt to any wireless environment and to engineer the wireless channel for maximum WPT (and WIPT) performance. Nevertheless, the adoption and design of large scale system and WPT-based future networks calls for a stronger integration of various expertise ranging from RF design, circuit, antenna, sensing, computing, communications, networking, and signal processing. 

A second conclusion is that the inter-disciplinary design of WPT and WIPT leads to new theoretical and experimental design challenges for engineers that are worth considering in future research. 
\par \textit{First}, various societies have an important role to play in WPT for future networks, but the techniques need to be developed in light of the physics and hardware constraints of WPT. Recall for instances that techniques like waveform, receive combiner, joint waveform and beamforming are deeply rooted in the harvester nonlinearity and only appeared to light once the nonlinearity is accounted for in the signal design. This calls communication engineers for abandoning naive and oversimplified models and accounting for various sources of nonlinearities and nonidealities in the signal and system optimization (e.g., \cite{Zhang:2021}) and RF engineers to account for multipath fading and advanced adaptive and closed-loop WPT architecture inspired by modern wireless communication systems. A genuine integration of RF, communication and signal processing is crucial for an end-to-end system optimization and efficiency maximization. Much work remains also to be done to prototype and experiment those end-to-end system optimization solutions in larger networks composed of multiple nodes (and therefore going beyond the single-user or point-to-point experiments discussed in this paper).

\par \textit{Second}, the prototyping and experimental results discussed in this paper are strongly rooted in a ``model and optimize'' design approach of WPT and WIPT.  Indeed, the approach relies on RF, optimization and communication theories to derive efficient and implementable strategies for WPT/WIPT. This has shown significant benefits as demonstrated by the prototyping and experimental results. Nevertheless such an approach also has some limitations: 1) difficulty to analytically account for various sources of nonlinearity and nonidealities, 2) mathematically extremely challenging and computationally intensive problems to solve. The lack of tractable mathematical models of the entire WPT chain (accounting for transmitter-receiver including HPA, RF impairments, rectenna, DC-to-DC converter, imperfect matching, non-ideal low pass filter, load, parasitics, etc), as well as algorithms to solve WPT/WIPT signal/system optimizations in general settings is a bottleneck towards efficient WPT/WIPT designs. In this regard, machine learning (ML) is instrumental since they can be used to circumvent these modelling and algorithmic challenges. This calls for a ``learning'' approach for WPT and WIPT advocated in \cite{Clerckx:2021,Varasteh:2020_TCOM}. A learning approach towards WPT and WIPT signal and system design, and an \emph{integrated} approach that leverages the complementarity and synergy of the learning and the model-and-optimize approaches, remain largely uncharted research territories, both on the theoretical and experimental sides.

\par \textit{Third}, the biggest hurdle in the widespread adoption of WPT remains its low power transfer efficiency. Although a large-sized transmit antenna arrays makes a smaller focal spot, it restricts the applicability of the WPT system and increases the manufacturing cost. The form factor of a sensor device is restricted by the application demands as well. Therefore, another opportunity to further increase the efficiency is to use higher frequency with shorter wavelength. Generally, high frequency circuits pose many design challenges, and more importantly, the number of antenna elements and RF paths in the power beacon is proportional to the frequency squared, which potentially leads to very high manufacturing cost. Thus, the prerequisite for the prevalence of wireless-powered networks is the development of high frequency (e.g., millimeter wave range) WPT systems with a moderate cost. Upcoming proliferation of millimeter wave communications and massive MIMO technologies is expected to push down the cost of high frequency WPT systems. Another direction can be developing a low cost alternative architecture for synthesizing a focused beam other than the phased antenna array, for example, a tunable metasurface-based power transmitter. 

\par \textit{Fourth}, long-term exposure of a human body to a strong radio wave potentially has a risk of adverse health effect even if it is not proven by concrete evidences. To prevent such a harmful effect, the regulatory bodies limit specific absorption rate (SAR), which measures the absorption rate of the radio wave by a human body. Strong electromagnetic fields at a focal point of the microwave beam is very likely to give rise to a violation of the SAR limit. Therefore,  WPT and WIPT can then be designed and optimized to account for such safety constraints, as in \cite{Zhang:2020}, and countermeasures could be prepared. One option is to actively detect the presence of a human body by using some external sensors such as vision, depth, and passive infrared (PIR) sensors. Then, the transmitter can suppress a radio wave emission towards the direction of the detected human body by completely turning off, steering a beam away, or forming a null.

\par \textit{Fifth}, more studies need to be conducted to reduce the power consumption and device complexity by combining the RF communications and WPT. In the downlink, the transmitter transfers power and data via the same radio wave by means of the SWIPT technique. To that end, a modulated source signal is fed into the phased antenna array to piggyback data on a power RF signal. Then, the sensor device splits the receive power into the rectifier and the decoding circuits, or simply detects the envelope of the receive signal for data decoding. In the uplink, the sensor device can use backscatter communications that reflect the incident wave to generate an uplink modulated signal. Backscatter communications consume very small power since a sensor device does not have to create an RF signal from its own energy. The interplay between closed-loop WPT/SWIPT and backscatter communication (for both channel acquisition and data transmission) need further investigation. However, short communication distance and instability are the shortcomings that need to be overcome.

\par \textit{Sixth}, the sensor device has a great impact on the power transfer efficiency due to the polarization and antenna gain. If linearly-polarized (LP) antennas are used, the mismatch in the polarization vectors of antennas deteriorates the polarization loss factor, resulting in low power transfer efficiency. To overcome this difficulty, circularly-polarized (CP) antennas are used or the transmit antenna array with the ability of controlling a polarization vector can be designed.

If a high gain antenna is used in the sensor device, it well receives the radio wave only from a specific direction. On the other hand, a low gain antenna such as a dipole antenna has a wide reception angle, but has a small effective antenna aperture. A high flexibility at the sensor device should be carefully taken into account to meet the application needs.

Moving to multiple antennas at the sensor, the experimental validation of the benefits and drawbacks of closed-loop MIMO WPT with RF and DC combining remains uncharted. Though RF combining appears appealing, loss may decrease its efficiency. 

\par \textit{Seventh}, multiple distributed antennas can be deployed to extend the WPT system coverage area. Sophisticated solutions are nevertheless needed to establish a means to attain frequency and phase synchronization across different antennas, and to adaptively control the phases of radio waves for making sure they constructively interfere at a sensor device. Scaling up the number of antennas, WPT DAS can leverage advances in joint transmission and cell-free massive MIMO in communications. 

\par \textit{Eighth}, a proper channel model for the theoretical WPT studies should be established based on the physical characteristics of RF WPT (i.e., {\it physics-based} channel model). Typically, the required receive power for WPT is several orders of magnitude higher than that for communications. This makes NLoS deployment particularly challenging and maybe less practical for WPT. Therefore, the {\it random} channel model that has been used for describing NLoS and rich scattering environment may not be as useful in WPT as in communications since the received power may be too low in such an environment. To achieve a reasonable power transfer efficiency, LoS is needed to be secured and the distance should not be too large. For multi-antenna WPT, the angular-domain channel of the LoS path should be properly modeled so that the microwave beam can be formed towards the receiver. Moreover, beam focusing is possible if the transmitter and receiver are within the radiative near field. In the radiative near-field region, the transmitted microwave beam can be focused onto the receiver to enhance the power transfer efficiency, which is called beam focusing. To describe this scenario, a radiative near-field channel model is required \cite{Park:2021}. Recently, the communications in mmWave or sub-THz frequencies are actively studied. These high frequency channels have very similar characteristics to the WPT channel, e.g., limited distance, dominance of LoS, importance of beamforming, and consideration for the radiative near field. Therefore, it is possible to use the channel model developed for mmWave or sub-THz, for example, the clustered delay line (CDL) channel model of 3GPP.

\par It is hoped that the techniques, prototypes, and challenges presented here will help inspire future research in this promising area and pave the way for designing and implementing efficient WPT, WIPT, and wireless-powered systems and networks in the future.

\ifCLASSOPTIONcaptionsoff
  \newpage
\fi

\begin{IEEEbiography}[{\includegraphics[width=1in,height=1.25in,clip,keepaspectratio]{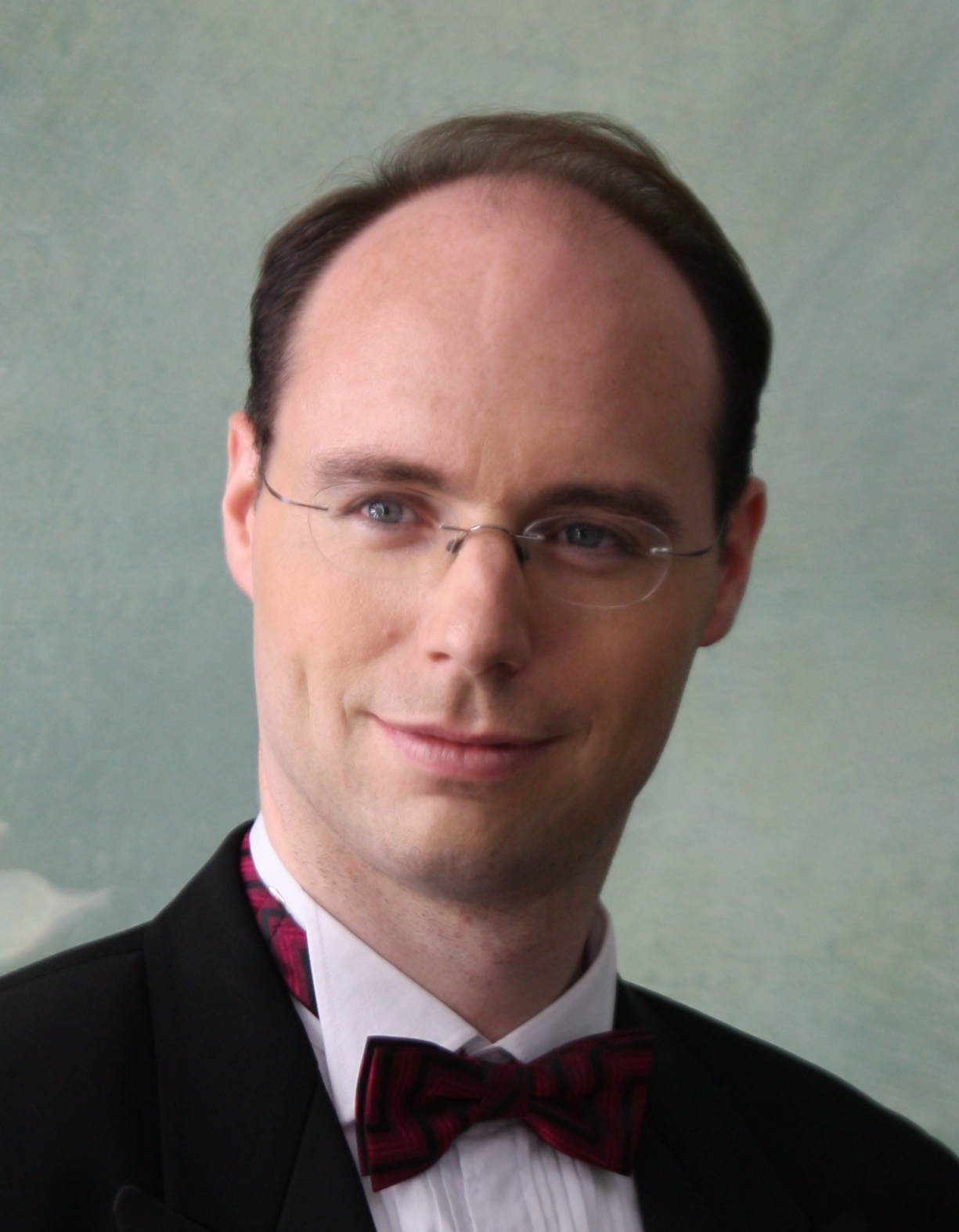}}]
{Bruno Clerckx} (Fellow, IEEE) is a (Full) Professor, the Head of the Wireless Communications and Signal Processing Lab, and the Deputy Head of the Communications and Signal Processing Group, within the Electrical and Electronic Engineering Department, Imperial College London, London, U.K. He received the M.S. and Ph.D. degrees in Electrical Engineering from the Université Catholique de Louvain, Louvain-la-Neuve, Belgium, in 2000 and 2005, respectively. From 2006 to 2011, he was with Samsung Electronics, Suwon, South Korea, where he actively contributed to 4G (3GPP LTE/LTE-A and IEEE 802.16m) and acted as the Rapporteur for the 3GPP Coordinated Multi-Point (CoMP) Study Item. Since 2011, he has been with Imperial College London, first as a Lecturer from 2011 to 2015, Senior Lecturer from 2015 to 2017, Reader from 2017 to 2020, and now as a Full Professor. From 2014 to 2016, he also was an Associate Professor with Korea University, South Korea, and from 2021 to 2022, he is a visiting Professor at Seoul National University, South Korea. He also held various long or short-term visiting research appointments at Stanford University, EURECOM, National University of Singapore, The University of Hong Kong, Princeton University, The University of Edinburgh, The University of New South Wales, and Tsinghua University. 
He has authored two books on “MIMO Wireless Communications” and “MIMO Wireless Networks”, 250 peer-reviewed international research papers, and 150 standards contributions, and is the inventor of 80 issued or pending patents among which 15 have been adopted in the specifications of 4G standards and are used by billions of devices worldwide. His research spans the general area of wireless communications and signal processing for wireless networks. He has been a TPC member, a symposium chair, or a TPC chair of many symposia on communication theory, signal processing for communication and wireless communication for several leading international IEEE conferences. He was an Elected Member of the IEEE Signal Processing Society “Signal Processing for Communications and Networking” (SPCOM) Technical Committee. He served as an Editor for the IEEE TRANSACTIONS ON COMMUNICATIONS, the IEEE TRANSACTIONS ON WIRELESS COMMUNICATIONS, and the IEEE TRANSACTIONS ON SIGNAL PROCESSING. He has also been a (lead) guest editor for special issues of the EURASIP Journal on Wireless Communications and Networking, IEEE ACCESS, the IEEE JOURNAL ON SELECTED AREAS IN COMMUNICATIONS, the IEEE JOURNAL OF SELECTED TOPICS IN SIGNAL PROCESSING, and the PROCEEDINGS OF THE IEEE. He was an Editor for the 3GPP LTE-Advanced Standard Technical Report on CoMP. He received the prestigious Blondel Medal 2021 for exceptional work contributing to the progress of Science and Electrical and Electronic Industries. He is a Fellow of the IEEE and an IEEE Communications Society Distinguished Lecturer 2021-2022.
\end{IEEEbiography}

\begin{IEEEbiography}[{\includegraphics[width=1in,height=1.25in,clip,keepaspectratio]{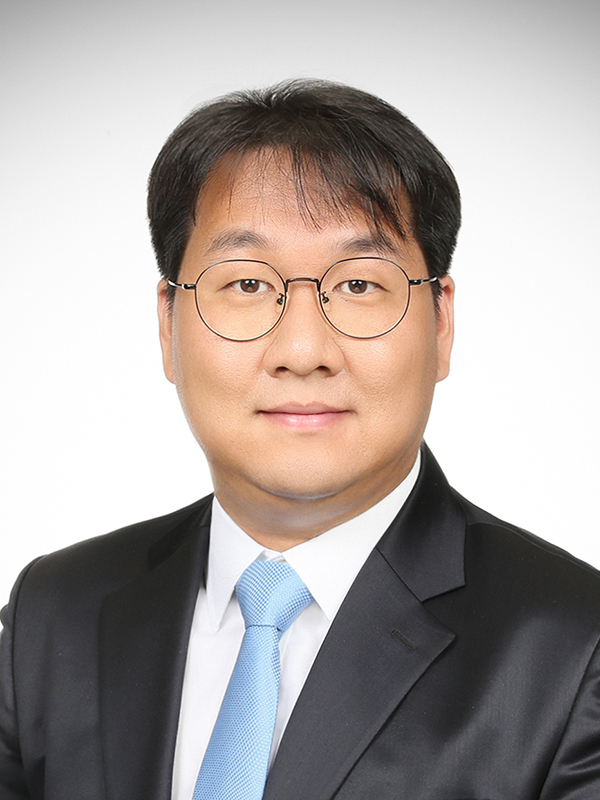}}]
{Junghoon Kim} (Member, IEEE) received the B.Sc. degree in electronic and electrical engineering from Hongik University, Seoul, South Korea, in 2008, and the M.Sc. degree in telecommunications from University College London, London, U.K., in 2015, and the Ph.D. degree in electrical and electronic engineering from Imperial College London, London, U.K., in 2020. 
He is currently a Postdoctoral Research Associate with Imperial College London. From 2008 to 2014, he was with Mobile Communication Division, Samsung Electronics, Suwon, South Korea, as an R\&D engineer.
His current research interests include RF energy harvesting, wireless information and power transfer, multiple-input and multiple-output systems, and Internet-of-Things.
\end{IEEEbiography}

\begin{IEEEbiography}[{\includegraphics[width=1in,height=1.25in]{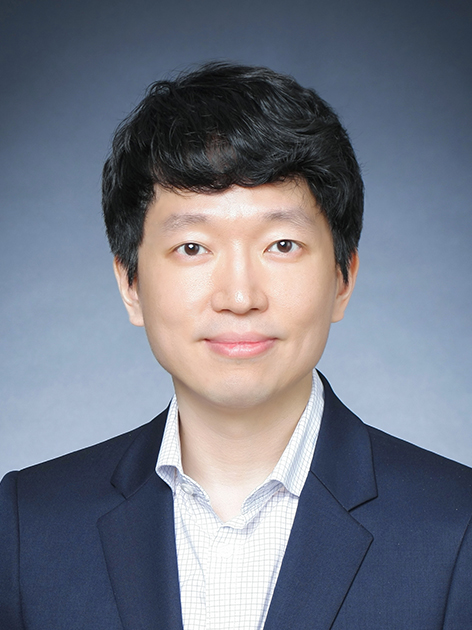}}]
{\textbf{Kae Won Choi}} (Senior Member, IEEE) received the B.S. degree in Civil, Urban, and Geosystem Engineering in 2001, and the M.S. and Ph.D. degrees in Electrical Engineering and Computer Science in 2003 and 2007, respectively, all from Seoul National University, Seoul, Korea. From 2008 to 2009, he was with Telecommunication Business of Samsung Electronics Co., Ltd., Korea. From 2009 to 2010, he was a postdoctoral researcher in the Department of Electrical and Computer Engineering, University of Manitoba, Winnipeg, MB, Canada. From 2010 to 2016, he was an assistant professor in the Department of Computer Science and Engineering, Seoul National University of Science and Technology, Korea. In 2016, he joined the faculty at Sungkyunkwan University, Korea, where he is currently an associate professor in the College of Information and Communication Engineering. His research interests include RF energy transfer, metasurface communication, visible light communication, cellular communication, cognitive radio, and radio resource management. He has served as an editor of IEEE Communications Surveys and Tutorials from 2014, an editor of IEEE Wireless Communications Letters from 2015, an editor of IEEE Transactions on Wireless Communications from 2017, and an editor of IEEE Transactions on Cognitive Communications and Networking from 2019.
\end{IEEEbiography}

\begin{IEEEbiography}[{\includegraphics[width=1in,height=1.25in,clip,keepaspectratio]{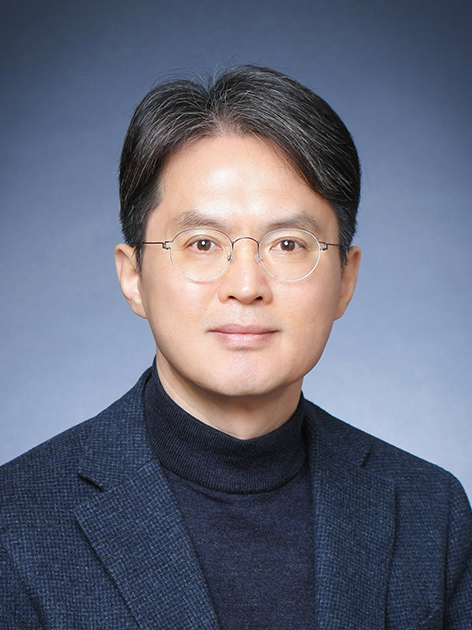}}]
{Dong In Kim} (Fellow, IEEE) received the Ph.D. degree in electrical engineering from the University of Southern California, Los Angeles, CA, USA, in 1990. He was a Tenured Professor with the School of Engineering Science, Simon Fraser University, Burnaby, BC, Canada. Since 2007, he has been an SKKU-Fellowship Professor with the College of Information and Communication Engineering, Sungkyunkwan University (SKKU), Suwon, South Korea. He is a Fellow of the Korean Academy of Science and Technology and a Member of the National Academy of Engineering of Korea. He has been a first recipient of the NRF of Korea Engineering Research Center in Wireless Communications for RF Energy Harvesting since 2014. He has been listed as a 2020 Highly Cited Researcher by Clarivate Analytics. From 2001 to 2020, he served as an editor and an editor at large of Wireless Communications I for the IEEE TRANSACTIONS ON COMMUNICATIONS. From 2002 to 2011, he also served as an editor and a Founding Area Editor of Cross-Layer Design and Optimization for the IEEE TRANSACTIONS ON WIRELESS COMMUNICATIONS. From 2008 to 2011, he served as the Co-Editor-in-Chief for the IEEE/KICS JOURNAL OF COMMUNICATIONS AND NETWORKS. He served as the Founding Editor-in-Chief for the IEEE WIRELESS COMMUNICATIONS LETTERS, from 2012 to 2015. He was selected the 2019 recipient of the IEEE Communications Society Joseph LoCicero Award for Exemplary Service to Publications. He is the General Chair for IEEE ICC 2022 in Seoul.
\end{IEEEbiography}

\end{document}